\documentclass[%
 aip,
amsmath,amssymb,
citeautoscript, 
reprint,%
]{revtex4-1}

\usepackage{hyperref}
\hypersetup{
    colorlinks=true,
    linkcolor=blue,
    citecolor=blue,
}

\usepackage{graphicx}
\usepackage{dcolumn}
\usepackage{bm}
\usepackage{esint}
\usepackage{dutchcal}
\usepackage{xcolor, soul}
\sethlcolor{yellow}
\usepackage[utf8]{inputenc}
\usepackage[T1]{fontenc}
\usepackage{dutchcal}
\usepackage{mathptmx}
\usepackage{etoolbox}
\allowdisplaybreaks

\newcommand{\blue}[1]{#1}

\newcommand{\violet}[1]{#1}

\newcommand{\bmcal}[1]{\boldsymbol{\mathcal{#1}}}

\newcommand{\fullFRAC}[2]{\frac{\mathrm{d}#1}{\mathrm{d}#2}}
\newcommand{\partialFRAC}[2]{\frac{\partial#1}{\partial#2}}
\newcommand{\rsub}[1]{_\mathrm{#1}}
\newcommand{\rsup}[1]{^\mathrm{#1}}
\newcommand{\dbar}[1]{\bar{\bar{#1}}}
\newcommand{\RE}[1]{\mathrm{Re}\left\{#1\right\}}
\newcommand{\IM}[1]{\mathrm{Im}\left\{#1\right\}}
\newcommand{\OF}[2]{(\mathbf{#1},{#2})}
\newcommand{\COPYRIGHT}[3]{\violet{Adapted with permission from #1.\cite{#2} Copyright #3}}

\makeatletter
\def\@email#1#2{%
 \endgroup
 \patchcmd{\titleblock@produce}
  {\frontmatter@RRAPformat}
  {\frontmatter@RRAPformat{\produce@RRAP{*#1\href{mailto:#2}{#2}}}\frontmatter@RRAPformat}
  {}{}
}%
\makeatother
\begin{document}


\title{Temporal coupled-mode theory in nonlinear resonant photonics: From basic principles to contemporary systems with 2D materials, dispersion, loss, and gain}
\author{Thomas Christopoulos}
    \altaffiliation[cthomasa]{@ece.auth.gr}
    \affiliation{School of Electrical and Computer Engineering, Aristotle University of Thessaloniki, 54124 Thessaloniki, Greece}
    \affiliation{Theoretical and Physical Chemistry Institute, National Hellenic Research Foundation, 11635 Athens, Greece}
\author{Odysseas Tsilipakos}
    \altaffiliation[otsilipakos]{@eie.gr}
    \affiliation{Theoretical and Physical Chemistry Institute, National Hellenic Research Foundation, 11635 Athens, Greece}
\author{Emmanouil E. Kriezis}
    \altaffiliation[Corresponding]{ author.}
    \affiliation{School of Electrical and Computer Engineering, Aristotle University of Thessaloniki, 54124 Thessaloniki, Greece}
    \email{mkriezis@auth.gr}

\date{\today}

\begin{abstract}

    Temporal coupled-mode theory (CMT) is an acclaimed and widely used theoretical framework for modeling the continuous wave (CW) response and temporal dynamics of any integrated or free-space photonic resonant structure. 
    It was initially employed to understand how energy is coupled into and out of a cavity and how it is exchanged between different resonant modes.
    In the 30 years that followed its establishment, CMT has been expanded to describe a broad range of nonlinear interactions as well (self- and cross-phase modulation, saturable absorption, frequency generation, gain, etc.).
    In this tutorial, we thoroughly present the basic principles and the evolution of CMT throughout the years, showcasing its immense capabilities for the analysis and design of linear and nonlinear resonant photonic systems. 
    Importantly, we focus on examples of modern, open nanophotonic resonators incorporating contemporary bulk or sheet (2D) materials that may be lossy and dispersive. 
    For each linear/nonlinear effect under study we follow a meticulous, step-by-step approach, starting from an accurate model of the physical phenomenon and proceeding to its introduction in the CMT framework all the way to the efficient solution of the resulting system of equations.  
    Our work highlights the merits of CMT as an efficient, accurate, and versatile theoretical tool. 
    We envision that it can serve both as an introductory reference for any reader, as well as a comprehensive handbook on how to incorporate a broad range of linear and nonlinear effects in the CMT framework.   
\end{abstract}

\maketitle

\section{Introduction}\label{sec:Intro}

    \emph{Temporal coupled-mode theory} (CMT) is a powerful theoretical tool in the field of optics and photonics, allowing to efficiently analyze and design resonant systems, as well as to probe the physics of (nonlinear) cavities, their mutual coupling, and the exchange of energy with the outside world.\cite{HausBook,Fan2002a,Bravo-abad2007} Photonic resonators are ubiquitous, they range from simple two-mirror cavities (Fabry-P\'erot resonators) to micro- and nano-scale structures with elaborate geometries (integrated rings, discs and toroids, nanocubes, etc.), and are essential components in a wide range of optical devices, including filters, switching and routing elements, lasers and sensors.\cite{VahalaBook}
    CMT offers a computationally effective, systematic, and intuitive approach for characterizing the interactions between different modes within these resonators and predicting their temporal dynamics. 
    Thus, it has become an indispensable tool in the design and optimization of optical resonators, enabling the development of novel photonic technologies with enhanced performance.

    The initial version of temporal CMT was developed in the 1980s.\cite{HausBook,Haus1991} In a nutshell, CMT provides the means to simplify the description of any resonant system with arbitrary material composition and geometric configuration. Through CMT, such a distributed (3D) system can be treated as a lumped (0D) harmonic oscillator, thereby reducing the mathematical representation to an ordinary differential equation (ODE) with respect to time, which can be efficiently solved to study the temporal response of the system. The physical quantities characterizing the cavity (e.g., resonant frequency and damping) are rigorously extracted from the physical system and become coefficients in the time-dependent ODE. 
    Importantly, multiple wavelength channels and additional linear or nonlinear phenomena 
    can be introduced by means of additional ODEs and extra terms; the resulting system is still readily solvable.\cite{Rodriguez2007,Yang2007,Maes2009,Ramirez2011,Tsilipakos2016} Apart from being computationally advantageous, CMT allows  for gaining interesting physical insight by inspecting the simpler ODEs that emerge and assessing in detail the effect of each term/phenomenon by artificially switching it on or off.\cite{Barclay2005} In addition, it allows for deriving intuitive design rules that can be followed to optimize the response of the resonant system.\cite{Tsilipakos2014} Finally, CMT can be used to describe resonant systems in different branches of wave physics, such as acoustic resonators.\cite{Koutserimpas2019}

    Over the years, CMT has proven quite flexible. Initially employed for bulky resonators in the early years of photonics,\cite{Haus1991} it has been progressively applied to micro- and nano-cavities\cite{Bravo-abad2007} with complex geometries and exotic material composition, including lossy and dispersive media,\cite{Tsilipakos2013} materials with non-instantaneous nonlinearities\cite{Wang2013a,Tsilipakos2016} or gain,\cite{Chua2011,Nousios2023} and 2D photonic materials.\cite{Christopoulos2016a} 
    In this tutorial, we aim to coherently present the development of CMT throughout the years, starting from the main elements of the technique and proceeding to showcase its massive capabilities for the analysis of contemporary linear and, importantly, nonlinear resonant photonic systems.
    
    The rest of the tutorial is organized as follows: 
    In Section~\ref{sec:LinearCMT}, we present the initial development of CMT, focusing on linear cavities and their coupling with each other and the outside world. We also establish the range of validity for CMT. 
    Section~\ref{sec:CMTParamsCalculations} includes methods and approaches that can be used to calculate key parameters that are necessary for building the CMT framework. 
    Section~\ref{sec:InstantNonLinearCMT} focuses on nonlinearities. Various nonlinear effects are examined and the means to include them into the CMT formalism is presented in detail, respecting in each case the underlying physics. 
    Section~\ref{sec:NoninstantNonLinearCMT} focuses on nonlinear phenomena as well, but those with non-instantaneous response that are described by additional dynamic equations and how they can be incorporated in the CMT framework. 
    Section~\ref{sec:Gain} is dedicated to systems with gain, i.e., to laser cavities, and a general methodology to treat such systems is presented, including the most general case of Class C lasers. 
    Section~\ref{sec:StabAnalysis} briefly discusses how the CMT framework can be used as a stability analysis tool. 
    In Section~\ref{sec:CMTValidation}, we verify the accuracy of nonlinear CMT by comparing with full-wave simulations. 
    Section~\ref{sec:Concl} is the concluding section.
    Furthermore, two Appendices are included:
    In Appendix~\ref{App:PertTheory}, we describe perturbation theory and how it can be used to introduce various \emph{linear} effects in the CMT formalism, such as loss, radiation damping, and coupling.
    In Appendix~\ref{App:DeltaOmegaVSa}, we use perturbation theory in order to introduce \emph{nonlinear} effects in the formalism.

\section{Coupled-Mode Theory for Linear Systems}\label{sec:LinearCMT}

    Temporal coupled-mode theory was initially developed as a perturbative theory to describe the energy exchange between a resonant system and the outside world (input/output channels or other, neighbouring resonators) using simple, solely time-dependent ordinary differential equations.\cite{HausBook,Haus1991} In this section, we present the core developments that are based on this simple, yet powerful, concept. 

    \subsection{Description of a single cavity}\label{subsec:LinerCMTsingleCavity}

\begin{figure*}
            \centering
            \includegraphics{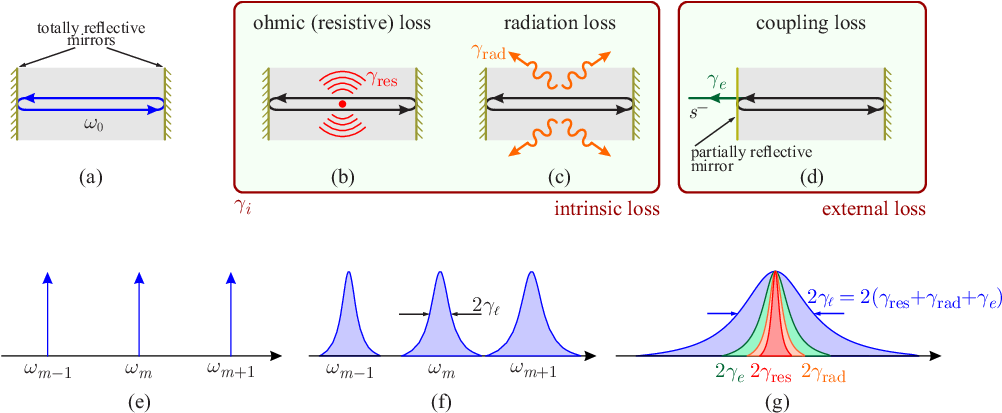}
            \caption{(a)~Isolated cavity supporting a resonant mode with  (angular) resonance frequency $\omega_0$. (b-d)~The same cavity with the addition of (b)~ohmic-loss-induced decay rate, $\gamma\rsub{res}$, (c)~radiation-induced decay rate, $\gamma\rsub{rad}$, and (d)~external-coupling-induced decay rate, $\gamma_e$. (e)~Spectral response of a lossless, isolated cavity supporting well-separated modes at resonance frequencies $\omega_m$. (f)~Spectral response of a lossy cavity. Losses lead to spectral broadening; when losses are sufficiently small each Lorentzian does not overlap with its neighbours and the CMT approximations hold. (g)~Collective contribution of the individual loss mechanisms in the perturbative regime. Each loss mechanism independently broadens the Lorentzian without spectrally shifting it.}
            \label{fig:CMTgeneral}
        \end{figure*}

        CMT is built upon the fact that the response of an isolated and lossless (i.e., Hermitian) resonant cavity [Fig.~\ref{fig:CMTgeneral}(a)] near a  resonance frequency $\omega_0$ can be described by the differential equation
        \begin{equation}
            \fullFRAC{a(t)}{t} = j\omega_0 a(t), \label{eq:CMTbasic}
        \end{equation}
        where $a(t)$ is the \emph{complex amplitude} of the resonant mode, normalized so that $|a|^2 \equiv W\rsub{res}$ equals the stored energy in the cavity (resonator).
        Note that Eq.~\eqref{eq:CMTbasic} is equivalent with the undamped harmonic oscillator ODE ($\mathrm{d}^2 x/\mathrm{d}t^2 = -\omega_0^2 x$). The second-order ODE can reduce to two, uncoupled first-order ODEs by defining the positive- and negative-frequency components (complex quantities) of the mode amplitude ($a_+$ and $a_-$), which are associated with the $\exp\{j\omega_0 t\}$ and $\exp\{-j\omega_0 t\}$ dependence, respectively.\cite{HausBook} Throughout this tutorial, $a(t)$ corresponds to $a_+(t)$ with the subscript being suppressed. Equation~\eqref{eq:CMTbasic} refers to a lumped (0D) resonator; the spatial dimensions are removed and the physical characteristics of the cavity (e.g., resonant frequency) are incorporated in the coefficients of the ODE. Naturally, this is significantly more efficient than solving a partial differential equation (PDE) involving the spatial dimensions as well. However, this approximation is valid only when the time light needs for a single full crossing of the cavity is negligible compared to the photon lifetime inside it. In lossless, isolated cavities this is always true since light is trapped inside them indefinitely. Despite their small photon lifetimes (typically in the $\mathrm{ps}$ regime), contemporary nanophotonic cavities also satisfy this restriction due to their very compact dimensions (\textmu$\mathrm{m}$ scale or smaller).

        Equation~\eqref{eq:CMTbasic} is not of much practical use, since it describes a lossless and isolated system that does not interact with the environment. Decay in the form of loss or outcoupling can be intuitively included through a complex resonance frequency $\tilde\omega_0=\omega_0+j\gamma$, which transforms Eq.~\eqref{eq:CMTbasic} into
        \begin{equation}
            \fullFRAC{a}{t} = j\omega_0 a - \gamma a, \label{eq:CMTloss}
        \end{equation}
        where the time dependence of the complex amplitude $a(t)$ has been dropped. The \emph{decay rate} $\gamma$ can describe any type of loss;\cite{HausBook} it renders the resonant system non-Hermitian and broadens its spectral response [Fig.~\ref{fig:CMTgeneral}(f)].\cite{Fan2003,Suh2004} Note that Eq.~\eqref{eq:CMTloss} is only approximately correct and valid when $\gamma \ll \omega_0$ (first order approximation). In the presence of loss, the reduction from a second-order to a single first-order ODE is not strictly possible. In general, second-order corrections on $\omega_0$ due to the presence of loss ($\gamma$) should be considered \cite{HausBook}. \blue{Note that a strategy to extend the applicability to cases where $\gamma\sim\omega_0$ is by adding the negative-frequency counterpart $\tilde{\omega}_0^-=\left(-\tilde{\omega}_0^+\right)^\ast$ of the positive-frequency pole $\tilde{\omega}_0^+$, thus recovering both poles of the second order differential equation.}
    
        As dictated by the conservation of energy, the rate of energy decay should equal the power loss dissipation, i.e., $\mathrm{d}W\rsub{res}/\mathrm{d}t=-P\rsub{loss}$. Using the formulation of Eq.~\eqref{eq:CMTloss}, we find
        \begin{equation}
            \fullFRAC{W\rsub{res}}{t} = \fullFRAC{|a|^2}{t} = -2\gamma|a|^2 = -2\gamma W\rsub{res} = -P\rsub{loss}, \label{eq:CMTenergy}
        \end{equation}
        meaning that energy decay rate is $2\gamma$, as expected, and $\gamma = P\rsub{loss}/2W\rsub{res}$. Using the definition of the quality factor as the stored energy in the cavity over the energy dissipation per optical cycle, $Q=\omega_0(W\rsub{res}/P\rsub{loss})$,\cite{Christopoulos2019} it is easily found that
        \begin{equation}
            Q = \frac{\omega_0}{2\gamma} = \frac{\omega_0\tau}{2}, \label{eq:Qfactor}
        \end{equation}
        where $\tau = 1/\gamma$ is the \emph{photon lifetime}, indicating the time that a photon is trapped inside the cavity before being dissipated. Equation~\eqref{eq:Qfactor} is customarily used to connect the photon lifetime $\tau$ with the respective $Q$-factor; the latter can be straightforwardly specified via simulations or measurements.\cite{Christopoulos2019} Note that when calculated correctly,\cite{Christopoulos2019,Lalanne2018} the quality factor encapsulates and correctly incorporates dispersion in the time-dependent ODEs, without the need for cumbersome terms involving convolutions. This topic is discussed in Sec.~\ref{sec:CMTParamsCalculations}. Having specified $\omega_0$ and $\tau$ (or $\gamma$), the spectral and/or temporal response of the cavity can be fully described through Eq.~\eqref{eq:CMTloss}, provided that $1/\tau \ll \omega_0$ holds; in effect, this implies that the accuracy of CMT becomes questionable for $Q$-factors below  $\sim1\,000$.\cite{Christopoulos2020OL} 
    
        Thus far, a single abstract decay mechanism was assumed. However, in realistic systems energy dissipation stems from various physical mechanisms, such as ohmic (resistive) loss, radiation, or coupling to an external port,
        Fig.~\ref{fig:CMTgeneral}(b-d). In the perturbative regime, all these decay mechanisms can be assumed to act independently and, thus, the total, \emph{loaded} decay rate $\gamma_\ell$ is simply the sum of the respective decay rates of each individual mechanism, i.e.,  $\gamma_\ell = \gamma\rsub{res} + \gamma\rsub{rad} + \gamma_e$ [Fig.~\ref{fig:CMTgeneral}(g)]. Therefore, Eq.~\eqref{eq:CMTloss} can be written as
        \begin{equation}
            \fullFRAC{a}{t} = j\omega_0 a - (\gamma\rsub{res}+\gamma\rsub{rad}+\gamma_e) a. \label{eq:CMTlossExpanded}
        \end{equation}
        In guided-wave systems, it is customary to consider resistive and radiation decay rates under a unified, \emph{intrinsic} decay parameter $\gamma_i=\gamma\rsub{res} + \gamma\rsub{rad}$, since they both refer to the ``unloaded'' (i.e., uncoupled) cavity.
    
\begin{figure}
            \centering
            \includegraphics{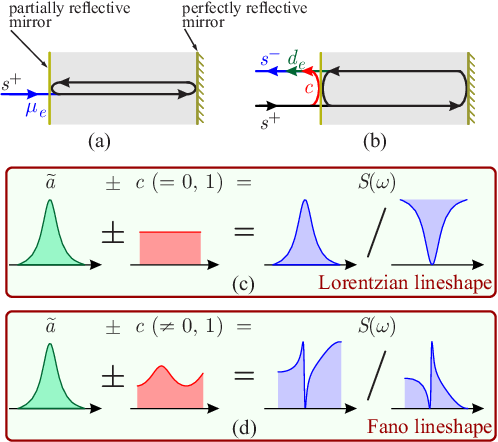}
            \caption{(a)~Coupling of an incident wave to the resonant cavity. (b)~Description of the two scattering mechanisms that compose the output wave: direct scattering $c$ (red arrow), indirect scattering $d_e$ mediated by the cavity response (green arrow), and their interference in the output port $s^-$ (blue arrow). (c)~Output lineshape when the direct scattering coefficient $c$ \blue{is absent or equals unity}. The output lineshape remains Lorentzian. (d)~Output lineshape \blue{in the general case of background interference ($c\neq\pm 1,\,0$).} The lineshape is of Fano type and its exact profile depends on the relative positioning/strength of the two individual contributions.}
            \label{fig:CMTgeneralCouplingSingle}
        \end{figure}

        To include the excitation of a cavity mode by an incident wave [Fig.~\ref{fig:CMTgeneralCouplingSingle}(a)], 
        Eq.~\eqref{eq:CMTlossExpanded} acquires an additional, feeding term
        \begin{equation}
            \fullFRAC{a}{t} = j\omega_0 a - (\gamma_i+\gamma_e) a + \mu_e s^+, \label{eq:CMTcoupling}
        \end{equation}
        where $\mu_e$ is a (complex) coupling coefficient  and $s^+$ is the complex amplitude of the incident wave, normalized so that $|s^+|^2 \equiv P\rsub{in}$, i.e., its norm squared equals the power carried by the incident wave. Although $\mu_e$ is arbitrary for now (apart from the condition $|\mu_e| \ll \omega_0$ due to the perturbative coupling considered), it is unsurprisingly related with $\gamma_e$. 
        Their connection is described in Sec.~\ref{subsec:LinerCMTCouplingParams}.\cite{HausBook,Fan2003,Suh2004,Zhao2019b}
    
        Equation~\eqref{eq:CMTcoupling} should be complemented with an additional algebraic equation that describes the complex amplitude of the output wave $s^-$,
        \begin{equation}
            s^- = c s^+ + d_e a,\label{eq:CMToutput}
        \end{equation}
         and allows to calculate the reflection or transmission coefficients. 
        Equation~\eqref{eq:CMToutput} consists of two terms [see Fig.~\ref{fig:CMTgeneralCouplingSingle}(b)]: The first describes the direct interaction of the input with the output channel in the absence of the cavity and is represented by the (complex) scattering coefficient $c$; the second describes indirect interaction mediated by the evanescent coupling of light from the cavity to the output channel and is represented by the (complex) coupling coefficient $d_e$. As can be expected, $d_e$ is related with both $\mu_e$ and $\gamma_e$ (more details in  Sec.~\ref{subsec:LinerCMTCouplingParams}). We stress that Eq.~\eqref{eq:CMToutput} is written using strictly complex amplitudes so that the phase of each coupling mechanism is correctly considered and interference effects are rigorously accounted for.

        Using Eqs.~\eqref{eq:CMTcoupling}~and~\eqref{eq:CMToutput}, it is now quite straightforward to calculate scattering coefficients under continuous-wave (CW) excitation ($\mathrm{d/d}t\rightarrow 0$), which are typically the quantities that are being measured experimentally. We assume that the incident field oscillates at an arbitrary (angular) frequency $\omega$, reasonably close to $\omega_0$, i.e., $s^+ = \tilde s^+ \exp\{j\omega t\}$, with the tilde in time-dependent quantities denoting a slowly varying 
        envelope. In the absence of any nonlinear effect, both the cavity amplitude and the output wave will oscillate at the same frequency, i.e., $a = \tilde a \exp\{j\omega t\}$ and $s^- = \tilde s^- \exp\{j\omega t\}$. Then, solving Eq.~\eqref{eq:CMTcoupling} for $\tilde a$ in CW (spectrally, a Lorentzian function), and returning to Eq.~\eqref{eq:CMToutput} we have
        \begin{equation}
            S = \frac{\tilde s^-}{\tilde s^+}  = c + \frac{d_e\mu_e}{j(\omega-\omega_0)+(\gamma_i+\gamma_e)}. \label{eq:CMTscattering}
        \end{equation}
        The form of Eq.~\eqref{eq:CMTscattering} is very intuitive: the first term describes the direct interaction of the input/output waves and the second introduces the contribution of the cavity, described by the parameters $\omega_0$, $d_e$, $\mu_e$, $\gamma_i$, and $\gamma_e$.\cite{Manolatou1999,Li2010} Naturally, as $\omega$ drifts from the neighborhood of $\omega_0$, the contribution of the cavity diminishes. Finally, the scattering parameter $c$ can be either constant or frequency-dependent. This is a very useful degree of freedom that allows for the description of more complex responses such as Fano lineshapes, \blue{obtained for $c\neq\pm 1,\,0$, i.e., in the presence of background interference} [Fig.~\ref{fig:CMTgeneralCouplingSingle}(d)].\cite{Fan2003,Suh2004,Fan2002a,Fan2002b,Tu2010} 

    \subsection{Coupling parameters of a single guided or free-space cavity}\label{subsec:LinerCMTCouplingParams}

        The CMT framework can be generalized for systems with multiple input/output channels.\cite{Fan2003} We shall use this general description to illustrate the connections between the coupling decay rate $\gamma_e$ and the respective coupling parameters $\mu_e$ and $d_e$. Assuming $n$ ports and a cavity supporting a single mode, Eqs.~\eqref{eq:CMTcoupling}~and~\eqref{eq:CMToutput} can be written in the compact matrix notation
        \begin{subequations}
            \begin{align}
                \fullFRAC{a}{t} &= (j\omega_0-\gamma)a + \mathbf{M}_e^T\mathbf{s}^+, \\
                \mathbf{s}^- &= \mathbf{C}\mathbf{s}^+ + \mathbf{D}_e a,  \label{eq:CMTsystemMultyPortsCoupling}
            \end{align}
            \label{eq:CMTsystemMultyPorts}
        \end{subequations}
        where $\mathbf{s}^\pm = [s_1^\pm\ s_2^\pm \ldots \ s_n^\pm]^T$ are column vectors containing the input/output complex wave amplitudes in each port ($T$ denotes the transpose matrix), $\mathbf{M}_e$ and $\mathbf{D}_e$ are column vectors containing the respective coupling coefficients $\mu_e$ and $d_e$ for each port, and $\mathbf{C}$ is the $n\times n$ direct scattering matrix in the absence of the resonant cavity. To connect $\mathbf{M}_e$ and $\mathbf{D}_e$ with $\gamma_e$, conservation of energy and the time-reversal symmetry that governs Maxwell's equations are typically applied.\cite{Fan2003,Suh2004,Asadchy2020} However, it has been shown recently that different constraints emerge when each principle is considered separately; this helps in the study of nonreciprocal systems where time-reversal symmetry is absent.\cite{Asadchy2020,Wang2018b,Zhao2019b,Sounas2013,Sounas2018} 
        We next present the relations between the matrices and focus on their physical interpretation. Detailed derivation and proof can be found in the literature.\cite{Fan2003,Wang2018b,Zhao2019b}
        Assuming only \emph{energy conservation} one finds that
        \begin{subequations}
            \begin{align}
                \mathbf{C}^\dagger\mathbf{C} &= \mathbf{I}_n, \label{eq:CMTconnectionsEnergy1} \\
                \mathbf{M}_e^\dagger\mathbf{M}_e = \mathbf{D}_e^\dagger\mathbf{D}_e &= 2\gamma_e, \label{eq:CMTconnectionsEnergy2} \\
                \mathbf{C}\mathbf{M}_e^* + \mathbf{D}_e = \mathbf{C}^T\mathbf{D}_e^* + \mathbf{M}_e &= \mathbf{0}, \label{eq:CMTconnectionsEnergy3}
            \end{align}
            \label{eq:CMTconnectionsEnergy}
        \end{subequations}
        where $\mathbf{I}_n$ is the $n\times n$ identity matrix. The dagger indicates the conjugate transpose matrix, meaning that matrix $\mathbf{C}$ is unitary [Eq.~\eqref{eq:CMTconnectionsEnergy1}], i.e., the direct interaction of input/output waves is lossless. Note that $\gamma_e$ represents the decay rate to all $n$ ports, i.e., $\gamma_e = \sum_m \gamma_{e,m}$ and, thus, Eq.~\eqref{eq:CMTconnectionsEnergy2} implies that energy is conserved under the coupling mechanism, as well. On the other hand, when the system obeys \emph{time-reversal symmetry}, one finds
        \begin{subequations}
            \begin{align}
                \mathbf{C}\mathbf{C}^* &= \mathbf{I}_n, \\
                \mathbf{M}_e^\dagger\mathbf{D}_e &= 2\gamma_e, \label{eq:CMTconnectionsTimeReversal2} \\
                \mathbf{C}\mathbf{D}_e^* + \mathbf{D}_e = \mathbf{C}^T\mathbf{M}_e^* + \mathbf{M}_e &= \mathbf{0}. \label{eq:CMTconnectionsTimeReversal3}
            \end{align}
            \label{eq:CMTconnectionsTimeReversal}
        \end{subequations}
        In either case, the coupling matrices $\mathbf{M}_e$ and $\mathbf{D}_e$ are connected with the coupling decay rate $\gamma_e$ [Eqs.~\eqref{eq:CMTconnectionsEnergy2}~and~\eqref{eq:CMTconnectionsTimeReversal2}]. Importantly, there always exists a connection between the direct scattering matrix $\mathbf{C}$ and the coupling matrices $\mathbf{M}_e$ and $\mathbf{D}_e$ [Eqs.~\eqref{eq:CMTconnectionsEnergy3}~and~\eqref{eq:CMTconnectionsTimeReversal3}] that allows for the correct consideration of the relative phase difference between the coupling mechanisms.
        \blue{Note that the phase of $c_{k\ell}$ is dictated by the reference planes chosen for the specific example and only the relative phase between $c_{k\ell}$, $\mu_{e,k}$, and $d_{e,\ell}$ is of practical interest.}
        Finally, when both energy conservation and time-reversal symmetry hold, the system is \emph{reciprocal} and the combination of Eqs.~\eqref{eq:CMTconnectionsEnergy}~and~\eqref{eq:CMTconnectionsTimeReversal} further implies that
        \begin{equation}
            \mathbf{M}_e = \mathbf{D}_e, \label{eq:CMTconnectionsReciprocity}
        \end{equation}
        i.e., coupling from the input port to the cavity is the same as coupling from the cavity to the output port.
        \blue{It should be noted here that the set of Eqs.~\eqref{eq:CMTconnectionsEnergy}-\eqref{eq:CMTconnectionsReciprocity} holds for Hermitian or quasi-Hermitian systems. Contemporary realizations of CMT that target open, non-Hermitian systems, such as the quasinormal mode coupled-mode theory of Ref.~\onlinecite{Benzaouia2021}, lead to different constraints.\cite{Benzaouia2021,Zhang2020} }
    
        \begin{figure}
            \centering
            \includegraphics{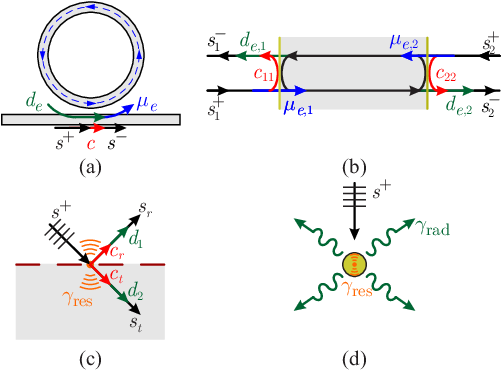}
            \caption{(a) Integrated ring resonator. The unidirectional propagation and interaction in the coupling region is clearly marked. (b) Fabry-P\'erot cavity. Due to the bidirectional propagation within the cavity a standing wave is formed. Light interacts with both ports, which coincide with the semi-transparent mirrors. (c) Resonant periodic structure (metasurface). Reflected and transmitted waves are dictated by the direct channel interaction and the scattering induced by the metasurface. Resistive loss can be interpreted as another ``port'' of the system. (d) Isolated free-space scatterer.}
            \label{fig:CMTcouplingPractical}
        \end{figure}
    
        We next demonstrate the use of Eqs.~\eqref{eq:CMTconnectionsEnergy}-\eqref{eq:CMTconnectionsReciprocity} through indicative guided-wave and free-space cavity examples.
        For the first example, a ring resonator, a typical integrated traveling-wave cavity, is used [Fig.~\ref{fig:CMTcouplingPractical}(a)]. Propagation of light in the ring is directional. Hence, when fed from the left side, light will travel in a counter-clockwise direction and exit only from the right port. Effectively, this renders the system a single-port cavity and, thus, the application of Eqs.~\eqref{eq:CMTconnectionsEnergy} results in $|d_e|=|\mu_e|=\sqrt{2\gamma_e}$. Furthermore, the direct scattering coefficient is intuitively set to $c = 1$, fixing the relative phase of the coupling mechanisms which can be found using Eq.~\eqref{eq:CMTconnectionsTimeReversal3}. For this example, Eq.~\eqref{eq:CMTconnectionsTimeReversal3} results in $d_e = -d_e^*$ and $\mu_e = -\mu_e^*$, i.e., they are both purely imaginary:  
        \begin{equation}
            d_e=\mu_e=j\sqrt{2\gamma_e}. \label{eq:CMTCouplingTraveling}
        \end{equation}
    
\begin{figure}
            \centering
            \includegraphics{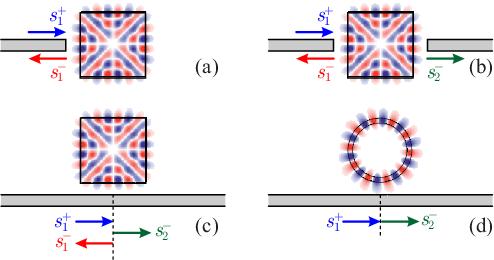}
            \caption{Schematics of archetypical resonators and coupling schemes, all fed through a single port. (a) Directly-coupled standing-wave resonator with a single output port. (b) Directly-coupled standing-wave resonator with two output ports. (c) Side-coupled standing-wave resonator with two output ports. (d) Side-coupled traveling-wave resonator with two output ports, effectively reduced to one due to directionality of the traveling wave.}
            \label{fig:CMTcouplingSchemes}
        \end{figure}

        The second example is a Fabry-P\'erot resonator [Fig.~\ref{fig:CMTcouplingPractical}(b)], a typical standing-wave cavity. Assuming that both mirrors of the cavity are partially reflective, light can leak to both directions and the cavity is effectively coupled to two ports. Due to geometric symmetry and when both mirrors have the same reflectivity, it holds $d_{e,1}=d_{e,2}$ and $\mu_{e,1}=\mu_{e,2}$. 
        Then, Eq.~\eqref{eq:CMTconnectionsEnergy2} gives  $|d_{e,1}|=|d_{e,2}|=|\mu_{e,1}|=|\mu_{e,2}|=\sqrt{\gamma_e}$. For the direct scattering matrix, it is again intuitive to write $c_{11}=c_{22}=-1$ and $c_{12}=c_{21}=0$, 
        \blue{since there is no direct coupling between input/output channels in the absence of the cavity.}
        Then, Eq.~\eqref{eq:CMTconnectionsTimeReversal3} implies that $d_{e,k} = d_{e,k}^*$ and $\mu_{e,k} = \mu_{e,k}^*$, i.e., they are both real and 
        \begin{equation}
            d_{e,1}=d_{e,2}=\mu_{e,1}=\mu_{e,2}=\sqrt{\gamma_e}. \label{eq:CMTCouplingFabryPerot}
        \end{equation}
        In the case of mirrors with different reflectivities, a similar analysis results in\cite{Wang2013b}
        \begin{subequations}
            \begin{align}
                d_{e,1}&=\mu_{e,1}=\sqrt{2\gamma_{e,1}},\\
                d_{e,2}&=\mu_{e,2}=\sqrt{2\gamma_{e,2}},
            \end{align}
            \label{eq:CMTCouplingFabryPerotAsymmetric}
        \end{subequations}
        which is intuitively anticipated since the two ports can be treated separately in the perturbative regime. For completeness, we compile in Table~\ref{tab:CMTCouplingCoefficients} some typical guided-wave resonators (traveling- or standing-wave) and the possible coupling schemes (direct or side coupling) with one or two ports, as schematically depicted in Fig.~\ref{fig:CMTcouplingSchemes}.

        \begin{table*}[t]
            \renewcommand*{\arraystretch}{1.2}
            \centering
            \begin{tabular}{cccccccc}
                \hline\hline
                ~ & \multicolumn{5}{c}{standing-wave cavity$^1$} & ~ & traveling-wave cavity$^2$ \\
                \cline{2-6} \cline{8-8}
                ~ & direct coupling, one port & ~ & direct coupling, two ports & ~ & side coupling, two ports & ~ & side coupling, two ports \\
                \hline
                \multicolumn{8}{c}{\tiny ~} \\
                $\mathbf{C}$ & $[-1]$ & ~ & $\left[\begin{matrix} -1 & 0 \\ 0 & -1 \end{matrix}\right]$ & ~ & $\left[\begin{matrix} 0 & 1 \\ 1 & 0 \end{matrix}\right]$ & ~ & $\left[\begin{matrix} 0 & 1 \\ 1 & 0 \end{matrix}\right]$ \\[12pt]
                $\mathbf{D}_e$ & $[\sqrt{2\gamma_e}]$ & ~ & $[\sqrt{\gamma_e}~\sqrt{\gamma_e}]^T$ & ~ & $[j\sqrt{\gamma_e}~j\sqrt{\gamma_e}]^T$ & ~ & $[j\sqrt{2\gamma_e}~j\sqrt{2\gamma_e}]^T$ \\[4pt]
                $\mathbf{M}_e$ & $[\sqrt{2\gamma_e}]$ & ~ & $[\sqrt{\gamma_e}~\sqrt{\gamma_e}]$ & ~ & $[j\sqrt{\gamma_e}~j\sqrt{\gamma_e}]$ & ~ &  $[j\sqrt{2\gamma_e}~j\sqrt{2\gamma_e}]$ \\[4pt]
                \hline\hline
                \multicolumn{8}{l}{\footnotesize$^1$ Fabry-P\'erot-like cavities, $^2$ Ring- or disk-like cavities}
            \end{tabular}
            \caption{CMT coupling parameters of typical guided-wave cavities and coupling schemes (see also Fig.~\ref{fig:CMTcouplingSchemes}).}
            \label{tab:CMTCouplingCoefficients}
        \end{table*}
    
        Another family of resonant structures that can be studied with CMT is free-space periodic systems (metasurfaces and gratings) [Fig.~\ref{fig:CMTcouplingPractical}(c)].\cite{Zhou2021,Zhang:2020,Zheng:2023} In this case, the input/output ports are plane waves with the allowed wavevectors (propagating diffraction orders).
        The key difference with guided-wave systems is that the cavity interacts with the ports through $\gamma\rsub{rad}$, i.e., through the radiation to different output channels, rendering $\gamma_e$ obsolete.
        It is also interesting that ohmic loss can be considered as an extra ``port'' carrying output power, but not interfering with the incident wave; thus phase is irrelevant for this interaction. Hence, it can be assumed that $s\rsub{abs} = d\rsub{abs}a$ and, through the energy conservation law, it can be found that $d\rsub{abs}=\sqrt{2\gamma\rsub{res}}$. On the contrary, $d_{e,k}$ and $\mu_{e,k}$ cannot be cast in closed form, as was the case with guided-wave systems, since the direct scattering matrix $\mathbf{C}$ may acquire any valid form (the only requisite is to be unitary) and Eqs.~\eqref{eq:CMTconnectionsEnergy3}~and~\eqref{eq:CMTconnectionsTimeReversal3} have to be applied to correctly consider the relative phase and the interference between direct and indirect scattering channels.
    
        The last example is a single scatterer [Fig.~\ref{fig:CMTcouplingPractical}(d)]. In this case, we can end up with closed-form relations for $d_{e}$ and $\mu_{e}$.\cite{Hamam2007,Ruan2010PRL,Ruan2010JoPC,Ruan2012,Bulgakov2023,Chen2023} The analysis is based on the expansion of the illuminating wave (typically a plane wave) into cylindrical or spherical waves (depending on the dimensionality of the system), which are the waveforms that the scattered field acquires in the far-field. Then, by appropriately treating each wave as an independent port one can find that the form of Eqs.~\eqref{eq:CMTsystemMultyPorts} is still valid and $\gamma_\ell=\gamma\rsub{res}+\gamma\rsub{rad}$, $c=\exp\{j\phi\}$, and $d_e=\mu_e=\sqrt{2\gamma\rsub{rad}}\exp\{j(\phi+\pi)/2\}$. \cite{Ruan2010JoPC} The additional phase $\phi$ introduced through the direct scattering coefficient allows for the description of complex, Fano-type responses\cite{Ruan2010JoPC} that appear when the size of the scatterer is comparable with the wavelength of the illuminating wave.

        The above analysis can be expanded in multimode cavities, where each mode is described by a single ODE. Mutual coupling between the modes should also be considered, entangling the system of ODEs. This analysis can be found in Ref.~\onlinecite{Suh2004} 
        including an elegant handling of dark modes, i.e., modes with $\gamma_e,\,\gamma\rsub{rad} \rightarrow \infty$ which cannot be externally excited or coupled to any physical port of the system.

    \subsection{Coupling of cavities and modes}\label{subsec:LinerCMTCouplingOfCavities}

        \begin{figure}[!b]
            \centering
            \includegraphics{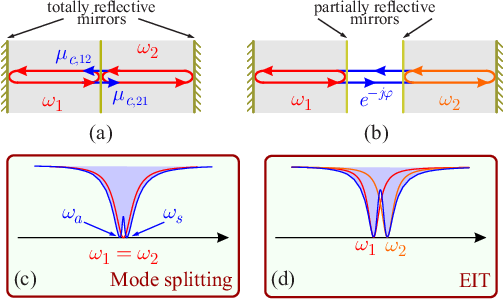}
            \caption{(a)~Direct mutual coupling between two cavities. (b)~Indirect coupling between two cavities through an external channel. (c)~Mode-splitting effect in the two identical, mutually-coupled cavities of panel (a). The common resonance frequency is split and two distinct dips appear in the spectrum. (d)~Spectral response of the two indirectly-coupled cavities with different resonance frequencies of panel (b). A response reminiscent of electromagnetically induced transparency is observed, with a peak emerging between two dips.}
            \label{fig:CMTgeneralCouplingMutual}
        \end{figure}
        
        The case of mutual coupling of modes that are supported either by the same or by neighbouring cavities [Fig.~\ref{fig:CMTgeneralCouplingMutual}(a)] can be also described within the CMT framework. Here, we will focus on two mutually coupled cavities that each support a single mode. They can be described by the equations
        \begin{subequations}
            \begin{align}
                \fullFRAC{\tilde a_1}{t} &= -j(\omega-\omega_1) \tilde a_1 + \mu_{c,12} \tilde a_2, \\
                \fullFRAC{\tilde a_2}{t} &= -j(\omega-\omega_2) \tilde a_2 + \mu_{c,21} \tilde a_1,
            \end{align}
            \label{eq:CMTmutualCoupling}
        \end{subequations}
        which have been expressed using the slowly varying envelope concept introduced in Eq.~\eqref{eq:CMTscattering}. In Eq.~\eqref{eq:CMTmutualCoupling}, we have allowed for coupling coefficients $\mu_{c,12}$, but have omitted self-terms ($\mu_{c,11}$ and $\mu_{c,22}$), which lead to a re-normalization of the resonant frequencies of the standalone cavities, or what is frequently called coupling-induced frequency shift (CIFS).\cite{Popovic2006} Calculation of self terms is covered in Appendix~\ref{AppSub:CouplingPertTheory}.
        In addition, in Eqs.~\eqref{eq:CMTmutualCoupling} the loss decay rates have been  momentarily ignored for the analysis.
        Due to the conservation of energy that in this case applies only to the exchange of energy between the two cavities, it is not difficult to show that the mutual coupling coefficients $\mu_{c,12}$ and $\mu_{c,21}$ are not independent but connected through \cite{HausBook,Haus1991}
        \begin{equation}
            \mu_{c,12} = -\mu_{c,21}^*, \label{eq:CMTmutualCouplingCoefficients}
        \end{equation}
        i.e., they are purely imaginary. Of course the framework is valid only when $|\mu_{c,ij}| \ll \omega_{i/j}$ (first order approximation), as with any other perturbative effect described thus far. 
        Moreover, to obtain a nontrivial solution of the system in Eqs.~\eqref{eq:CMTmutualCoupling}, its determinant should be zero. Thus, the resonance frequencies of the two supported supermodes (symmetric and antisymmetric superposition of the modes supported by the two cavities) can be easily found\cite{HausBook,Haus1991}
        \begin{equation}
            \omega_{s/a} = \frac{\omega_1+\omega_2}{2} \pm \sqrt{ \left(\frac{\omega_1-\omega_2}{2}\right)^2 + |\mu_{c,12}|^2 } = \frac{\omega_1+\omega_2}{2} \pm \Omega_0, \label{eq:CMTmutualCouplingFrequencies}
        \end{equation}
        revealing that energy is exchanged between the two cavities every $\pi/(2\Omega_0$). Equation~\eqref{eq:CMTmutualCouplingFrequencies} can be also used to calculate the amplitude of the mutual coupling coefficient when the resonance frequencies of the two supermodes are known (e.g., by simulations or measurements). The phase is then determined from Eq.~\eqref{eq:CMTmutualCouplingCoefficients}.
        An alternative approach to calculate $\mu_c$ utilizing perturbation theory is discussed in Appendix~\ref{AppSub:CouplingPertTheory}.
    
        From the so far discussion, it is evident that mutual coupling induces a mode-splitting effect [Fig.~\ref{fig:CMTgeneralCouplingMutual}(c)], even when identical resonators ($\omega_1 = \omega_2$) are considered.\cite{HausBook,Li2010} This effect can be used in applications like filtering where, e.g., a flat response over a large frequency span is required.\cite{Little1997,Song2009,Lin2015} Such a response is typically achieved by mutually coupling multiple resonators; in the weakly/evanescent coupling regime, the described treatment under CMT remains the same,\cite{Little1997} although additional terms might emerge to fully describe coupling.\cite{Chen2022} Ultimately and for appropriate conditions Fano-like responses may also appear.\cite{Song2009,Du2019,Lu2022}
        Coupled cavities have also been used as delay lines, with temporal CMT being a valuable tool for their analysis.\cite{Sandhu2006}
        Finally, the presented treatment of mode coupling can be used to describe parity-time (PT) symmetry in cavities with gain and loss.\cite{Zhou2016,Kominis2018,Ramezanpour2021,Xiao2021}
    
        Another approach to achieve frequency splitting albeit without actually spectrally splitting a single mode, is the indirect coupling of two detuned cavities (i.e., cavities with different resonance frequencies) through a secondary path,\cite{Li2010,Lin2015,Maes2005,AlQubaisi2020} see Fig.~\ref{fig:CMTgeneralCouplingMutual}(b). The only  prerequisite here is the correct consideration of the phase $\phi$ that the wave accumulates as it propagates between the two cavities.\cite{Maes2005} Such an effect is typically introduced in the formalism of CMT through a simple and intuitive term of the form $\exp\{-j\beta L\}$ in the appropriate wave amplitudes involved, capturing the accumulated phase of a wave propagating with propagation constant $\beta$ between the two, $L$-distanced cavities, i.e., $\phi=\beta L$. With this simple treatment, it has been shown that CMT can reproduce effects such as an equivalent to the electromagnetically induced transparency in optical cavities where high transmission is obtained inside a spectral dip [Fig.~\ref{fig:CMTgeneralCouplingMutual}(d)],\cite{Kekatpure2010,Ketzaki2013} or symmetry breaking of Fano modes in geometrically symmetric systems of two cavities.\cite{Maes2006,Maes2008}
    
        Before concluding this section,
        we should mention that CMT does not make any assumption regarding the physical implementation of the resonant system itself. The presented framework is universally applicable, provided that the restrictions  regarding roundtrip time,  losses, and coupling strength are met. This key advantage of CMT allowed its application in a vast range of optical/photonic structures over the years: integrated dielectric and hybrid-plasmonic cavities, photonic crystal systems, dielectric nanoparticles, dielectric and plasmonic metasurfaces and, more recently, cavities embedding contemporary 2D materials (graphene, transition metal dichalcogenides, black phosphorus, etc.), to name but a few.
        More recently, a discussion has opened regarding the further expansion of the CMT formulation in systems with even higher losses, where $\gamma$ and $\omega_0$ are comparable,\cite{Kristensen2017,Christopoulos2020OL,Tao2020,Zhang2020,Benzaouia2021,Zhou2021,Wu2024} or systems with nonlocal responses.\cite{Lin2020,Overvig2024}
        To date, such efforts have led to CMT implementations with increased complexity compared to the original version. This next milestone for CMT will certainly further enhance its already wide applicability. 

\section{Retrieving Coupled-Mode Theory Parameters from the Physical System} \label{sec:CMTParamsCalculations}

    As discussed in the Sec.~\ref{sec:LinearCMT}, the main advantage of CMT that makes the theory so attractive is that it can be used to accurately calculate both the temporal and spectral response of a resonant system, requiring only the knowledge of a few key parameters, namely, the resonance frequency, the quality factor [which is connected with the cavity lifetime through Eq.~\eqref{eq:Qfactor}], and, when applicable, the coupling strength between cavities. 
    It should be stressed that all required parameters can be rigorously \emph{calculated} from the physical system under study. CMT does not rely on fitting as is the case with other models that can only offer a qualitative understanding of the response.  

\begin{figure}[!t]
        \centering
        \includegraphics{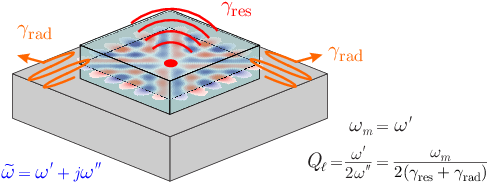}
        \caption{Square dielectric resonator on a substrate. The spatial $\mathbf{E}$-field distribution on resonance (eigenvector) and the complex resonance frequency (eigenvalue) are shown. The extraction of the resonance frequency ($\omega_m$) and the loaded quality factor ($Q_\ell$) is also included.}
        \label{fig:3DResonatorGeneral}
    \end{figure}

    A typical computational approach to calculate the parameters  appearing the in temporal CMT ODEs is to use a modal technique to retrieve the eigenmodes and eigenvalues of the physical resonant system. An eigenmode corresponds to the spatial field distribution on resonance (mode profile), while the respective (complex) eigenvalue ($\tilde{\omega}$) holds information regarding the resonance frequency ($\omega'$) and the quality factor of this mode [$Q_\ell = \omega'/(2\omega''$)].\cite{Lalanne2018,Christopoulos2019} An example is shown in Fig.~\ref{fig:3DResonatorGeneral}. This approach is meaningful in the sense that a single (or a few) modal simulations are used to extract a handful of parameters that can populate the coefficients of the CMT equations. Then, it becomes possible to readily study the broadband spectral response or temporal dynamics with a first-order ODE, rather than resorting to full-wave time-domain simulations, which are time consuming. 

    In resonant systems consisting of dispersionless materials, the real part of the eigenvalue corresponds to the resonance frequency of the cavity mode and the respective quality factor is given directly by the ratio of the real over the imaginary part of the eigenvalue.\cite{Christopoulos2019} This is not as straightforward when materials with strong dispersion are involved since dispersion is typically not captured by conventional eigensolvers. Alternatively, one can use the eigenmode, i.e., the spatial field profile on resonance, to calculate the stored energy in the resonator, $W\rsub{res}$, and the power loss dissipation, $P\rsub{loss}$, and then resort to the definition of the quality factor [$Q=\omega_0(W\rsub{res}/P\rsub{loss})$] to specify it. Reference~\onlinecite{Christopoulos2019} discusses in detail the topic of quality factor calculation, including the most general case where dispersion, loss, and leakage are present.
    Note that it is possible to rigorously include dispersion in the eigenvalue problem by introducing appropriate auxiliary fields, as has been demonstrated in the context of the finite-element method (FEM).\cite{Raman2010,Raman2011} Then, both the resonance frequency and the respective quality factor can be directly calculated correctly in dispersive systems. \cite{Lalanne2018}

    In systems where multiple loss channels are present, it is typical to use appropriate ancillary problems to determine the individual quality factors and, thus, the respective decay rates of each independent channel. In a guided-wave resonator for example, one can use an ancillary problem where the feeding waveguide is omitted to calculate the intrinsic quality factor, $Q_i$, then return to the coupled system to calculate the loaded quality factor, $Q_\ell$, and finally find the external $Q$-factor as $Q_e^{-1}=Q_\ell^{-1}-Q_i^{-1}$. The same approach can be used when $Q_i$ needs to be decomposed in $Q\rsub{res}$ and $Q\rsub{rad}$ (e.g., in free-space systems); losses are momentarily ignored to calculate $Q\rsub{rad}$ and then the restive $Q$-factor is retrieved through $Q\rsub{res}^{-1}=Q_i^{-1}-Q\rsub{rad}^{-1}$. The above decomposition, although intuitive, is only accurate in the perturbative regime. When, for instance, there is relatively strong coupling between the cavity and the feeding ports, the respective intrinsic and radiation $Q$-factors should be corrected accordingly.\cite{Christopoulos2019} 

    Another approach that can be followed to calculate the contributions of the perturbative loss and coupling mechanisms is the application of the \emph{first-order perturbation theory}.\cite{Joannopoulos} First, a simple, uncoupled and lossless cavity is assumed for the modal simulation. Then, the contribution of each petrurbative effect is evaluated through appropriate integrals involving the unpertubed eigenmode and its spatial interaction with the perturbation. There are realizations of perturbation theory in the frequency domain for material perturbations\cite{Joannopoulos} (useful for the introduction of losses,\cite{Christopoulos2016} nonlinear effects,\cite{Bravo-abad2007} or gain\cite{Nousios2023}), shape perturbations,\cite{Pozar,Johnson2002} and coupling phenomena either between cavities (mutual coupling)\cite{Tasolamprou2016} or between a cavity and an adjacent waveguide.\cite{Haus1991,Popovic2006} Furthermore, perturbation theory can be developed for the time domain as well.\cite{Daniel2011} Finally, higher-order realizations also exist,\cite{Sehmi2020,Both2022} targeting higher accuracy in systems with stronger perturbations. In Appendix~\ref{App:PertTheory} we present the derivation of perturbation theory for general material perturbations that results in Eq.~\eqref{eq:GeneralPertTheory} below. Perturbations due to coupling are also discussed in Appendix~\ref{App:PertTheory}.
    \begin{widetext}
        \begin{equation}
            \frac{\Delta\tilde\omega}{\omega_0} = -\frac{\displaystyle\iiint_V \Delta\mathbf{P}\cdot\mathbf{E}_0^* \mathrm{d}V - j\displaystyle\frac{1}{\omega_0}\displaystyle\iiint_V \Delta\mathbf{J}\cdot\mathbf{E}_0^* \mathrm{d}V}{\displaystyle\iiint_V \varepsilon_0\left.\partialFRAC{\{\omega\RE{\dbar{\varepsilon}_r}\}}{\omega}\right|_{\omega=\omega_0}\mathbf{E}_0^*\cdot\mathbf{E}_0 \mathrm{d}V + \displaystyle\iiint_V \mu_0\left.\partialFRAC{\{\omega\RE{\dbar{\mu}_r}\}}{\omega}\right|_{\omega=\omega_0}\mathbf{H}_0^*\cdot\mathbf{H}_0 \mathrm{d}V + \displaystyle\iiint_V \left.\partialFRAC{\IM{\dbar{\sigma}}}{\omega}\right|_{\omega=\omega_0}\mathbf{E}_0^*\cdot\mathbf{E}_0 \mathrm{d}V }. 
            \label{eq:GeneralPertTheory}
        \end{equation}
    \end{widetext}

    Equation~\eqref{eq:GeneralPertTheory} allows us to calculate the  complex resonance frequency shift that results from small perturbations (linear or nonlinear) either on the polarization or on the induced current density of a cavity though appropriate integrals of the unperturbed fields. Furthermore, as revealed from the normalization term in the denominator (corresponds to the stored energy), it also rigorously includes anisotropy and material dispersion. Importantly, through the inclusion of the induced current density and the respective (complex and anisotropic) conductivity, Eq.~\eqref{eq:GeneralPertTheory} is capable of rigorously handling 2D materials by using a surface current/surface conductivity description that respects their true two-dimensional nature. For comparison, Eq.~\eqref{eq:InitialPertTheory} that follows is one of the first realizations of linear material perturbation, accurate for a non-dispersive, linear dielectric cavity.\cite{Bravo-abad2007} 
    \begin{equation}
        \frac{\Delta\omega}{\omega_0} = -\frac{1}{2}\frac{\displaystyle\iiint_V \Delta\varepsilon |\mathbf{E}_0|^2 \mathrm{d}V}{\displaystyle\iiint_V \varepsilon_0\varepsilon_r |\mathbf{E}_0|^2 \mathrm{d}V }. 
        \label{eq:InitialPertTheory}
    \end{equation}
    Note that in Eq.~\eqref{eq:InitialPertTheory} only the electric field appears in the denominator since the stored electric and magnetic energies are equal on resonance; this not true in the general case described by Eq.~\eqref{eq:GeneralPertTheory} as is seen by the form of the denominator.\cite{Christopoulos2016a}

    \blue{Although it has been a long way from Eq.~\eqref{eq:InitialPertTheory} to Eq.~\eqref{eq:GeneralPertTheory}, the latter can still be restrictive in capturing the behavior of some contemporary leaky and lossy systems like dielectric and plasmonic nanocavities.\cite{Christopoulos2020OL,Yang2015,Lalanne2018,Both2019,Cognee2019,Wu2021} As it is discussed in more detail in Appendix~\ref{App:PertTheory}, Eqs.~\eqref{eq:GeneralPertTheory}~and~\eqref{eq:InitialPertTheory} are strictly accurate only for Hermitian and approximately accurate for quasi-Hermitian systems with low ohmic and radiation loss. However, they have the advantage of involving a physical quantity for the normalization of the electric and magnetic field [the denominator of Eq.~\eqref{eq:GeneralPertTheory}], namely the on resonance stored energy of the cavity. This representation positively impacts the strategy of incorporating nonlinear effects in CMT, as it will become evident in the sections that follow. Nonetheless, it is not the only available strategy to incorporate nonlinearities in the CMT framework.\cite{Christopoulos2020OL} }

    \blue{For the more general case of  non-Hermitian systems, perturbation theory can be developed following a seemingly similar but essentially different approach, using unconjugated fields to lift the restriction of ohmic loss and radiation leakage absence.\cite{Christopoulos2020OL,Yang2015,Lalanne2018,Wu2021} This approach sometimes is referred to as \emph{quasinormal modes perturbation theory},\cite{Lalanne2018,Wu2021} since it involves the concept of quasinormal modes (QNMs). To extract the respective expression for the complex resonance frequency shift, one can follow the steps presented in Appendix~\ref{App:PertTheory}, although using solely unconjugated fields (see for example Ref.~\onlinecite{Christopoulos2020OLSuppl}, the supplementary document of Ref.~\onlinecite{Christopoulos2020OL}, for a detailed proof). For reference, in Eq.~\eqref{eq:GeneralPertTheoryUnconj} we provide with the QNM perturbation theory expression for polarization-induced material perturbation in a nonmagnetic medium.\cite{Yang2015,Christopoulos2020OL} Due to the absence of conjugated fields, the denominator of Eq.~\eqref{eq:GeneralPertTheoryUnconj} is a complex number, obviously with a different physical interpretation\cite{Sauvan2013} than the stored energy of the conjugated version. Furthermore,  its calculation is nontrivial because of the involvement of QNMs which spatially diverge.\cite{Sauvan2013} However, there have been developed techniques to compensate this spatial divergence like the so-called ``PML normalization'' which uses perfectly matched layers (PMLs) to suppress the divergence of the QNMs\cite{Lalanne2018} or by including an additional surface integral for the normalization which counteracts the diverging terms.\cite{Both2022}
    Unlike Hermitian systems and their modes (referred to as normal modes), the treatment of the natural modes of non-Hermitian systems, i.e., the quasinormal modes, is, in general, an elegant matter that lies outside the scope of this Tutorial. Nevertheless, it seems to fit with the concept of coupled-mode theory and there is some progress and discussion towards this direction.\cite{Kristensen2017,Tao2020,Zhang2020,Benzaouia2021,Wu2024} The interested reader on QNMs is referred to some recent reviews and the references therein.\cite{Lalanne2018,Kristensen2020,Both2022} }
    \blue{\begin{equation}
        \frac{\Delta\tilde\omega}{\omega_0} = -\frac{\displaystyle\iiint_V \Delta\mathbf{P}\cdot\mathbf{E}_0 \mathrm{d}V}{\displaystyle\iiint_V \varepsilon_0\partialFRAC{\{\omega\dbar{\varepsilon}_r\}}{\omega}\mathbf{E}_0\cdot\mathbf{E}_0 \mathrm{d}V - \displaystyle\iiint_V \mu_0\mathbf{H}_0\cdot\mathbf{H}_0 \mathrm{d}V}.
        \label{eq:GeneralPertTheoryUnconj}
    \end{equation}}
    \blue{Appropriate versions of Eq.~\eqref{eq:GeneralPertTheoryUnconj} have been used to capture both the resonance frequency shift and the change in the linewidth (i.e., the quality factor) of various leaky and lossy systems, plasmonic or dielectric,\cite{Christopoulos2020OL,Yang2015,Lalanne2018,Wu2021,Both2019} even in experimental set-ups.\cite{Cognee2019} There are also realizations of QNMs perturbation theory for the spatial coupling of cavities.\cite{Popovic2006,Tasolamprou2016} }
    
\section{Coupled-Mode Theory for Nonlinear Systems}\label{sec:InstantNonLinearCMT}

    Thus far, we have covered the treatment of linear systems with CMT. In Sections~\ref{sec:InstantNonLinearCMT}-\ref{sec:Gain}, we focus on nonlinear systems, which exhibit a wealth of interesting phenomena. To introduce the nonlinear response of a system within the CMT framework, its contribution should be perturbative, as was the case with the linear effects (loss, coupling) that have been discussed so far. This is in most cases true,\cite{BoydBook,ButcherCotter} meaning that CMT can accurately model most practical nonlinearities in contemporary material platforms and systems.
    In this section, we focus on instantaneous nonlinearities, which typically stem from the response of the bound electrons in dielectric media.\cite{ButcherCotter} Although not strictly instantaneous, their response time is in the fs range, much faster than the typical ps time-scales in photonics. Non-instantaneous nonlinear effects will be discussed in Sec.~\ref{sec:NoninstantNonLinearCMT}. Finally, Sec.~\ref{sec:Gain} is dedicated to the treatment of gain.
    
    Based on their perturbative nature, nonlinearities are typically treated through a Taylor expansion with respect to the electric field, meaning that the polarization or the induced current density in a nonlinear medium is described by a power series, with each order becoming significant for gradually higher power levels. 
    Even-order nonlinearities vanish in media with inversion symmetry in their molecular structure.\cite{ButcherCotter} 
    Thus, we dedicate a large part of this section to odd-order nonlinearities, which exist regardless of the molecular structure, and, more specifically, on third-order nonlinear effects which manifest first (lowest available order). We focus on single-channel effects (Kerr effect and two-photon absorption) and multi-channel interactions (third-harmonic generation, cross-phase modulation, four-wave mixing). Finally, part of the section is dedicated to saturable absorption, i.e., the intensity-dependent saturation of losses.

    The usual approach to introduce nonlinearities in the CMT framework is through a version of Eq.~\eqref{eq:GeneralPertTheory}, which quantifies the contribution of the nonlinearity on the (complex) resonance frequency of the perturbed system, considering both phase effects ($\RE{\Delta\tilde\omega}$) and losses ($\IM{\Delta\tilde\omega}$). This requires deriving an appropriate expression for $\Delta\mathbf{P}\rsub{NL}$ and/or $\Delta\mathbf{J}\rsub{NL}$ for each nonlinear phenomenon. 
    The expression for $\Delta\tilde\omega$ should be of the form $\Delta\tilde\omega \equiv \Delta\tilde\omega(a) \propto a(t)$, since a nonlinear effect depends on the amplitude of the respective cavity mode. 
    The actual expression depends on the specific nonlinear effect; some notable perturbative nonlinearities are discussed in Appendix~\ref{App:DeltaOmegaVSa}. 
    Then, $\Delta\tilde\omega$ is introduced in the framework by replacing $\omega_0$ with $\omega_0 + \Delta\tilde\omega$ and obtaining a nonlinear ODE (or a system of coupled nonlinear ODEs). This ODE governs quite accurately the response of the system when reasonably close to its (unperturbed) resonance frequency.

    \blue{Perturbation theory arguments can be applied to include not only deterministic but also perturbative stochastic processes in CMT, such as thermal noise\cite{Zhu2013,Khandekar2015,Braeckeveldt2023} or spontaneous emission.\cite{Hamel2015} In this case, a Langevin-type equation emerges. }
    
    \subsection{Single-channel nonlinearities: The Kerr effect, two-photon absorption, and optical bistability}\label{subsec:SPM}

        The first endeavor to include self-induced nonlinear phenomena in the CMT framework was with the Kerr effect in photonic crystals (PhCs),\cite{Soljacic2002,Yanik2003,Maes2005,Maes2006,Maes2008} which later extended to silicon-on-insulator (SOI),\cite{Wang2013a,Sethi2014} hybrid-plasmonic,\cite{Tsilipakos2014,Christopoulos2016} and 2D material platforms.\cite{Christopoulos2016a,Christopoulos2017} In Appendix~\ref{AppSub:DeltaOmegaVSaKerrTPA}, the process of transforming the electromagnetic description of the Kerr effect into CMT-compatible terms is presented, focusing mostly on 2D materials, which constitute the most recent development. The expressions for bulk dielectric materials are also provided; the treatment of both cases is ultimately quite similar.
    
        Here, we focus on how CMT is  used to obtain the nonlinear optical response response when self-induced nonlinearities are involved. The nonlinear resonance frequency shift is of the general form
        \begin{equation}
            \Delta\tilde\omega_3(a) = \tilde\gamma_3 |a|^2 = (-\gamma\rsub{SPM}+j\gamma\rsub{TPA})|a|^2,
            \label{eq:DeltaOmegaKerrTPACMT}
        \end{equation}
        where $\gamma\rsub{SPM}$ describes the contribution of self-phase modulation (SPM) induced by the Kerr effect, and $\gamma\rsub{TPA}$ describes two-photon absorption (TPA). Both parameters are measured in $\mathrm{1/Js}$. Depending on the involved materials, different expressions to calculate $\tilde\gamma_3$ should be used, as presented in Appendix~\ref{AppSub:DeltaOmegaVSaKerrTPA}. We will only consider here the implication of Eq.~\eqref{eq:DeltaOmegaKerrTPACMT} in the CMT ODE~\eqref{eq:CMTcoupling}. By replacing $\omega_0$ with $\omega_0 + \Delta\tilde\omega_3$, i.e., by introducing the nonlinearity as a perturbation, Eq.~\eqref{eq:CMTcoupling} becomes
        \begin{equation}
            \fullFRAC{a}{t} = j(\omega_0-\gamma\rsub{SPM}|a|^2) a - (\gamma_i+\gamma_e+\gamma\rsub{TPA}|a|^2) a + \mu_e s^+. 
            \label{eq:CMTKerrTPAFULLFIELD}
        \end{equation}
        \blue{For simplicity, we assume here that the considered mode is excited through a single input port. However, we allow for the general case of multiple output ports, as it will be shown later.}
        The slowly-varying envelope representation of Eq.~\eqref{eq:CMTKerrTPAFULLFIELD} is more useful and is numerically handled more easily.\cite{Soljacic2002} We will adopt this form for the rest of the tutorial, assuming that $a = \tilde a \exp\{j\omega t\}$ and $s = \tilde s \exp\{j\omega t\}$, so that
        \begin{equation}
            \fullFRAC{\tilde a}{t} = j[-(\omega-\omega_0)-\gamma\rsub{SPM}|\tilde a|^2] \tilde a - (\gamma_i+\gamma_e+\gamma\rsub{TPA}|\tilde a|^2) \tilde a + \mu_e \tilde s^+. 
            \label{eq:CMTKerrTPA}
        \end{equation}
        It is readily seen that the strength of the nonlinear terms increases as a function of $|\tilde a|^2$, i.e., stronger frequency shifting is induced via SPM and higher TPA losses. The parameter $\gamma\rsub{SPM}$ may acquire positive or negative values (self-focusing or defocusing Kerr nonlinearity), whereas $\gamma\rsub{TPA}>0$ since it corresponds to losses rather than gain. Note that to be consistent with perturbation theory, it should hold  $|\gamma\rsub{SPM}||\tilde a|^2 \ll \omega_0$ and $\gamma\rsub{TPA}|\tilde a|^2 \ll \omega_0$. 
        
        For purposes of numerical stability, it is quite common to normalize Eq.~\eqref{eq:CMTKerrTPA} so that the mode amplitudes, the input/output wave amplitudes, and the respective coefficients are comparable in magnitude.\cite{Chen2012} The choice of the appropriate normalization is not unique but depends on the dominant nonlinear effect. For now, we will assume that the Kerr effect dominates and use $\gamma\rsub{SPM}$ for the normalization. Specifically, we introduce the (dimensionless) normalized cavity amplitude and the normalized input/output wave amplitudes through\cite{Chen2012,Christopoulos2017}
        \begin{subequations}
            \begin{align}
                \tilde u(t) &= \sqrt{\tau_1|\gamma\rsub{SPM}|} \tilde a(t), \\
                \tilde \psi(t) &= \sqrt{\tau_2^2|\gamma\rsub{SPM}|} \tilde s(t), \label{eq:CMTNormalizationKerrPsi}
            \end{align}
            \label{eq:CMTNormalizationKerr}
        \end{subequations}
        where  $\tau_{1,2}$ are generic photon lifetime parameters, proportional to either  $\tau_e$ or $\tau_i$, depending on the cavity type (standing/traveling-wave) and the coupling scheme (direct/side coupling), see Table~\ref{tab:CMTNLKerrCoefficients}.
        Under the normalization of Eqs.~\eqref{eq:CMTNormalizationKerr}, the main nonlinear CMT ODE~\eqref{eq:CMTKerrTPA} becomes
        \begin{equation}
            \fullFRAC{\tilde u}{t^\prime} = j(-\delta-\mathbcal{s}|\tilde u|^2) \tilde u - (1+r_Q+r\rsub{TPA}|\tilde u|^2) \tilde u + \varpi_e \tilde \psi^+,
            \label{eq:CMTKerrTPANormalized}
        \end{equation}
        where  $t^\prime$ is the normalized time, $\delta$ is the normalized detuning, $r_Q$ is the ratio of intrinsic/external losses, $\varpi_e$ is a normalized coupling coefficient, $\mathbcal{s} = \pm 1$ is the sign of $\gamma\rsub{SPM}$ ($+1$ for self-focusing and $-1$ for defocusing Kerr nonlinearity), and $r\rsub{TPA}=\gamma\rsub{TPA}/|\gamma\rsub{SPM}|$ is the ratio of TPA intensity versus SPM. 
        For the definition of each normalized quantity depending on the specific cavity type and coupling scheme, see Table~\ref{tab:CMTNLKerrCoefficients}.
        The normalization of Eq.~\eqref{eq:CMTKerrTPANormalized} should be introduced in the output coupling equation(s) [Eq.~\eqref{eq:CMTsystemMultyPortsCoupling}] as well
        \begin{equation}
            \tilde\psi^-_k = c_{k\ell}\tilde\psi^+_\ell + \vartheta_{e,k}\tilde u,
            \label{eq:CMTcouplingNormal}
        \end{equation}
        where we have written the coupling equation for a single input ($\tilde\psi^+_\ell$) and output ($\tilde\psi^-_k$) wave and $k,\,\ell$ run from 1 to the total number of involved ports. For practical reasons, Table~\ref{tab:CMTNLKerrCoefficients} includes configurations with \blue{one input port, one or two output ports,} and a single cavity (Fig.~\ref{fig:CMTcouplingSchemes}). Parameters $\vartheta_{e,k}$ (or $\boldsymbol{\Theta}_e$ in matrix format) are also normalized coupling coefficients, similar to $\varpi_e$.


        \begin{table*}[t]
            \renewcommand*{\arraystretch}{1.2}
            \centering
            \begin{tabular}{cccccccc}
                \hline\hline
                ~ & \multicolumn{5}{c}{standing-wave cavity$^1$} & ~ & traveling-wave cavity$^2$ \\
                \cline{2-6} \cline{8-8}
                ~ & direct coupling, one port & ~ & direct coupling, two ports & ~ & side coupling, two ports & ~ & side coupling, two ports \\
                \hline
                $\tau_1$ & $\tau_e$ & ~ & $\tau_e$ & ~ & $\tau_e$ & ~ & $\tau_i$ \\
                $\tau_2$ & $\tau_e/\sqrt{2}$ & ~ & $\tau_e$ & ~ & $\tau_e$ & ~ & $\tau_i/\sqrt{2}$ \\
                $t^\prime$ & $t/\tau_e$ & ~ & $t/\tau_e$ & ~ & $t/\tau_e$ & ~ & $t/\tau_i$ \\
                $\delta$ & $(\omega-\omega_0)\tau_e$ & ~ & $(\omega-\omega_0)\tau_e$ & ~ & $(\omega-\omega_0)\tau_e$ & ~ & $(\omega-\omega_0)\tau_i$ \\
                $r_Q$ & $Q_e/Q_i$ & ~ & $Q_e/Q_i$ & ~ & $Q_e/Q_i$ & ~ & $Q_i/Q_e$ \\
                $\varpi_e$ & $2$ & ~ & $1$ & ~ & $j$ & ~ & $j2\sqrt{r_Q}$ \\[4pt]
                $\mathbf{C}$ & $[-1]$ & ~ & $\left[\begin{matrix} -1 & 0 \\ 0 & -1 \end{matrix}\right]$ & ~ & $\left[\begin{matrix} 0 & 1 \\ 1 & 0 \end{matrix}\right]$ & ~ & $\left[\begin{matrix} 0 & 1 \\ 1 & 0 \end{matrix}\right]$ \\[10pt]
                $\boldsymbol{\Theta}_e$ & $[1]$ & ~ & $[1~~1]^T$ & ~ & $[j~~j]^T$ & ~ & $[0~~j\sqrt{r_Q}]^T$ \\[4pt]
                $|\tilde u|^2$ & $p_1^+ - p_i^\ddag$ & ~ & $p_2^-$ & ~ & $p_1^-$ & ~ & $p_1^+ - p_2^-$ \\[2pt]
                \hline\hline
                \multicolumn{8}{l}{\footnotesize$^1$ Fabry-P\'erot-like cavities, $^2$ Ring- or disk-like cavities, \blue{$^\ddag p_i = P_i/P\rsub{0,SPM}$ represents the normalized intrinsic power loss of the cavity.}}
            \end{tabular}
            \caption{Normalized nonlinear CMT parameters of instantaneous nonlinear effects for typical guided-wave cavities and coupling schemes (see also Fig.~\ref{fig:CMTcouplingSchemes}).}
            \label{tab:CMTNLKerrCoefficients}
        \end{table*}

        The normalization of Eq.~\eqref{eq:CMTNormalizationKerrPsi} implies that SPM becomes significant when the power of the feeding wave is of the order
        \begin{equation}
            P\rsub{0,SPM} = \frac{1}{\tau_2^2|\gamma\rsub{SPM}|}.
            \label{eq:CharPowerKerr}
        \end{equation}
        The parameter $P\rsub{0,SPM}$ is typically termed the SPM \emph{characteristic power}.\cite{Soljacic2002,Yanik2003} 
        A characteristic power quantity, $P_0$, can be defined similarly for any nonlinear effect, e.g., $P\rsub{0,TPA} = 1/\tau_2^2\gamma\rsub{TPA}$ for TPA; we will define $P_0$ for other nonlinear effects in the sections that follow, as they can be used to efficiently compare the power levels where each nonlinearity is expected to manifest. The characteristic power is typically used to normalize the input/output power flow, i.e., $p = P/P_0$ (lower-case $p$ is used to indicate the normalization) or, returning in Eq.~\eqref{eq:CMTNormalizationKerrPsi}, $p = |\tilde\psi|^2 = |\tilde s|^2/P_0 = P/P_0$.

        The normalized version of the CMT ODE [Eq.~\eqref{eq:CMTKerrTPANormalized}] together with the output coupling expressions [Eqs.~\eqref{eq:CMTcouplingNormal}] are readily solvable and can be used to describe the temporal evolution of a cavity experiencing SPM.\cite{Soljacic2002,Yanik2003,Tsilipakos2014,Christopoulos2016} Furthermore, they are quite general and include an arbitrary level of linear loss through $r_Q$.\cite{Tsilipakos2014,Christopoulos2016} However, to further understand the impact of third-order nonlinearity in a cavity one should examine a CW version of the CMT equations, which reveals the phenomenon of \emph{optical bistability},\cite{GibbsBook} i.e., a hysteresis loop curve that allows to potentially access two different output states for the same input power. Such a response can always appear when nonlinearity is combined with positive feedback provided by the cavity. 
        Optical bistability was first studied theoretically in the 1980s and realized in free-space interferometric cavities.\cite{GibbsBook,HausBook} Later, in the early 2000s the PhC platform was used as the basis to obtain Kerr-induced optical bistability,\cite{Dharanipathy2014} with CMT being suitable for the modeling due to the small dimensions of the involved cavity.\cite{Soljacic2002,Yanik2003} More recently, bistability was investigated in integrated silicon photonics,\cite{Zhang2018a} as well as  hybrid plasmonic,\cite{Tsilipakos2014,Christopoulos2016} and 2D material-based cavities.\cite{Christopoulos2016a,Christopoulos2017} 

        For the mathematical simplification of the CMT ODEs, we will momentarily ignore TPA by setting $r\rsub{TPA} = 0$ and focus on CW conditions to find the scattering matrix of the system with elements
        \begin{equation}
            S_{k\ell} = \frac{\tilde \psi^-_k}{\tilde \psi^+_\ell}  = c_{k\ell} + \frac{\varpi_e\vartheta_{e,k}}{j(\delta+\mathbcal{s}|\tilde u|^2)+(1+r_Q)}, \label{eq:CMTscatteringSPM}
        \end{equation}
        which  depend on the stored energy in the cavity; they reduce to Eq.~\eqref{eq:CMTscattering} as $|\tilde u|^2 \rightarrow 0$. To better examine the behavior of Eq.~\eqref{eq:CMTscatteringSPM} we shall focus on a specific example of  a side-coupled traveling-wave resonator [Fig.~\ref{fig:CMTcouplingSchemes}(d)]. Using the respective quantities of Table~\ref{tab:CMTNLKerrCoefficients}, we reach\cite{Tsilipakos2014}
        \begin{equation}
            \frac{p^-_2}{p^+_1} = |S_{21}|^2 = \frac{(\delta+\mathbcal{s}|\tilde u|^2)^2+(1-r_Q)^2}{(\delta+\mathbcal{s}|\tilde u|^2)^2+(1+r_Q)^2}. \label{eq:CMTTransmissiontravelingwithenergy}
        \end{equation}
        Equation~\eqref{eq:CMTTransmissiontravelingwithenergy} is not imediately exploitable since it involves $|\tilde u|^2$ on the right-hand side; using only input/output power quantities is preferable for studying the behavior of the system. This can be achieved through the following approach\cite{Tsilipakos2014}
        \begin{equation}
            |\tilde u|^2 = \tau_i|\gamma\rsub{SPM}| |\tilde a|^2 = \frac{P^+_1 - P^-_2}{2/(\tau_i^2|\gamma\rsub{SPM}|)} = p_1^+ - p_2^-,
        \end{equation}
        utilizing the fact that the intrinsic quality factor is given by\cite{Christopoulos2019} $Q_i = \omega_0\tau_i/2 =  \omega_0(W\rsub{res}/P_i) = \omega_0(|\tilde a|^2/P_i)$ and that intrinsic power dissipation is $P_i = P^+_1 - P^-_2$, as conservation of energy dictates in this traveling wave configuration. Different expressions for $|\tilde u|^2$ emerge for other cavity types/coupling schemes; they are compiled in Table~\ref{tab:CMTNLKerrCoefficients} as well. Now, it is straightforward to write Eq.~\eqref{eq:CMTTransmissiontravelingwithenergy} as
        \begin{equation}
            \frac{p^-_2}{p^+_1} = \frac{[\delta+\mathbcal{s}(p^+_1 - p^-_2)]^2+(1-r_Q)^2}{[\delta+\mathbcal{s}(p^+_1 - p^-_2)]^2+(1+r_Q)^2}, \label{eq:CMTTransmissiontraveling}
        \end{equation}
        which is clearly a third-order polynomial of $p^-_2$ for a given input power level $p^+_1$. The simple polynomial form of Eq.~\eqref{eq:CMTTransmissiontraveling} cannot be reached for all possible cavity types and coupling schemes. For instance, for the side-coupled standing-wave configuration of Fig.~\ref{fig:CMTcouplingSchemes}(c), a system of two third order polynomials with two unknowns emerges.\cite{Christopoulos2016} 

\begin{figure}
            \centering
            \includegraphics[scale=1]{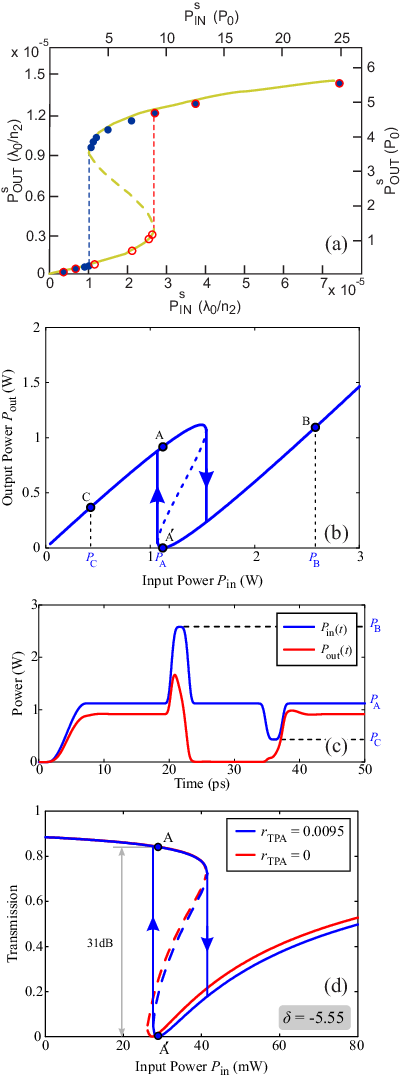}
            \caption{(a)~Bistable hysteresis loop realized in a directly coupled standing-wave nonlinear resonator. 100\% transmission (zero insertion loss) is achievable in the absence of losses. \COPYRIGHT{Solja{\v{c}}i{\'{c}} \emph{et al.}, Phys. Rev. E \textbf{66}, 055601(R) (2002)}{Soljacic2002}{2002 the American Physical Society}. (b)~Bistable hysteresis loop realized in a side coupled traveling-wave nonlinear resonator. Zero transmission (infinite extinction ratio) is achievable in the absence of losses. (c)~Self-controlled memory operation, realized using the CW curve of panel (b) and appropriate input pulses. \COPYRIGHT{Tsilipakos and Kriezis, J. Opt. Soc. Am. B \textbf{31}, 1698 (2014)}{Tsilipakos2014}{2014 Optica}. (d)~Impact of TPA in the bistablility loop. Marginal modifications are obtained since $r\rsub{TPA} \approx 0.01$. \COPYRIGHT{Tsilipakos \emph{et al.}, J. Lightwave Technol. \textbf{34}, 1333 (2016)}{Tsilipakos2016}{2016 IEEE}.}
            \label{fig:BistabilitySPM}
        \end{figure}

        As a third-order polynomial, Eq.~\eqref{eq:CMTTransmissiontraveling} may acquire one or three physically-acceptable (real and positive) solutions for a given $p^+_1$. Thus, multiple solutions can be potentially accessed with the actually obtained one depending on the initial conditions of the system. It can be shown (e.g., via a simple stability analysis, Sec.~\ref{sec:StabAnalysis}) that from the three possible solutions only two are stable and practically obtainable,\cite{Kaplan1982,GibbsBook} justifying the term \emph{bistability}. Bistable hysteresis loops of a directly-coupled and a side-coupled cavity are depicted in Fig.~\ref{fig:BistabilitySPM}(a),(b), respectively.\cite{Soljacic2002,Tsilipakos2014} A hysteris loop is clearly observed, with the two possible outputs indicated by the solid lines and the third unstable solution denoted by the dashed lines. Each configuration exhibits different characteristics. In the directly-coupled cavity, 100\% transmission can be achieved (in the absence of losses), allowing for zero insertion loss. On the other hand, in the side-coupled cavity a zero transmission is achieved (again in the absence of losses), setting the respective extinction ratio between the two possible outputs to infinity. 
        The presence of the bistable regime can be exploited to demonstrate all-optical switching or memory operations, as shown in Fig.~\ref{fig:BistabilitySPM}(c) where set/reset pulses of appropriate amplitudes are used to change states between points $\mathrm{A}$ and $\mathrm{A^\prime}$, following the paths $\mathrm{ABA^\prime}$ (set) and $\mathrm{A^\prime CA}$ (reset) in the CW curve of Fig.~\ref{fig:BistabilitySPM}(b).
        
        The polynomial form of Eq.~\eqref{eq:CMTTransmissiontraveling} also allows to determine the conditions required for optical bistability to appear and, more specifically, the relation between detuning and cavity loss. For any of the considered coupling schemes and regardless of the cavity type, it can be shown that optical bistability appears when\cite{Christopoulos2016,Kaplan1982} $\delta<-(1+r_Q)\sqrt{3}$ or $\delta>(1+r_Q)\sqrt{3}$, with the first inequality achieved when self-focusing nonlinear materials are involved and the second for defocusing materials. Then, to access the bistable part of the curve an appropriate level of input power is  required, typical in the order of few $P_0$. 

        Two-photon absorption can also be considered by allowing $r\rsub{TPA}\neq 0$. Since TPA constitutes a power-dependent loss mechanism, an additional polynomial equation is needed to correctly capture the behavior of the resonant system, leading to some extra complexity. However, a closed-form bistability limit can still be calculated, now involving  $r\rsub{TPA}$ as well.\cite{Tsilipakos2016} TPA may or may not have detrimental effects on the resonant cavity, either directly by the additional power-dependent losses that suppress bistability or indirectly by the influence of the generated free carriers. For example, in Fig.~\ref{fig:BistabilitySPM}(d) the primary effect of TPA losses is demonstrated for a configuration with $r\rsub{TPA}\approx 0.01$; larger values of $r\rsub{TPA}$ in the order of $\sim 0.1$ have more pronounced effects on the bistability curve but can be mitigated by, e.g., appropriately pre-shifting the operating frequency.\cite{Christopoulos2017} 

    \subsection{Single-channel nonlinearities: Saturable absorption}\label{subsec:SA}

        The second nonlinear effect that we will focus on is \emph{saturable absorption} (SA), i.e., the nonlinear saturation of the losses of a material when it is illuminated with light of high intensity. In terms of CMT and perturbation theory, SA is described by a decay rate term that depends on the stored energy in the cavity; the derivation of this term utilizing perturbation theory is discussed in Appendix~\ref{AppSub:DeltaOmegaVSaSA}. In this section, we will focus on the simplest possible form of SA (more realistic models will be discussed in Sec.~\ref{subsec:DynSA}). Specifically, we will examine a medium with instantaneous saturable absorption that also has a uniform field distribution inside it so that we can use Eq.~\eqref{eq:DeltaOmegaGrapheneSASimplified} of Appendix~\ref{AppSub:DeltaOmegaVSaSA}, i.e., \cite{Ma2014,Ataloglou2018}
        \begin{equation}
            \gamma\rsub{SA}(a) = \frac{\gamma\rsub{SA,0}}{1+|a|^2/W\rsub{sat}}.
            \label{eq:DeltaGammaSACMT}
        \end{equation}
        Parameter $\gamma\rsub{SA,0}$ is the (linear) loss decay rate induced by the SA medium for low intensities (i.e., in the absences of loss saturation). The linear losses of other materials in the resonator configuration can be captured by introducing an extra $\gamma\rsub{res}$ term.
        Furthermore, $W\rsub{sat}$ is the saturation energy, i.e., the value of the total stored energy in the cavity for which the SA decay rate is halved. $W\rsub{sat}$ (measured in $\mathrm{J}$) can be considered equivalent to, e.g., $\gamma\rsub{SPM}$ (measured in $1/\mathrm{Js}$), quantifying the threshold for the nonlinear phenomenon to manifest.
    
        Equation~\eqref{eq:DeltaGammaSACMT} can be readily incorporated in the CMT ODE, which when only SA is present reads\cite{Ataloglou2018}
        \begin{align}
            \fullFRAC{\tilde a}{t} = &-j(\omega-\omega_0) \tilde a - \left(\frac{\gamma\rsub{SA,0}}{1+|\tilde a|^2/W\rsub{sat}}+\gamma\rsub{res}+\gamma\rsub{rad}+\gamma_e\right) \tilde a \nonumber \\
            &+ \mu_e \tilde s^+. 
            \label{eq:CMTSA}
        \end{align}
        In Eq.~\eqref{eq:CMTSA} we have decomposed the intrinsic decay rate into its radiation and resistive counterparts, with the latter further decomposed into linear and nonlinear (saturable) terms. 
        As with the Kerr effect, it is beneficial to normalize the CMT ODE. In this case, this is performed with respect to the saturation energy, i.e., by defining
        \begin{subequations}
            \begin{align}
                \tilde u(t) &=  \tilde a(t)/\sqrt{W\rsub{sat}}, \\ 
                \tilde \psi(t) &= \tilde s(t)/\sqrt{P\rsub{0,SA}}, 
            \end{align}
            \label{eq:CMTNormalizationSA}
        \end{subequations}
        with the SA characteristic power being
        \begin{equation}
            P\rsub{0,SA} = \frac{W\rsub{sat}}{\tau_2}.
            \label{eq:CharPowerSA}
        \end{equation}
        The normalized version of the CMT ODE [Eq.~\eqref{eq:CMTSA}] is
        \begin{equation}
            \fullFRAC{\tilde u}{t^\prime} = -j\delta \tilde u - \left(\frac{1}{1+|\tilde u|^2}+r\rsub{res}+r\rsub{rad}+r_e\right) \tilde u + \varpi_e \tilde\psi^+, 
            \label{eq:CMTSANormal}
        \end{equation}
        where $\tau\rsub{SA,0}$ has been used for the normalizations, i.e., $\delta = (\omega-\omega_0)\tau\rsub{SA,0}$, $t^\prime = t/\tau\rsub{SA,0}$, $r\rsub{res} = \tau\rsub{SA,0}/\tau\rsub{res}$, $r\rsub{rad} = \tau\rsub{SA,0}/\tau\rsub{rad}$, and $r_e = \tau\rsub{SA,0}/\tau_e$. Furthermore, $\tau_2$ is set to $\tau\rsub{SA,0}$ [configurations of Figs.~\ref{fig:CMTcouplingSchemes}(b),(c)] or $\tau\rsub{SA,0}/2$ [configurations of Figs.~\ref{fig:CMTcouplingSchemes}(a),(d)]. Note that it is straightforward to include additional nonlinear terms in Eq.~\eqref{eq:CMTSANormal}. For example, if SPM is also present, the detuning term is simply modified to $\delta + r\rsub{SPM}|\tilde u|^2$, with $r\rsub{SPM} = \gamma\rsub{SPM}\tau\rsub{SA,0}W\rsub{sat}$ quantifying (in normalized terms) the strength of SPM compared to SA.\cite{Ataloglou2018} Similarly, TPA can be included as an additional decay rate of the form $r\rsub{TPA}|\tilde u|^2 = \gamma\rsub{TPA}\tau\rsub{SA,0}W\rsub{sat}|\tilde u|^2$.\cite{Ataloglou2018}

\begin{figure}
            \centering
            \includegraphics[scale=0.96]{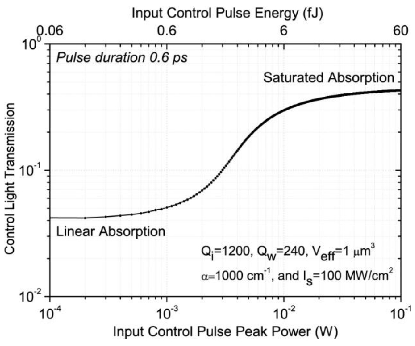}
            \caption{Influence of saturable absorption in the output of a single cavity system. Low transmission is obtained for low input intensities while high transmission due to loss saturation is achieved for higher input intensities. \COPYRIGHT{Ma \emph{et al.}, IEEE J. Quantum Elect. \textbf{50}, 1019 (2014)}{Ma2014}{2014 IEEE}.}
            \label{fig:SATransmission}
        \end{figure}

        Equation~\eqref{eq:CMTSANormal} can be also transformed in a polynomial form with the sole assumption of low radiation, i.e., $r\rsub{rad}\rightarrow 0$.\cite{Ataloglou2018} The obtained polynomial will be of third order once again, resulting in a possible bistable behavior as well. The duality of the Kerr effect and SA is well known and described in the literature.\cite{HausBook} However, in contemporary integrated resonant systems SA is mostly utilized for switching operation:\cite{Ma2014,Ataloglou2018} a low transmission state is achieved for a low intensity input due to high linear absorption and a high transmission state is achieved for high intensities where resistive loss are saturated and, thus, suppressed (Fig.~\ref{fig:SATransmission}). More elaborate realizations for light routing exploiting the critical coupling condition in add-drop filter configurations have been also suggested recently, utilizing graphene as the saturable absorber.\cite{Christopoulos2020JAP, Nousios2022}

    \subsection{Multi-channel nonlinearities: Cross-phase modulation, cross saturable absorption, high-harmonic generation, and four-wave mixing}\label{subsec:MultiChannel}

        Having examined two important singe-channel nonlinear effects, in this section we will focus on multi-frequency schemes. 
        Let us assume two waves at two distinct wavelengths, both lying close to a different resonance mode of the same cavity. In linear cavities, the two modes are orthogonal and the two waves do not interact. However, in the presence of nonlinearity each wave influences the other by shifting its resonance frequency, a phenomenon known as \emph{cross-phase modulation} (XPM) in systems with third-order nonlinearities. In Appendix~\ref{AppSub:DeltaOmegaVSaXPMTHGFWM}, we apply perturbation theory to derive detailed expressions for  these frequency shifts. Here, we will introduce this shift in the CMT framework. 
        
        The  shift experienced by the $k$-th mode due to the presence of the $\ell$-th mode ($k,\,\ell=\{1,2\}$ and $k\neq\ell$) is of the form
        \begin{equation}
            \Delta\omega_{\mathrm{XPM},k}(a_\ell) = -2\gamma_{\mathrm{XPM},k\ell}|a_\ell(t)|^2.
            \label{eq:DeltaOmegaXPMCMT}
        \end{equation}
       As expected, it holds that $\gamma_{\mathrm{XPM},k\ell} = \gamma_{\mathrm{XPM},\ell k}$ [see Eqs.~\eqref{eq:DeltaOmegaGrapheneXPM}~and~\eqref{eq:kappaGrapheneXPM} in Appendix~\ref{AppSub:DeltaOmegaVSaXPMTHGFWM}]. 
       SPM is also present, acting on each wave separately, i.e., the total resonance frequency shift of the $k$-th mode is the sum of the quantities in Eqs.~\eqref{eq:DeltaOmegaKerrTPACMT}~and~\eqref{eq:DeltaOmegaXPMCMT}. Assuming only self- and cross-phase modulation, and a cavity of which only two modes are excited, the normalized CMT ODEs that describe these modes become
        \begin{align}
            \fullFRAC{\tilde u_k}{t^\prime} = &j(-\delta_k-r_{\mathrm{SPM},k}|\tilde u_k|^2-2r_{\mathrm{XPM},k\ell}|\tilde u_\ell|^2) \tilde u_k \nonumber \\
            &- (r_{i,k}+r_{e,k}) \tilde u_k + \varpi_{e,k} \tilde \psi^+_k.
            \label{eq:CMTXPMNormalized}
        \end{align}
        To reach the normalized CMT ODE, we have used $\gamma_{\mathrm{XPM},12}$ for the normalization, so that the XPM characteristic power is defined through 
        \begin{equation}
            P\rsub{0,XPM} = \frac{1}{\tau_2^2|\gamma\rsub{XPM,12}|},
            \label{eq:CharPowerXPM}
        \end{equation}
        and the normalized cavity amplitudes and input/output wave amplitudes are $\tilde u_k = \sqrt{\tau_1|\gamma\rsub{XPM,12}|} \tilde a$ and $\tilde \psi_k = \sqrt{\tau_2^2|\gamma\rsub{XPM,12}|} \tilde s_k$, respectively. It should be noted that the same normalization must be applied to all equations and, thus, $\tau_1$ and $\tau_2$ correspond to the lifetimes of one mode, e.g., mode 1 here and, consequently, $r_{i,k} = \tau_1\gamma_{i,k}$ and $r_{e,k} = \tau_1\gamma_{e,k}$. Similarly, $r_{\mathrm{XPM},k\ell} = \gamma_{\mathrm{XPM},k\ell}/|\gamma_{\mathrm{XPM},12}|$ and for the examined scenario with only 2 waves, it either equals $+1$ or $-1$; its introduction is more meaningful in systems with more than two waves where it deviates from $\pm 1$ and represents how strongly the $k$-th wave is affected by the $\ell$-th wave in comparison with how wave 2 affects wave 1.\cite{Christopoulos2018}
        \blue{Although Eqs.~\eqref{eq:CMTXPMNormalized} can be cast in polynomial form as with SPM or SA, one understands that as the number of modes and thus the coupled polynomial equations increase, the practical usefulness of this approach becomes questionable. Nevertheless, following the steps of Sec.~\ref{subsec:SPM}, the derivation of polynomial equations should be straightforward.}

\begin{figure}
            \centering
            \includegraphics[scale=1]{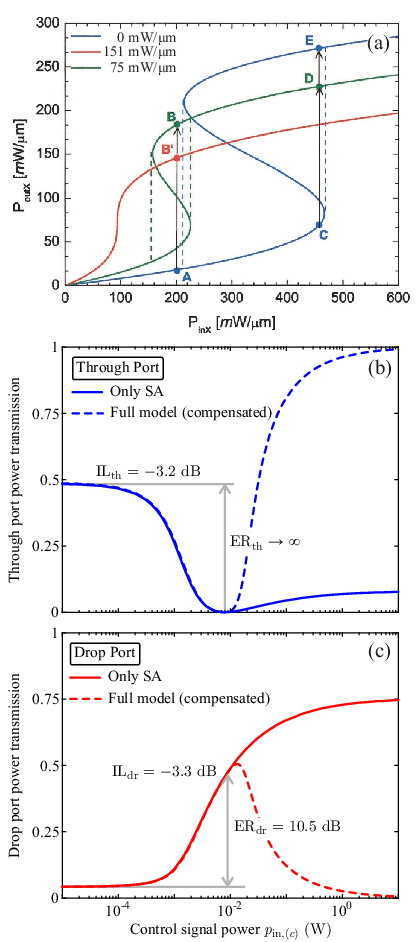}
            \caption{(a)~Control of bistable operation through XPM. When a pump wave is inserted in the cavity, the hysteresis loop is modified (red and green curves), allowing for switching between bistable states (passing from point C to E through D). \COPYRIGHT{Yanik \emph{et al.}, Opt. Lett. \textbf{28}, 2506 (2003)}{Yanik2003a}{2003 Optica}. (b)~Through- and (c)~drop-port transmission of an add-drop nonlinear disk cavity operating based on cross-SA. In the absence of pumping, the weak probe signal is mostly routed in the through port while for a specific pumping intensity, it can be completely rerouted to the drop port. \COPYRIGHT{Christopoulos \emph{et al.}, J. Appl. Phys. \textbf{127}, 223102 (2020)}{Christopoulos2020JAP}{2020 AIP Publishing}.}
            \label{fig:XPMBistability}
        \end{figure}

        Cross-phase modulation allows for controlling the response of a weak probe signal with a strong pump one. XPM has been utilized to demonstrate all-optical control of bistable PhC cavities allowing for  sophisticated memory operations,\cite{Yanik2003a,Yanik2004,Fasihi2014} switching via phase manipulation in silicon rings,\cite{Daniel2012,Daniel2012a} and nonreciprocal transmission in glass cylindrical cavities.\cite{DelBino2017} For example, Fig.~\ref{fig:XPMBistability}(a) demonstrates how XPM modifies the  hysteresis loop of a nonlinear cavity via  a strong pump wave.\cite{Yanik2003a} This action allows for the switching between two bistable points (e.g., passing from point C to E through D in this example), acting as an optical memory component.

        The idea of cross-wave control is applicable in systems with SA as well. Specifically, the presence of a strong pump wave can saturate the losses that are experienced by a weak probe signal, leading to high or low transmission, depending on the configuration. In CMT terms, this behavior can be captured by the respective decay rates
        \begin{subequations}
            \begin{align}
                \gamma_{\mathrm{SA},s}(a_p) &= \frac{\gamma_{\mathrm{SA},s,0}}{1+|a_p|^2/W\rsub{sat}}, \\
                \gamma_{\mathrm{SA},p}(a_p) &= \frac{\gamma_{\mathrm{SA},p,0}}{1+|a_p|^2/W\rsub{sat}},
            \end{align}
            \label{eq:DeltaGammaXSACMT}
        \end{subequations}
        which both depend on the energy of the pump mode $|a_p|^2$. We have seemingly arbitrarily introduced these terms here but they can emerge rigorously using perturbation theory.\cite{Ataloglou2018,Christopoulos2020JAP} The described approach is only accurate when the signal wave is significantly weaker than the pump and other cross-induced nonlinearities can be safely ignored.\cite{Christopoulos2020JAP} Cross saturable absorption has  been exploited to demonstrate more elaborate operations, such as a routing scheme achieved in a disk cavity in an add-drop configuration [Fig.~\ref{fig:XPMBistability}(b,c)].\cite{Christopoulos2020JAP,Nousios2022}

        Up to this point, we have not discussed a completely different type of interaction that appears when at least two waves are present in the same nonlinear medium, namely wave-mixing effects. For example, in a scenario involving two waves, the nonlinear interplay between the waves at $\omega_1$ and $\omega_2$ creates two new (idler) waves at $2\omega_1-\omega_2$ and $2\omega_2-\omega_1$. This effect is called \emph{degenerate four-wave mixing} (DFWM) and appears in media with third-order nonlinearity. 
        When $\omega_1$ and $\omega_2$ are well separated in frequency (as in the XPM and cross-SA scenarios that have been examined so far), the emerging waves do not fall in the vicinity of either of the two resonances. That is why we have ignored the respective terms in, e.g., Eqs.~\eqref{eq:CMTXPMNormalized}.

        Before discussing DFWM [or the more general case of four-wave mixing (FWM)] and how to treat it with CMT, we will first talk about nonlinear harmonic generation. Harmonic generation appears in any nonlinear medium regardless of its nonlinearity order; second-harmonic generation (SHG) in $\chi^{(2)}$ media, third-harmonic generation (THG) in $\chi^{(3)}$ media, etc, and it describes the generation of a new wave at the respective harmonic frequency of the fundamental wave. Here, we will focus on THG,\cite{Hashemi2009,Christopoulos2018} noting that there also exist versions of the framework for SHG in the literature.\cite{Rodriguez2007,Li2018a} 

        In Appendix~\ref{AppSub:DeltaOmegaVSaXPMTHGFWM} we discuss perturbation theory for THG and how it differs from the so-far examined nonlinear phenomena. To describe THG, two CMT ODEs are necessary, one to capture the pump wave at $\omega_1$ and another for the produced (signal) wave at $\omega_3 = 3\omega_1$. Thus, a cavity with two resonant modes close to these frequencies is desired. The effect of THG is described by the quantities\cite{Rodriguez2007}
        \begin{subequations}
            \begin{align}
                \Delta\tilde\omega_{\mathrm{THG},1}(a_1,a_3) &= -3\tilde\beta_{\mathrm{THG},1}\frac{(a_1^*)^2 a_3}{a_1}, \\
                \Delta\tilde\omega_{\mathrm{THG},3}(a_1,a_3) &= - \tilde\beta_{\mathrm{THG},3}\frac{a_1^3}{a_3},
            \end{align}
            \label{eq:DeltaOmegaTHGCMT}
        \end{subequations}
        which can be incorporated in the CMT ODEs. Note that not only the mode at $\omega_3$ is generated by the nonlinear process but also the mode at $\omega_1$ is affected through the DFWM interaction $\omega_3-2\omega_1 = \omega_1$. The quantities $\Delta\tilde\omega_{\mathrm{THG},k}$ are complex in general, describing the process of pump depletion to produce the signal wave or the backward process of frequency mixing between pump and signal waves, as well as possible nonlinear shifts of each resonance frequency. Finally, the nonlinear parameters $\tilde\beta_{\mathrm{THG},1}$ and $\tilde\beta_{\mathrm{THG},3}$ are related, 
        see Appendix~\ref{AppSub:DeltaOmegaVSaXPMTHGFWM} for more details.

        Using the normalized CMT form of Eq.~\eqref{eq:CMTXPMNormalized}, it is straightforward to write the respective equations for THG including SPM and XPM contributions as well\cite{Rodriguez2007}
        \begin{align}
            \fullFRAC{\tilde u_k}{t^\prime} = &j(-\delta_k-r_{\mathrm{SPM},k}|\tilde u_k|^2-2r_{\mathrm{XPM},k\ell}|\tilde u_\ell|^2) \tilde u_k - (r_{i,k}+r_{e,k}) \tilde u_k \nonumber \\
            & - j\tilde r_{\mathrm{THG},k}\Phi_k(\tilde u_1, \tilde u_3) + \varpi_{e,k} \tilde \psi^+_k,
            \label{eq:CMTTHGNormalized}
        \end{align}
        where $k,\,\ell=\{1,3\}$, $k\neq\ell$, and we have introduced the function
        \begin{equation}
            \Phi_k(\tilde u_1, \tilde u_3) = \left\{
                \begin{array}{ll}
                    3 (\tilde u_1^*)^2 \tilde u_3, & k = 1 \\
                      \tilde u_1^3               , & k = 3
                \end{array}
            \right.
            \label{eq:CMTTHGPhiFunction}
        \end{equation}
        for compactness of the notation.
        In Eq.~\eqref{eq:CMTTHGNormalized},  $\tilde\beta_{\mathrm{THG},3}$ has been used for the normalization, i.e., $\tilde u_k = \sqrt{\tau_1|\tilde\beta\rsub{THG,3}|} \tilde a$ and $\tilde \psi_k = \sqrt{\tau_2^2|\tilde\beta\rsub{THG,3}|} \tilde s_k$, and $\tau_1$, $\tau_2$ correspond to lifetimes of the mode at $\omega_3$, i.e., $\tau_1$ equals either $\tau_{i,3}$ or $\tau_{e,3}$, in accordance with Table~\ref{tab:CMTNLKerrCoefficients} (similarly for $\tau_2$). This impacts all the involved nonlinear parameters which become $r_{\mathrm{SPM},k} = \gamma_{\mathrm{SPM},k}/|\tilde\beta\rsub{\mathrm{THG},3}|$, $r_{\mathrm{XPM},k\ell} = \gamma_{\mathrm{XPM},k\ell}/|\tilde\beta\rsub{\mathrm{THG},3}|$, and $\tilde r_{\mathrm{THG},k} = \tilde\beta_{\mathrm{THG},k}/|\tilde\beta\rsub{\mathrm{THG},3}|$, so that the THG characteristic power is defined as
        \begin{equation}
            P\rsub{0,THG} = \frac{1}{\tau_2^2|\beta\rsub{THG,3}|}.
            \label{eq:CharPowerTHG}
        \end{equation}

        \begin{figure}
            \centering
            \includegraphics[scale=1]{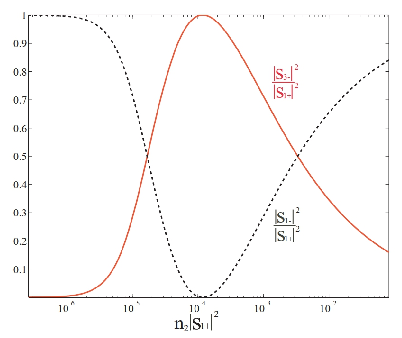}
            \caption{Third-harmonic generation conversion efficiency (red solid line) of a PhC cavity resonant system. 100\% CE can be achieved for a specific pump power lever, above which the backwards mixing process from $\omega_3$ to $\omega_1$ dominates and suppresses CE. \COPYRIGHT{Rodriguez \emph{et al.}, Opt. Express \textbf{15}, 7303 (2007)}{Rodriguez2007}{2007 Optica}.}
            \label{fig:THGCE}
        \end{figure}

        Harmonic generation in resonant systems has been shown to possess two important advantages compared with more conventional demonstrations in optical fibers or integrated waveguides. Firstly, to obtain efficient THG in waveguides, a phase-matching condition should be met: The propagation constant $\beta$ of the signal wave should be three times that of the pump wave, i.e., $\beta_3 = 3\beta_1$.\cite{BoydBook} This condition is not required in resonant cavities.\cite{Rodriguez2007} Rather, two resonant modes, one at $\omega_1$ and one at $\omega_3$ which will accommodate and resonantly enhance the pump and signal waves, are typically desired to boost conversion efficiency. This conditions is more easily met by tuning the dimensions of the cavity. Furthermore, complete conversion of the pump wave is achievable in resonant systems, requiring input powers of the order of $P\rsub{0,THG}$.\cite{Rodriguez2007,Hashemi2009} This 100\% conversion efficiency (CE) is achieved for a specific power lever (see Fig.~\ref{fig:THGCE}) and beyond that level the backwards mixing process from $\omega_3$ to $\omega_1$ dominates and suppresses CE. As expected, resistive and radiation loss limit the conversion efficiency, as do SPM and XMP processes. The latter, however, can be compensated by appropriately pre-shifting the operating frequencies to accommodate for the SPM- and XPM-induced nonlinear resonance frequency shifts.\cite{Hashemi2009}

        Next, we choose the degenerate four-wave mixing process to illustrate how wave-mixing effects are included in the CMT framework. Any other multi-wave interaction (e.g., stimulated Raman scattering\cite{Kippenberg2004,Yang2007}) can be included in a similar manner. In DFWM, a strong pump wave at $\omega_1$ and a weaker signal wave at $\omega_2$ interact to produce an idler wave at $\omega_3 = 2\omega_1 - \omega_2$. In CMT terms, the contribution of the nonlinear process in each of the three involved cavity mode amplitudes is given by\cite{Ramirez2011,Zeng2014,Lin2014,Christopoulos2018,Christopoulos2020JOSAB}
        \begin{subequations}
            \begin{align}
                \Delta\tilde\omega_{\mathrm{DFWM},1}(a_1,a_2,a_3) &= -2\tilde\beta_{\mathrm{DFWM},1}\frac{a_1^*a_2a_3}{a_1}, \\
                \Delta\tilde\omega_{\mathrm{DFWM},2}(a_1,a_2,a_3) &= -\tilde\beta_{\mathrm{DFWM},2}\frac{a_1^2a_3^*}{a_2}, \\
                \Delta\tilde\omega_{\mathrm{DFWM},3}(a_1,a_2,a_3) &= -\tilde\beta_{\mathrm{DFWM},3}\frac{a_1^2a_2^*}{a_3}.
            \end{align}
            \label{eq:DeltaOmegaDFWMCMT}
        \end{subequations}
        The approach to extract Eqs.~\eqref{eq:DeltaOmegaDFWMCMT} using electromagnetic arguments is included in Appendix~\ref{AppSub:DeltaOmegaVSaXPMTHGFWM}. In the same Appendix the relation between the $\tilde\beta\rsub{DFWM}$ parameters is also included.
        As with THG, it is not difficult to reach the final normalized CMT ODEs
        \begin{align}
            \fullFRAC{\tilde u_k}{t^\prime} &= j(-\delta_k-r_{\mathrm{SPM},k}|\tilde u_k|^2-2r_{\mathrm{XPM},k\ell}|\tilde u_\ell|^2-2r_{\mathrm{XPM},km}|\tilde u_m|^2) \tilde u_k \nonumber \\
            & - (r_{i,k}+r_{e,k}) \tilde u_k - j\tilde r_{\mathrm{DFWM},k}\Phi_k(\tilde u_1, \tilde u_2, \tilde u_3) + \varpi_{e,k} \tilde \psi^+_k,
            \label{eq:CMTDFWMNormalized}
        \end{align}
        where now $k,\,\ell,m=\{1,2,3\}$, $k\neq\ell\neq m$, and we have introduced the function
        \begin{equation}
            \Phi_k(\tilde u_1, \tilde u_2, \tilde u_3) = \left\{
                \begin{array}{ll}
                    2\tilde u_1^*\tilde u_2\tilde u_3, & k = 1 \\
                    \tilde u_1^2\tilde u_3^*         , & k = 2 \\
                    \tilde u_1^2\tilde u_2^*         , & k = 3 \\
                \end{array}
            \right. 
            \label{eq:CMTDFWMPhiFunction}
        \end{equation}
        for compactness of the notation. For the normalization of Eqs.~\eqref{eq:CMTDFWMNormalized} the nonlinear parameter $\tilde\beta\rsub{DFWM,3}$ and the respective lifetimes of the $\omega_3$ mode are used, so that $\tilde u_k = \sqrt{\tau_1|\tilde\beta\rsub{DFWM,3}|} \tilde a_k$, $\tilde \psi_k = \sqrt{\tau_2^2|\tilde\beta\rsub{DFWM,3}|} \tilde s_k$, and
        \begin{equation}
            P\rsub{0,DFWM} = \frac{1}{\tau_2^2|\tilde\beta\rsub{DFWM,3}|}.
            \label{eq:CharPowerDFWM}
        \end{equation}

        The advantages of realizing DFWM in cavities rather than waveguides are more or less the same as with THG: mainly, a strict phase-matching condition is not required.\cite{Ramirez2011} The optimum conversion efficiency in this case is 50\%, obtained in ideal cavities without losses and with compensated SPM and XPM-induced resonance frequency shifts.\cite{Zeng2014,Christopoulos2020JOSAB} Finally, we should highlight the rich dynamics of a nonlinear system supporting three modes that interact with each other. Multiple instabilities and limit cycles can be obtained,\cite{Ramirez2011,Lin2014} allowing for additional functionality. The presented CMT framework allows to study such nonlinear dynamics via an appropriate stability analysis, see Sec.~\ref{sec:StabAnalysis} for a brief introduction.

        Before concluding this section, we shall once again mention that the multi-channel nonlinearities presented in here (XPM, THG, DFWM) consist only a small fraction of the phenomena that can be described with the CMT. Notable other multi-channel nonlinear effects that have been described with CMT using a similar approach are second-harmonic generation,\cite{Rodriguez2007} cascaded sum frequency generation,\cite{McLaughlin2022} as well as stimulated\cite{Yang2007} and cascaded Raman scattering\cite{Kippenberg2004}.
        \blue{Kerr frequency combs deserve a special mention, since CMT was initially used to describe their response as well as to extract metrics and thresholds.\cite{Chembo2010,Chembo2016} However, due to the involvement of a very large number of modes (and thus coupled ODEs) that should be used to fully capture the spectrum of the comb, a more efficient framework was later developed, namely, the Lugiato–Lefever equation (LLE). It is numerically more efficient compared to CMT,\cite{Chembo2016}, but lacks the ability to extract metrics and thresholds.\cite{Chembo2016,Pitilakis2024} }
            
\section{Coupled-Mode Theory for Systems with Non-Instantaneous Nonlinear Phenomena}\label{sec:NoninstantNonLinearCMT}

    In this section, we will introduce in the CMT formalism important nonlinear phenomena that do not exhibit an instantaneous response but, rather, their response times are  comparable (or slower) with the ps time-scales of photonics. First, we will focus on silicon to present the \emph{free-carrier effect} (FCE) and the \emph{thermo-optic effect} (TOE). 
    We will also present a more physically accurate, non-instantaneous and spatially dependent model for saturable absorption, valid for any material system. The approach followed to describe the aforementioned three effects should be considered as a guideline for including any other similar perturbative nonlinearity in the CMT formalism.

    \subsection{Non-instantaneous nonlinearities in silicon}\label{subsec:FCEsTOE}

        In silicon, the density of free carriers in the conduction band dictates the absorption and refractive index of the material.  \emph{Free-carrier absorption} (FCA) and \emph{free-carrier dispersion} (FCD) are collectively termed free-carrier effects. 
        FCE may be induced either electrically, by injecting carriers via a $pn$-junction, or optically by photo-generating free carriers via linear or nonlinear absorption. Since Si is transparent at infrared wavelengths, we will assume TPA as the carrier generation mechanism. The physics of FCE are discussed in Appendix~\ref{AppSub:DeltaOmegaVSaFCEsTOE}; here, we will only examine how it is introduced in CMT through the induced complex resonance frequency shift
        \begin{equation}
            \Delta\tilde\omega\rsub{FCE}(\bar N_c) = \tilde\gamma_{N,\mathrm{FCE}} \bar N_c(t) = (\gamma_{N,\mathrm{FCD}}+j\gamma_{N,\mathrm{FCA}})\bar N_c(t).
            \label{eq:DeltaOmegaFCEs}
        \end{equation}
        To describe FCE, we have introduced the spatially averaged carrier density $\bar N_c(t)$ (measured in $1/\mathrm{m}^3$), i.e., a new time-dependent variable quantifying, in a weighted manner, the density of carriers in the conduction band of Si. Its time evolution is governed by a separate differential equation which, in its simplest form, is\cite{Zhang2013a}
        \begin{equation}
            \fullFRAC{\bar N_c(t)}{t} = -\frac{\bar N_c(t)}{\tau_c} + \gamma_N|\tilde a(t)|^4.
            \label{eq:CarrierRateEquationCMT}
        \end{equation}
        In Eq.~\eqref{eq:CarrierRateEquationCMT}, $\tau_c$ is the effective carrier lifetime, which describes the recombination lifetime of free carriers in Si, while phenomenologically accounting for carrier diffusion or other relevant phenomena.\cite{Moille2016} Additionally, the nonlinear parameter $\gamma_N$ (measured in $\mathrm{1/J^2sm^3}$) introduced here should not be confused with $\gamma_{N,\mathrm{FCD/FCA}}$ (measured in $\mathrm{m^3/s}$) of Eqs.~\eqref{eq:DeltaOmegaFCEs}.

\begin{figure}
            \centering
            \includegraphics[scale=1]{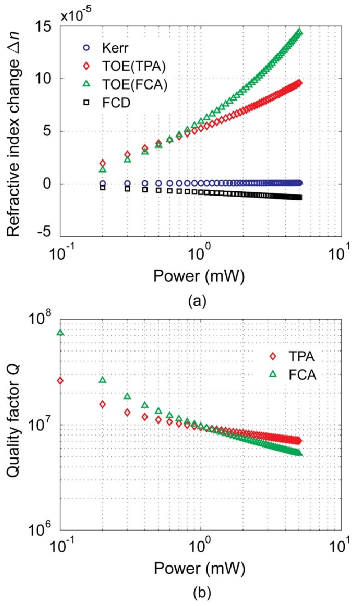}
            \caption{(a) Refractive index change and (b) quality factor reduction of an integrated SOI ring cavity, associated with the various nonlinearities of Si as functions of input power. Refractive index change is a measure of the nonlinear resonance frequency shift while $Q$-factor reduction a measure of nonlinear losses. \COPYRIGHT{Wang \emph{et al.}, J. Lightwave Technol. \textbf{31}, 313 (2013)}{Wang2013a}{2013 IEEE}.}
            \label{fig:AllNLinSIcomparison}
        \end{figure}

        Equation~\eqref{eq:CarrierRateEquationCMT} complements the basic CMT ODE~\eqref{eq:CMTcoupling}, leading to the following system of coupled ODEs
        \begin{subequations}
            \begin{align}
                \fullFRAC{\tilde a}{t} = &j[-(\omega-\omega_0)-\gamma\rsub{SPM}|\tilde a|^2 + \gamma_{N,\mathrm{FCD}}\bar N_c] \tilde a \nonumber \\
                &- (\gamma_i+\gamma_e+\gamma\rsub{TPA}|\tilde a|^2 + \gamma_{N,\mathrm{FCA}}\bar N_c) \tilde a + \mu_e \tilde s^+, \label{eq:CMTFCEsAmplitude} \\
                \fullFRAC{\bar N_c}{t} = &-\frac{\bar N_c}{\tau_c} + \gamma_N|\tilde a|^4,
            \end{align}
            \label{eq:CMTFCEs}
        \end{subequations}
        which are solved simultaneously. In Eq.~\eqref{eq:CMTFCEsAmplitude}, both the Kerr effect and TPA are included in the cavity-amplitude ODE. 
        TPA and FCA both induce losses, which cumulatively decrease the $Q$-factor of the cavity. On the other hand, 
        note that FCD induces a blue-shift on the resonance frequency [$\gamma_{N,\mathrm{FCD}}>0$, see Eq.~\eqref{eq:DeltaOmegaDynamicFCEs}] and may or may not counteract the Kerr effect which can be of either sign (self-focusing and defocusing materials). Silicon is a self-focusing material ($n_2>0$) so FCD and SPM oppose each other [see black and blue markers in Fig.~\ref{fig:AllNLinSIcomparison}(a)].

        To better understand the impact of FCE, let us examine the case where $\tau_c \ll \tau_\ell$ meaning that FCE acts quite faster than the time-scale at which the cavity responds (i.e., can be considered instantaneous). For integrated SOI cavities, $\tau_c$ is in the ns time-scale so this scenario is accurate for cavities with very large $Q$-factors ($>10^7$). The carrier density is then $\bar N_c = \tau_c\gamma_N|\tilde a|^4$ and the FCD/FCA terms in Eq.~\eqref{eq:CMTFCEsAmplitude} become $\Delta\omega\rsub{FCD/FCA} = \tau_c\gamma_{N,\mathrm{FCD/FCA}}\gamma_N |\tilde a|^4 = \gamma\rsub{FCD/FCA} |\tilde a|^4$, indicating that FCEs scale with the square of the stored energy ($W\rsub{res} \equiv |a|^2$). 
        Parameters $\gamma\rsub{FCD/FCA} = \tau_c\gamma_{N,\mathrm{FCD/FCA}}\gamma_N$, measured in $\mathrm{1/J^2s}$, are introduced here to better indicate this fact. 
        In Si, TPA is strong ($r\rsub{TPA}\approx 0.2$) and FCD typically dominates over the Kerr effect (Fig.~\ref{fig:AllNLinSIcomparison}). Even in different material systems with much smaller $r\rsub{TPA}$, e.g., $\sim 0.01$, FCE can still be detrimental.\cite{Christopoulos2017} In some cases, FCE is suppressed by reducing the actual carrier lifetime $\tau_c$ through, e.g., a carrier sweeping mechanism.\cite{Tsilipakos2016}

        \begin{table*}[!t]
            \renewcommand*{\arraystretch}{1.2}
            \centering
            \begin{tabular}{cccccccc}
                \hline\hline
                ~ & \multicolumn{5}{c}{standing-wave cavity$^1$} & ~ & traveling-wave cavity$^2$ \\
                \cline{2-6} \cline{8-8}
                ~ & direct coupling, one port & ~ & direct coupling, two ports & ~ & side coupling, two ports & ~ & side coupling, two ports \\
                \hline
                $\tau_3$ & $\tau_e/\sqrt[3]{4}$ & ~ & $\tau_e$ & ~ & $\tau_e$ & ~ & $\tau_i/\sqrt[3]{4}$ \\
                $\tau_4$ & $\tau_e/\sqrt[5]{8}$ & ~ & $\tau_e$ & ~ & $\tau_e$ & ~ & $\tau_i/\sqrt[5]{8}$ \\
                $\tau_c^\prime$ & $\tau_c/\tau_e$ & ~ & $\tau_c/\tau_e$ & ~ & $\tau_c/\tau_e$ & ~ & $\tau_c/\tau_i$ \\
                $\tau_\theta^\prime$ & $\tau_\theta/\tau_e$ & ~ & $\tau_\theta/\tau_e$ & ~ & $\tau_\theta/\tau_e$ & ~ & $\tau_\theta/\tau_i$ \\
                $\varpi_{e,\mathrm{FCE}}$ & $2\sqrt[4]{1/\tau_c^\prime}$ & ~ & $\sqrt[4]{1/\tau_c^\prime}$ & ~ & $j\sqrt[4]{1/\tau_c^\prime}$ & ~ & $j2\sqrt{r_Q}\sqrt[4]{1/\tau_c^\prime}$ \\[2pt]
                $\boldsymbol{\Theta}_{e,\mathrm{FCE}}$ & $[\sqrt[4]{\tau_c^\prime}]$ & ~ & $[\sqrt[4]{\tau_c^\prime}~~\sqrt[4]{\tau_c^\prime}]^T$ & ~ & $[j\sqrt[4]{\tau_c^\prime}~~j\sqrt[4]{\tau_c^\prime}]^T$ & ~ & $[0~~j\sqrt{r_Q}\sqrt[4]{\tau_c^\prime}]^T$ \\[2pt]
                $\varpi_{e,\mathrm{TOE}}$ & $2\sqrt[6]{1/\tau_\theta^\prime}$ & ~ & $\sqrt[6]{1/\tau_\theta^\prime}$ & ~ & $j\sqrt[6]{1/\tau_\theta^\prime}$ & ~ & $j2\sqrt{r_Q}\sqrt[6]{1/\tau_\theta^\prime}$ \\[2pt]
                $\boldsymbol{\Theta}_{e,\mathrm{TOE}}$ & $[\sqrt[6]{\tau_\theta^\prime}]$ & ~ & $[\sqrt[6]{\tau_\theta^\prime}~~\sqrt[6]{\tau_\theta^\prime}]^T$ & ~ & $[j\sqrt[6]{\tau_\theta^\prime}~~j\sqrt[6]{\tau_\theta^\prime}]^T$ & ~ & $[0~~j\sqrt{r_Q}\sqrt[6]{\tau_\theta^\prime}]^T$ \\[2pt]
                \hline\hline
                \multicolumn{8}{l}{\footnotesize$^1$ Fabry-P\'erot-like cavities, $^2$ Ring- or disk-like cavities}
            \end{tabular}
            \caption{Normalized nonlinear CMT parameters of non-instantaneous nonlinear effects for typical guided-wave cavities and coupling schemes (see also Fig.~\ref{fig:CMTcouplingSchemes}).}
            \label{tab:CMTNLFCEsTOECoefficients}
        \end{table*}

        For the numerical handling of Eqs.~\eqref{eq:CMTFCEs}, one may follow different normalization strategies based on the nonlinear effect that is expected to dominate or the one that they want to focus on. Here, we will normalize all equations with respect to FCD; the normalized cavity amplitude $\tilde u$, the normalized carrier density $\bar n_c$, and the normalized input/output wave amplitudes $\tilde\psi$ are given by
        \begin{subequations}
            \begin{align}
                \tilde u(t) &= \sqrt[4]{\tau_1^2\gamma_{N,\mathrm{FCD}}\gamma_N} \tilde a(t), \\
                \bar n_c(t) &= \tau_1\gamma_{N,\mathrm{FCD}} \bar N_c(t), \\
                \tilde \psi(t) &= \sqrt[4]{\tau_3^3\tau_c\gamma_{N,\mathrm{FCD}}\gamma_N} \tilde s(t),
            \end{align}
            \label{eq:CMTNormalizationFCD}
        \end{subequations}
        and therefore the FCD characteristic power is defined through
        \begin{equation}
            P\rsub{0,FCD} = \sqrt{\frac{1}{\tau_3^3\tau_c\gamma_{N,\mathrm{FCD}}}}.
            \label{eq:CharPowerFCD}
        \end{equation}
        It is then straightforward to express Eqs.~\eqref{eq:CMTFCEs}, as well as the input/output coupling equations, in normalized terms as
        \begin{subequations}
            \begin{align}
                \fullFRAC{\tilde u}{t^\prime} = &j(-\delta-r\rsub{SPM}|\tilde u|^2+\bar n_c) \tilde u \nonumber \\
                &- (r_i+r_e+r\rsub{TPA}|\tilde u|^2+r\rsub{FCA}\bar n_c) \tilde u + \varpi_{e,\mathrm{FCE}} \tilde \psi^+, \\
                \fullFRAC{\bar n_c}{t^\prime} = &-\frac{\bar n_c}{\tau_c^\prime} + |\tilde u|^4, \\
                \tilde\psi^-_k = &c_{k\ell}\tilde\psi^+_\ell + \vartheta_{e,\mathrm{FCE},k}\tilde u.
            \end{align}
            \label{eq:CMTnormalizedFCEs}
        \end{subequations}
        The expressions for some newly-introduced parameters that depend on the cavity type/coupling scheme, namely $\tau_3$, $\tau_c^\prime$, $\varpi_{e,\mathrm{FCD}}$, and $\vartheta_{e,\mathrm{FCD}}$ (elements of $\boldsymbol{\Theta}_{e,\mathrm{FCD}}$), are included in Table~\ref{tab:CMTNLFCEsTOECoefficients}. The definitions of the nonlinear intensity ratios for each of the involved nonlinear effect are given by $r\rsub{SPM/TPA} = \gamma\rsub{SPM/TPA}/\sqrt{\gamma_{N,\mathrm{FCD}}\gamma_N}$ and $r\rsub{FCA} = \gamma\rsub{FCA}/\gamma\rsub{FCD}$, regardless of the cavity type or coupling scheme. We note once again that the proposed normalization is not unique; any other nonlinear effect can be used as the basis. For example, the respective expressions for SPM-based normalization can be found in Ref.~\onlinecite{Tsilipakos2016}. In the same work or in Ref.~\onlinecite{Chen2012}, a more complex and physically accurate framework is presented where the FCD-induced change of the refractive index nonlinearly depends on $N_c$\cite{Soref1987,Nedeljkovic2011} (here, we have assumed that $\Delta n = -\sigma_n N_c$ for simplicity, see Appendix~\ref{AppSub:DeltaOmegaVSaFCEsTOE}). 

        Following a similar approach, we will also examine the TOE, i.e., the dependence of a material refractive index with temperature. In Si and for temperatures close to 300~K this dependence is linear and, thus, easily handled by CMT. Any heating source, internal or external, may affect the refractive index and, thus, shift the resonance frequency of the cavity. In fact, external heating is widely used to tune  practical ring/disk cavities.\cite{Grillanda2014}   Here, we will only examine internal heating mechanisms, i.e., heating due to linear and/or nonlinear absorption. In a first order approximation, TOE does not induce additional losses; we can thus write
        \begin{equation}
            \Delta\omega\rsub{TOE}(\Delta\bar T) = -\gamma_{T,\mathrm{TOE}} \Delta\bar T(t),
            \label{eq:DeltaOmegaTOE}
        \end{equation}
        where $\Delta\bar T$ is the spatially averaged temperature change, $\gamma_{T,\mathrm{TOE}}$ is purely real and so is $\Delta\omega\rsub{TOE}$. The sign of $\gamma_{T,\mathrm{TOE}}$ depends on the material, as briefly described in Appendix~\ref{AppSub:DeltaOmegaVSaFCEsTOE}. Thus, TOE may red- (e.g. Si) or blue-shift (e.g. SiO$_2$) the resonance frequency.
        
        An ODE similar to Eq.~\eqref{eq:CarrierRateEquationCMT} is typically used to capture the temporal evolution of $\Delta\bar T$\cite{Zhang2013a,Iadanza2020}
        \begin{align}
            \fullFRAC{\Delta\bar T(t)}{t} = -\frac{\Delta\bar T(t)}{\tau_\theta} &+ \gamma_{T,\mathrm{J}}|\tilde a(t)|^2 + \gamma_{T,\mathrm{TPA}}|\tilde a(t)|^4 \nonumber \\
            &+ \gamma_{T,\mathrm{FCA}} \bar N_c(t) |\tilde a(t)|^2.
            \label{eq:TemperatureRateEquationCMT}
        \end{align}
        The thermal lifetime $\tau_\theta$ is an implicit way to capture the effect of heat diffusion and typically acquires values in the \textmu$\mathrm{s}$ range meaning that
        fast pulses with $\mathrm{ps}$ duration are not sufficient to change the background temperature despite their potential high peak power. Importantly, ODE~\eqref{eq:TemperatureRateEquationCMT} is driven by three terms since we have allowed for three different heating mechanisms, namely Joule heating (Ohmic loss), TPA, and FCA, respectively. Although Si is transparent at communication wavelengths, Joule heating can be induced by neighbouring lossy materials.\cite{Gao2017}

        TOE is introduced in the CMT cavity amplitude equation  through Eq.~\eqref{eq:DeltaOmegaTOE}, complemented by Eq.~\eqref{eq:TemperatureRateEquationCMT}. We present here only a normalized set of the ODEs that govern such a general system, which are
        \begin{subequations}
            \begin{align}
                \fullFRAC{\tilde u}{t^\prime} = &j(-\delta-r\rsub{SPM}|\tilde u|^2+\bar n_c-\delta\bar\tau) \tilde u \nonumber \\
                &- (r_i+r_e+r\rsub{TPA}|\tilde u|^2+r\rsub{FCA}\bar n_c) \tilde u + \varpi_{e,\mathrm{TOE}} \tilde \psi^+, \\
                \fullFRAC{\bar n_c}{t^\prime} = &-\frac{\bar n_c}{\tau_c^\prime} + r_N|\tilde u|^4, \\
                \fullFRAC{\delta\bar\tau}{t^\prime} = &-\frac{\delta\bar\tau}{\tau_\theta^\prime} + r_{T,\mathrm{J}}|\tilde u|^2 + r_{T,\mathrm{TPA}}|\tilde u|^4 + r_{T,\mathrm{FCA}} \bar n_c |\tilde u|^2, \\
                \tilde\psi^-_k = &c_{k\ell}\tilde\psi^+_\ell + \vartheta_{e,\mathrm{TOE},k}\tilde u,
            \end{align}
            \label{eq:CMTnormalizedTOE}
        \end{subequations}
        where we have used the FCA-induced part of TOE for the normalization, meaning that
        \begin{subequations}
            \begin{align}
                \tilde u(t) &= \sqrt[6]{\tau_1^3\gamma_{T,\mathrm{TOE}}\gamma_{T,\mathrm{FCA}}\gamma_N} \tilde a(t), \\
                \bar n_c(t) &= \tau_1\gamma_{N,\mathrm{FCD}} \bar N_c(t), \\
                \delta\bar\tau(t) &= \tau_1\gamma_{T,\mathrm{TOE}} \Delta\bar T(t), \\
                \tilde \psi(t) &= \sqrt[6]{\tau_4^5\tau_\theta\gamma_{T,\mathrm{TOE}}\gamma_{T,\mathrm{FCA}}\gamma_N} \tilde s(t),
            \end{align}
            \label{eq:CMTNormalizationTOE}
        \end{subequations}
        The parameters appearing in Eqs.~\eqref{eq:CMTnormalizedTOE} are defined in Table~\ref{tab:CMTNLFCEsTOECoefficients}. The definitions of the newly-introduced intensity ratios for each nonlinear effect are omitted for brevity; specifying them should be routine by now.

        In terms of nonlinear applications, FCE and TOE have been used to demonstrate all-optical control of light in PhCs\cite{Barclay2005} as well as in Si-based integrated ring and disk cavities.\cite{Johnson2006,Xu2006,Chen2012,Zhang2013a,Wang2013a,Sethi2014,Long2015} Through CMT, it is straightforward to predict and compare the intensity of each nonlinear effect and better understand how to enhance or suppress their respective contribution. In Fig.~\ref{fig:AllNLinSIcomparison} for example, the intensity of each nonlinearity in an integrated SOI ring resonator is depicted.\cite{Wang2013a} Due to the finite response times of both FCE and TOE, other dynamic phenomena such as self-pulsation\cite{Malaguti2011,Chen2012,Zhang2013a,Cazier2013,DiLauro2017,Tamura2022} and excitability\cite{VanVaerenbergh2012,VanVaerenbergh2012a} can be observed and modeled by CMT, as briefly discussed in Sec.~\ref{sec:StabAnalysis}.

        Obviously, the framework presented in this section for Si is valid for any material that exhibits FCE and/or TOE. It can also be expanded to include other effects that are governed by one or more additional ODEs. For instance, a similar dynamic framework for the non-instantaneous Kerr effect (typically appearing in liquids) is developed in Ref.~\onlinecite{Armaroli2011}. The aim of this section was to highlight how CMT can be combined with additional ODEs that govern the physics of linear or nonlinear effects and not to extensively discuss all the effects that it can describe.

    \subsection{Non-instantaneous saturable absorption}\label{subsec:DynSA}
        For the second part of this section, we will discuss the dynamics of saturable absorption, as SA in graphene and other contemporary 2D materials has been intensively utilized recently in nonlinear applications. Beginning from the analysis of Sec.~\ref{subsec:SA}, we will present here a more accurate description of SA in graphene taking into account carrier diffusion.\cite{Nousios2022} A similar approach has been followed to capture free carrier diffusion in semiconductor cavities.\cite{Moille2016}

        Carrier dynamics in graphene (including carrier diffusion) are captured by a PDE of the form\cite{Chatzidimitriou2020,Nousios2022}
        \begin{align}
            \partialFRAC{N_c\OF{r}{t}}{t} = &-\frac{N_c\OF{r}{t}}{\tau_c} + \frac{1}{2}\frac{\RE{\sigma_1(N_c)}|\mathbf{E}\rsub{ref,\parallel}(\mathbf{r})|^2}{\hbar\omega} |a(t)|^2 \nonumber \\
            &+ D\nabla^2N_c\OF{r}{t},
            \label{eq:CarrierRateEquationCMTSA}
        \end{align}
         as discussed in Appendix~\ref{AppSub:DeltaOmegaVSaSA}. The driving term of Eq.~\eqref{eq:CarrierRateEquationCMTSA} models carrier excitation in the conduction band of graphene due to linear absorption using a carrier-dependent $\RE{\sigma_1}$ (the real part of the surface conductivity). The 2D nature of graphene is captured by using the tangential (reference) electric field $\mathbf{E}\rsub{ref,\parallel}$, instead of the full (reference) field $\mathbf{E}\rsub{ref}$. $\tau_c$ is the actual carrier recombination lifetime (without effective contributions), since the last term directly accounts for carrier diffusion with a diffusion coefficient $D$. Obviously, due to carrier diffusion one cannot use a  spatially-averaged carrier density and Eq.~\eqref{eq:CarrierRateEquationCMTSA} should be solved concurrently with the CMT equations utilizing a spatially-dependent PDE solution approach, such as FEM. 
         Although this diverges from the treatment of a cavity as a point (0D) oscillator, it still allows for computational benefits compared to a full wave approach for all electromagnetic quantities.\cite{Nousios2022,Moille2016} In some cases, the impact of diffusion can be captured implicitly by using an effective carrier lifetime $\tau_{c,\mathrm{eff}}$ and then the simpler ODE-based approach of Sec.~\ref{subsec:FCEsTOE} suffices. Such an equivalence was demonstrated in Ref.~\onlinecite{Nousios2022}. Alternatively, in semiconductor cavities a second ODE utilizing effective quantities as well might be necessary to capture the initial stages of carrier diffusion.\cite{Moille2016}

\begin{figure}
            \centering
            \includegraphics[scale=.9]{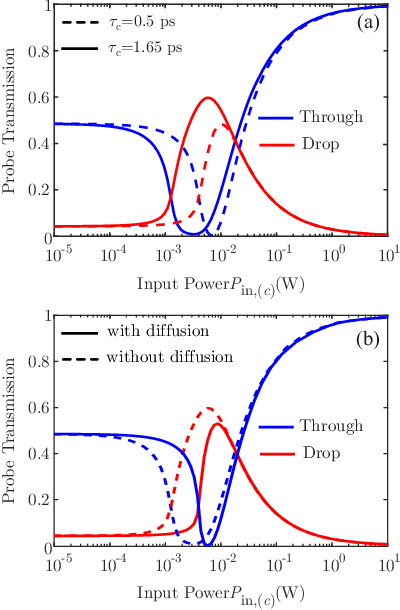}
            \caption{CW through and drop transmission of a SOI disk cavity overlaid with graphene in an add-drop configuration. (a) Influence of carrier diffusion. A significant change in CW transmission is observed. (b) CW response utilizing different carrier lifetimes. An appropriate effective lifetime value can emulate the effect of diffusion on the response. \COPYRIGHT{Nousios \emph{et al.}, J. Appl. Phys. \textbf{131}, 053104 (2022)}{Nousios2022}{2022 AIP Publishing}.}
            \label{fig:SADiffusion}
        \end{figure}

        To connect Eq.~\eqref{eq:CarrierRateEquationCMTSA} with CMT, the spatially and temporally dependent nonlinear SA parameter
        \begin{equation}
            \gamma\rsub{SA}(N_c) = \frac{1}{4}\iint_{S_p} \RE{\sigma_0}\left[1-\frac{N_c\OF{r}{t}}{2N\rsub{sat}}\right]|\mathbf{E}\rsub{ref,\parallel}(\mathbf{r})|^2\mathrm{d}S,
            \label{eq:GammaSA}
        \end{equation}      
        should be used, derived through perturbation theory.\cite{Ataloglou2018,Christopoulos2020JAP,Nousios2022} Then, the CMT cavity-amplitude ODE can be written as
        \begin{align}
            \fullFRAC{\tilde a(t)}{t} = &-j(\omega-\omega_0) \tilde a(t) - [\gamma\rsub{SA}(N_c)+\gamma\rsub{res}+\gamma\rsub{rad}+\gamma_e] \tilde a(t) \nonumber \\
            &+ \mu_e \tilde s^+(t). 
            \label{eq:CMTSAfull}
        \end{align}
        Although in Eq.~\eqref{eq:GammaSA} $\gamma\rsub{SA}$ involves a spatial integration of $N_c\OF{r}{t}$, any numerical integration scheme can be used to calculate the integral in Eq.~\eqref{eq:GammaSA} concurrently with the solution of the PDEs (or ODEs when diffusion is handled through an effective lifetime), since $\mathbf{E}\rsub{ref,\parallel}(\mathbf{r})$ is constant with time. As an example of the effect of diffusion in the response of a nonlinear system with SA, in Fig.~\ref{fig:SADiffusion}(a) the CW through and drop transmission of a SOI disk cavity overlaid with graphene in an add-drop configuration is depicted.\cite{Nousios2022} Evidently, transmission in both ports is significantly affected by carrier diffusion. A similar response is recovered when an effective carrier lifetime is used to implicitly account for diffusion [Fig.~\ref{fig:SADiffusion}(b)], with the effective value being specified heuristically.

\section{Coupled-Mode Theory for Systems with Gain}\label{sec:Gain}

    On a different topic, in this section we will present the treatment of gain with temporal coupled-mode theory. Gain is the opposite of loss and intuitively one can argue that an appropriate gain coefficient $g$ (depending, amongst others, on the level of pumping), similar to the decay rate $\gamma$ but of opposite sign, can model gain, i.e., we can write $\mathrm{d}a/\mathrm{d}t = j\omega_0 a - \gamma a + g a$.\cite{HausBook} This crude description of gain is only accurate for relatively small pumping levels and can be used to predict the lasing threshold, reached when losses and gain are balanced, i.e., $g\rsub{th} = \gamma$. At higher pumping levels where, e.g., gain saturates, a more complex model should be used. We will focus on such an accurate model here that can describe the most general case of Class~C lasers, i.e., lasers that exhibit carrier and polarization dynamics which cannot be ignored.\cite{Siegman} These dynamics are briefly discussed in Appendix~\ref{AppSub:DeltaOmegaVSaGain} from a physics point of view.

    Gain is typically described by a homogeneously-broadened Lorentzian polarization field which, following the treatment presented in Appendix~\ref{AppSub:DeltaOmegaVSaGain}, can be written as the first order ODE~\eqref{eq:Pfinal}.\cite{Chua2011,Nousios2023,Drong2022} This polarization is driven by the physical gain mechanism, which is typically the excitation of carriers to a higher energy level and the population inversion achieved in a metastable level of the gain medium.\cite{Siegman}  In literature, such a set of equations are referred to as the semiclassical Maxwell-Bloch equations. Finally, the population-inversion-driven polarization field is treated as a perturbation and in such a way it can be coupled with our typical CMT ODE through a complex resonance frequency shift, as with any other linear or nonlinear effect discussed thus far. This shift is given by\cite{Chua2011,Nousios2023}
    \begin{equation}
        \Delta\tilde\omega_g (\tilde p) = j\xi\frac{1}{\tilde a(t)}\left[ j\omega\rsub{ref} \tilde p(t) + \fullFRAC{\tilde p(t)}{t} \right],
        \label{eq:DeltaOmegaGainCMT}
    \end{equation}
    and is significantly different from any other perturbative nonlinearity that has been presented so far in this tutorial. As dictated by the physics of gain, the polarization field amplitude $\tilde p(t)$ \emph{and} its time derivative appear in $\Delta\tilde\omega_g$.\cite{Chua2011,Nousios2023} Furthermore, $\omega\rsub{ref}$ is an arbitrarily chosen frequency, reasonably close to the ``cold'' resonance frequency $\omega_0$ of the unperturbed system. This is another subtle handling that should be performed since the actual lasing frequency $\omega_L$ (perturbed system) is unknown in the general case of a cavity that does not resonate exactly at the atomic transition frequency $\omega_m$ of the gain material.\cite{Nousios2023} Based on this treatment, we can get a quite accurate estimation of $\omega_L$, as shown below.

    Ultimately, one can introduce Eq.~\eqref{eq:DeltaOmegaGainCMT} in the CMT ODE to include gain. This will couple the main CMT equation with the respective polarization ODE, which is also coupled to the carrier dynamics of the gain medium. A system of at least three coupled ODEs is eventually constructed, with the actual number of necessary ODEs depending on the energy levels of the gain material.\cite{Chua2011a,Chua2014} Here, we will present the final set of equations for an optically pumped three-level gain medium, i.e., a gain material with a single metastable energy level that can be used to achieve population inversion, stimulated emission, and, ultimately, lasing. Three-level systems represent a wide variety of gain materials in optics and photonics, such as Erbium in fibers or integrated optics,\cite{Lei2014} III-V semiconductors for integrated photonics,\cite{Chua2011} or the recently emerged 2D TMD heterobilayers.\cite{Nousios2023} Regarding the various physical parameters appearing in the equations below, the reader is referred to Appendix~\ref{AppSub:DeltaOmegaVSaGain} or in Ref.~\onlinecite{Nousios2023}.
    \begin{subequations}
        \begin{align}
            \fullFRAC{\tilde a}{t} &= -j(\omega\rsub{ref}-\omega_0) \tilde a - \gamma_\ell \tilde a - \xi\left( j\omega\rsub{ref}\tilde p + \fullFRAC{\tilde p}{t} \right), \label{eq:CMTGainfullAmplitude} \\
            \fullFRAC{\tilde p}{t} &= - \frac{\omega_m^2-\omega\rsub{ref}^2+j\omega\rsub{ref}\Gamma_m}{\Gamma_m+j2\omega\rsub{ref}}\tilde p -\frac{\sigma_m}{\Gamma_m+j2\omega\rsub{ref}} (\bar N_2 - \bar N_1) \tilde a, \\
            \fullFRAC{\bar N_3}{t} &=  W_p(\bar N_1 - \bar N_3) -\frac{\bar N_3}{\tau_{32}}, \\
            \fullFRAC{\bar N_2}{t} &=  \frac{\bar N_3}{\tau_{32}} -\frac{\bar N_2}{\tau_{21}} + \frac{\xi_N}{\hbar\omega_m}\frac{1}{2}\RE{\left[j\omega\rsub{ref}\tilde p+\fullFRAC{\tilde p}{t}\right]\tilde a^*}, \\
            \fullFRAC{\bar N_1}{t} &=  \frac{\bar N_2}{\tau_{21}} - \frac{\xi_N}{\hbar\omega_m}\frac{1}{2}\RE{\left[j\omega\rsub{ref}\tilde p+\fullFRAC{\tilde p}{t}\right]\tilde a^*} - W_p(\bar N_1 - \bar N_3).
        \end{align}
        \label{eq:CMTGainfull}
    \end{subequations}

    The system of ODEs~\eqref{eq:CMTGainfull} is readily solvable and can be used to calculate either the CW or the pulsed lasing response of a resonant system with gain. Here, it is given in natural units but typically it is normalized for numerical stability, as with any other nonlinear phenomenon presented in this tutorial. Interestingly, Eqs.~\eqref{eq:CMTGainfull} can be used to estimate fundamental lasing characteristics, such as the lasing threshold or the lasing frequency, and extract useful design directives.\cite{Nousios2023} For example, the lasing threshold (minimum pumping level to achieve lasing), retrieved when loss equals gain is\cite{Nousios2023}
    \begin{equation}
        W_{p,\mathrm{th}} = \frac{1}{\tau_{21}}\frac{\bar N\rsub{tot} - \Delta\bar N\rsub{th}}{\bar N\rsub{tot} + \Delta\bar N\rsub{th}},
        \label{eq:PumpingThresholdGain}
    \end{equation}
    where $\Delta\bar N\rsub{th} = \Gamma_m\gamma_\ell/\sigma_m\xi$ is the population inversion threshold, and typically $\bar N\rsub{tot} = \bar N_1 + \bar N_2 + \bar N_3 \gg \Delta\bar N\rsub{th}$ so that $W_{p,\mathrm{th}}\approx1/\tau_{21}$. Importantly, the lasing frequency can be estimated through\cite{Nousios2023}
    \begin{equation}
        \omega_L = \frac{\Gamma_m\omega_0  \pm \sqrt{(\Gamma_m\omega_0)^2 + 4\gamma_\ell(\gamma_\ell+\Gamma_m)\omega_m^2}}{2(\gamma_\ell+\Gamma_m)},
        \label{eq:LasingFrequencyGain}
    \end{equation}
    retrieved from the analytic solution of the CW version of Eqs.~\eqref{eq:CMTGainfull} by requiring a real and positive (i.e., physically meaningful) population inversion threshold. When $\omega_0 = \omega_m$, Eq.~\eqref{eq:LasingFrequencyGain} obviously dictates that $\omega_L = \omega_0 = \omega_m$ which is intuitively expected. However, when $\omega_0\neq\omega_m$, $\omega_L$ does not coincide either with $\omega_0$ or with $\omega_m$ and the actual lasing frequency can be predicted through Eq.~\eqref{eq:LasingFrequencyGain}, which has proven quite accurate.\cite{Nousios2023} When estimated via Eq.~\eqref{eq:LasingFrequencyGain}, $\omega_L$ can be used in Eqs.~\eqref{eq:CMTGainfull} instead of $\omega\rsub{ref}$, significantly expediting the numerical solution of the system. When the chosen $\omega\rsub{ref}$ deviates from the actual lasing frequency, nonphysical oscillations in the real and imaginary parts of $\tilde a$ and $\tilde p$ appear with an oscillation frequency $|\omega\rsub{ref}-\omega_L|$, which can be recovered by a simple Fourier transform.\cite{Nousios2023} This behavior stresses the Runge-Kutta scheme typically used for the solution. However, their norms are constant and with any reasonable choice of $\omega\rsub{ref}$ we can recover the correct lasing response since $p\rsub{out}\propto|\tilde a|^2$.\cite{Nousios2023}

    The framework of Eqs.~\eqref{eq:CMTGainfull} describes the general case of Class~C lasers. However, it can be easily simplified to describe Class~B (i.e., lasers with gain materials that exhibit short polarization lifetimes compared to the lifetime of the carriers in the population inversion level, such as erbium\cite{Lei2014,Hu2019}), or Class~A lasers (i.e., lasers with gain materials having carrier lifetimes much smaller than $\tau_\ell$). In a Class~A lasers the gain term can be  simplified to\cite{Nousios2023} $g_0(W_p)/[1+|\tilde a|^2/W\rsub{sat}(W_p)]$. This form corresponds to the simplest intuitive description of gain discussed in the first paragraph of Sec.~\ref{sec:Gain} with the addition of gain saturation.

    In terms of applications, the coupled CMT/semiclassical Maxwell-Bloch framework of Eqs.~\eqref{eq:CMTGainfull} (or simplified versions of it) has been used to calculate fundamental laser characteristics and also to efficiently model CW and pulsed operation.\cite{Chua2011,Chua2011a,Yacomotti2013,Chua2014,Lei2014,Hu2019} What is even more interesting, however, is that this set of equations can be used as the basis for other, more complex phenomena. For example, if an external wave is also included in Eq.~\eqref{eq:CMTGainfullAmplitude} the very same resonant system can act as an amplifier rather than a lasing cavity, albeit with a very limited bandwidth. More interestingly, when gain is combined with SA, a pulsed output (for CW pumping) with tunable characteristics can be obtained. This phenomenon emerges due to the $Q$-switching mechanism and has been analyzed with CMT in various physical implementations.\cite{Rasmussen2017,Rasmussen2018,Benzaouia2022,Nousios2024} Finally, the presence of gain is naturally connected with PT-symmetry, and CMT has been used to describe such systems as well.\cite{Zhou2016,Ramezanpour2021}

\section{Stability analysis with the Coupled-Mode Theory}\label{sec:StabAnalysis}

    The simple yet quite accurate form in which the CMT framework can model nonlinear resonant systems has yet another significant advantage. Throughout this tutorial, we have discussed that some solutions of the CMT equations are physically unacceptable (the unstable branch of a bistable system) and others exhibit rich dynamics, showing either sinusoidal (self-pulsation) or, in general, periodic (excitability, $Q$-switching) output patterns. Such complex dynamics are an integral part of resonant nonlinear systems.\cite{Strogatz} The CMT ODEs allow to study and understand such nonlinear dynamics through a \emph{stability analysis}. Stability analysis of nonlinear systems is a deeply studied topic that is outside the scope of this tutorial. Here, we will only discuss very simple approaches that can be followed to examine the stability of nonlinear systems and comment on some applications that have been demonstrated over the years utilizing CMT.

    The simplest approach to examine the stability of a nonlinear CMT ODE (or a system of ODEs which is more meaningful) is \emph{linear stability analysis}, which has been proven quite accurate for relatively simple, low-dimensional systems. In this context, a small perturbation is introduced into a CW solution (also called fixed point) of the CMT equations and a linearized version of the system in the form $\mathrm{d}\boldsymbol{\epsilon}/\mathrm{d}t = \mathbf{J}\boldsymbol{\epsilon}$ is obtained, where $\boldsymbol{\epsilon}$ is the induced perturbation vector. The eigenvalues of $\mathbf{J}$, i.e., the Jacobian matrix of the system evaluated at the fixed point, characterize its stability. Specifically, if all eigenvalues have negative real parts, the perturbation eventually vanishes and the fixed point is \emph{linearly stable}; otherwise, the perturbation dominates and the fixed point becomes unstable. The type of instability depends on the respective eigenvalues (real, complex conjugate, etc.). However, the eigenvalues cannot always reveal the true type of the instability (after all, they stem from the linearized system); in that case, a complete \emph{bifurcation analysis} is required.

    In nonlinear dynamics terminology, the optical bistability that was extensively discussed in Sec.~\ref{subsec:SPM} is a classic example of a \emph{supercritical pitchfork bifurcation} in an one-dimensional system, i.e., a system described by a single ODE. In that type of bifurcation,  a stable fixed point splits into three trajectories with the two outer ones being stable and the middle one unstable.\cite{Strogatz,TaborChaos} These trajectories correspond to the three branches of the bistability loop (Fig.~\ref{fig:BistabilitySPM}). A linear stability analysis reveals real eigenvalues that are either all negative (stable solutions) or at least one positive (unstable solution).\cite{Wang1996}
    
    In two-dimensional systems, \emph{Hopf bifurcations} (characterized by at least a pair of complex conjugate eigenvalues with positive real parts) lead to a limit cycle.\cite{Strogatz} In semiconductor cavities, for example, \emph{supercritical Hopf bifurcations} were extensively studied using a linear stability analysis of the coupled CMT ODEs and were found to exhibit (under certain conditions) a sinusoidal periodic output due to the interplay of FCD-induced nonlinearities with the cavity.\cite{Armaroli2011,Malaguti2011,Chen2012,Cazier2013,Hamerly2015,Tsilipakos2016,Abdollahi2017,Ataloglou2018,Tamura2022} This effect is termed \emph{self-pulsation} and appears due to the interplay between the linear cavity response and the finite carrier lifetime, leading to a   sinusoidal or quasi-sinusoidal output for CW input. When the TOE is also considered and the system under study becomes three-dimensional (cavity amplitude, carriers, temperature), a \emph{subcritical Hopf bifurcation} can emerge, which may drive multi-dimensional systems into chaos.\cite{Strogatz,TaborChaos} Self-pulsation due to the TOE has been also reported in the literature\cite{DiLauro2017}, as well as excitability,\cite{Yacomotti2006,VanVaerenbergh2012,VanVaerenbergh2012a} i.e., a pulsed periodic output emerging due to the competing interplay of FCD, TOE and the linear response of the cavity.

    The concept of pulsed output in systems with CW excitation has also been investigated in lasers, where a cavity with gain can exhibit a pulsed output for CW pumping when a loss saturation effect is present. This behavior stems from the $Q$-switching mechanism, namely, the quality factor of the cavity is modified due to loss saturation and consequently the  critical coupling condition is met for a specific level of the stored energy. Such an effect has been demonstrated in PhC lasers\cite{Rasmussen2017,Kaminski2019} as well as silicon nitride cavities,\cite{Nousios2024} with their dynamic behavior being characterized using coupled CMT equations and linear stability analysis. \blue{Similarly, pulsed output can also be obtain in multi-cavity systems, with or without nonlinearities.\cite{Yacomotti2013,Kominis2020} }

    Finally, self-pulsation or other instabilities have been reported in Kerr nonlinear cavities, either mutually coupled,\cite{Abdollahi2014,Dumeige2015,Armaroli2018} or in cavities designed for wave generation or wave mixing (multi-resonant cavities to accommodate the different input/output wavelengths).\cite{Hashemi2009,Ramirez2011,Lin2014} Again such multi-dimensional systems exhibit rich dynamics and multistability or limit cycles may appear for different sets of parameters. Interestingly, in such complex systems a limit cycle does not necessarily appear  due to a Hopf bifurcation but other alternatives exist. An indicative example is a \emph{saddle node-homoclinic bifurcation} which appears when $\mathbf{J}$ possesses a zero eigenvalue and may lead to a limit cycle due to the annihilation of two fixed points (one stable and one unstable) connected with at least two heteroclinic orbits.\cite{Strogatz}

\section{Full-wave validation of nonlinear CMT}\label{sec:CMTValidation}

    Although temporal coupled-mode theory has been routinely  used to reproduce the response of complex linear systems, its validity regarding nonlinear cavities was initially questionable. Thus, in various works the validity of the nonlinear CMT framework has been verified, by comparing the retrieved results with full-wave nonlinear simulations. 

\begin{figure*}[!t]
        \centering
        \includegraphics{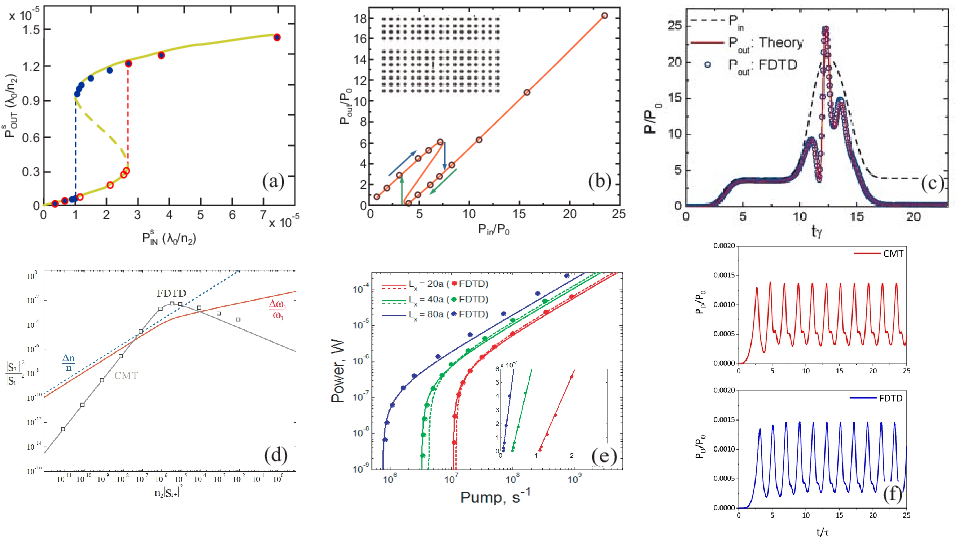}
        \caption{Validation of nonlinear CMT (solid lines) using FDTD (point markers). (a)~Optical bistability (CW curve) on a directly coupled PhC cavity. \COPYRIGHT{Solja{\v{c}}i{\'{c}} \emph{et al.}, Phys. Rev. E \textbf{66}, 055601(R) (2002)}{Soljacic2002}{2002 the American Physical Society}. (b)~Optical bistability (CW curve) on a side coupled PhC cavity (see inset), and (c)~temporal evolution of the output power; switching between the two bistable states is obvious. \COPYRIGHT{Yanik \emph{et al.}, Appl. Phys. Lett. \textbf{83}, 2739 (2003)}{Yanik2003}{2003 AIP Publishing}. (d)~Conversion efficiency of THG in a nonlinear 1D PhC cavity; normalized resonance frequency shift and modification of the refractive index due to nonlinearity are also included to highligh the validity range of CMT. \COPYRIGHT{Rodriguez \emph{et al.}, Opt. Express \textbf{15}, 7303 (2007)}{Rodriguez2007}{2007 Optica}. (e)~Light-light curves of a 1D PhC laser for different geometric configurations of the cavity. \COPYRIGHT{Chua \emph{et al.}, Opt. Express \textbf{19}, 1539 (2011)}{Chua2011}{2011 Optica}. (f)~Self-pulsing behavior of two nonlinear coupled ring cavities. \COPYRIGHT{Jebali \emph{et al.}, J. Opt. Soc. Am. B \textbf{37}, 2557 (2020)}{Jebali2020}{2020 Optica}.}
        \label{fig:CMTValidationFDTD}
    \end{figure*}

    Chronologically, the first method that was utilized to validate nonlinear CMT was the finite-difference time-domain (FDTD) method.\cite{TafloveHagness} Being a time-domain method, FDTD is naturally suited to nonlinear calculations. As discussed in Sec.~\ref{subsec:SPM}, the PhC platform was initially utilized to demonstrate optical bistability in integrated systems. In some of these pioneering works, FDTD was used to validate the obtained response; comparisons for the phenomenon of optical bistability are included in Fig.~\ref{fig:CMTValidationFDTD}(a-c).\cite{Soljacic2002,Yanik2003} In a bistable system, the ascending bistablility branch can be easily revealed with FDTD by performing a simulation with quasi-CW input and obtaining the respective output. To reveal the second (descending) branch, the desired input power must be approached from higher power levels, see Fig.~\ref{fig:CMTValidationFDTD}(c). This allows to first surpass the bistable region and then relax to the second branch of the hysteresis loop. As expected, the unstable branch cannot be traced with full-wave simulations. Realizations of both direct- and side-coupled cavities showed excellent agreement between nonlinear CMT and FDTD, not only for the CW hysteresis loop [Fig.~\ref{fig:CMTValidationFDTD}(a,b)] but also for the entire temporal evolution [Fig.~\ref{fig:CMTValidationFDTD}(c)].

    FDTD has also been utilized to validate frequency generation as described with CMT and more specifically the THG process.\cite{Rodriguez2007} In the results shown in Fig.~\ref{fig:CMTValidationFDTD}(d), a 1D PhC with a nonlinear cavity exhibiting two resonances centered at the fundamental ($\omega_1$) and third harmonic ($\omega_3$) frequencies was excited at $\omega_1$. The reflected wave at $\omega_3$ was recorded, with the results of Fig.~\ref{fig:CMTValidationFDTD}(d) depicting the conversion efficiency of the process. For low to moderate input power levels, CMT and FDTD coincide; the agreement in CE starts to deteriorate only for very high input powers,  due to second order corrections that have not been taken into account in CMT.\cite{Rodriguez2007}
    A similar 1D PhC has been also utilized to validate the results of CMT in a system with gain.\cite{Chua2011} To computationally analyze such a system, the standard FDTD method was expanded to include the polarization field and the carrier rate equations, as is done with CMT. Figure~\ref{fig:CMTValidationFDTD}(e) shows three light-light curves for CW pumping of the PhC laser, assuming three different realizations of the cavity. The agreement between CMT and full-wave simulations is excellent.
    \blue{Lasing can also be studied using computationally less expensive (compared to FDTD), semi-analytic approaches, such as the steady-state \emph{ab initio} laser theory (SALT).\cite{Chua2011,Benzaouia2022,Ge2010,Pick2015} SALT is a frequency-domain approach in which the laser equations are converted into a set of coupled nonlinear wave equations and are then solved self-consistently to obtain laser properties such as the lasing frequency or the output power of the cavity. Figure~\ref{fig:CMTValidationFDTD}(e) also includes SALT calculations (dashed lines) which agree well with CMT and FDTD.}

    Finally, more recently FDTD has been applied in a system of two nonlinear coupled ring cavities that support bistability and self-pulsation.\cite{Jebali2020} Nonlinear CMT routinely reproduced the results of FDTD both by reproducing the CW bistability curve and, importantly, the dynamic quasi-sinusoidal response induced by self-pulsation [Fig.~\ref{fig:CMTValidationFDTD}(f)].

\begin{figure*}
        \centering
        \includegraphics[width=0.9\linewidth]{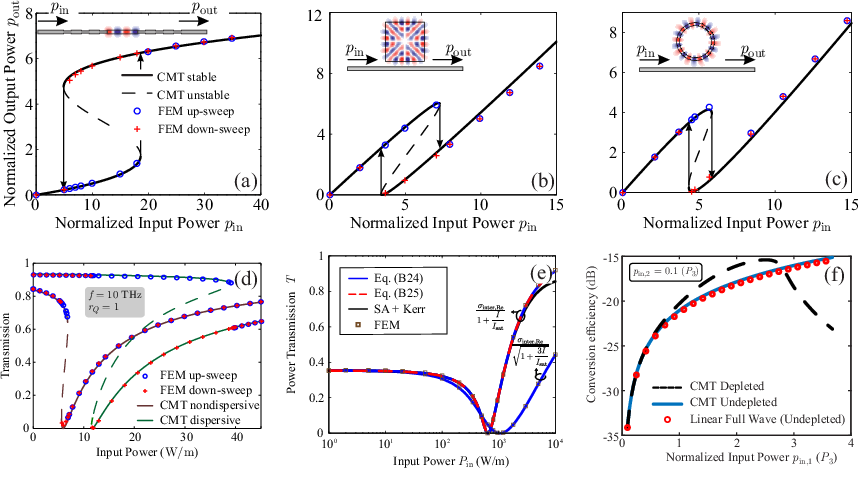}
        \caption{Validation of nonlinear CMT (solid lines) using an appropriate nonlinear version of FEM (point markers). Optical bistability utilizing: (a)~a directly coupled standing-wave cavity realized with two Bragg reflectors, (b)~a side coupled standing-wave square resonator with ultra-high $Q$-factor, and (c)~a side coupled traveling-wave ring cavity. (d)~Optical bistability in a graphene plasmonic traveling-wave resonator with or without considering graphene's dispersion. \COPYRIGHT{Christopoulos \emph{et al.}, Phys. Rev. E \textbf{94}, 062219 (2016)}{Christopoulos2016a}{2016 the American Physical Society}. (e)~Transmission of a side coupled traveling-wave disk resonator enhanced with graphene to induce SA. Two different SA models are examined both using CMT and full-wave FEM. The simplified model of Eq.~\eqref{eq:DeltaOmegaGrapheneSASimplified} is also compared with the full one of Eq.~\eqref{eq:DeltaOmegaGrapheneSANormalized}. \COPYRIGHT{Ataloglou \emph{et al.}, Phys. Rev. A \textbf{97}, 063836 (2018)}{Ataloglou2018}{2018 the American Physical Society}. (f)~Conversion efficiency of the DFWM process in a graphene plasmonic standing-wave cavity. Full-wave analysis only models the undepleted pump scenario where back-conversion from $\omega_3$ to $\omega_2$ and $\omega_1$ is ignored. \COPYRIGHT{Christopoulos \emph{et al.}, Phys. Rev. B \textbf{98}, 235421 (2018)}{Christopoulos2018}{2018 the American Physical Society}.}
        \label{fig:CMTValidationFEM}
    \end{figure*}

    The (frequency-domain) finite-element method\cite{Jin} has been also utilized to validate CMT in nonlinear integrated systems. Firstly, SOI cavities in standing- and traveling-wave configurations were examined [Fig.~\ref{fig:CMTValidationFEM}(a-c)].\cite{Christopoulos2015OWTNM} For their full-wave analysis, the conventional (linear) version of FEM was modified, utilizing an iterative approach where in each step a linear problem is solved and then the refractive index of the nonlinear material is refreshed through $n^{i+1}(\mathbf{r}) = n^{i}(\mathbf{r})+n_2|\mathbf{E}^i(\mathbf{r})|/(2\eta)$, as the Kerr effect dictates.\cite{Suryanto2003} The new refractive index distribution is used for the solution of a new linear problem, and the process is repeated until convergence (typically within ten iterations or less).\cite{Suryanto2003,Christopoulos2016} This iterative method can be applied for the full-wave solution of a nonlinear problem at a singe input power level and should be repeated for several input power levels to reveal the full bistability curve. For each next point on the power scan, the refractive index distribution obtained from the previous point should be used in the initial step, to help with convergence.\cite{Suryanto2003} Additionally, to reveal both bistablility branches, two power scans should be performed, one with ascending and one with descending power steps. The described method of using the refractive index distribution of the previous power point at the initial step of the next point is crucial to reveal the second branch of the bistability curve.\cite{Christopoulos2016} The same full-wave approach has been used to validate  CMT in a nonlinear graphene ring cavity that supports graphene surface plasmons (GSPs) in the THz frequency band [Fig.~\ref{fig:CMTValidationFEM}(d)], with the notable difference that now the Kerr effect influences the imaginary part of graphene's surface conductivity.\cite{Christopoulos2016} For this system, it is crucial to accurately incorporate the dispersion of the linear electromagnetic properties of graphene in order to get correct results; their introduction in the CMT framework relates with  the correct calculation of the $Q$-factor and the SPM nonlinear parameter $\kappa\rsub{SPM}$ [see the discussion in Appendix~\ref{AppSub:MaterialPertTheory} regarding the correct energy calculation in cavities with dispersive materials and Ref.~\onlinecite{Christopoulos2019} for the correct calculation of the quality factor].

    A very similar approach has been utilized to validate CMT in a system exhibiting saturable absorption.\cite{Ataloglou2018} A side-coupled Si disk cavity overlaid with graphene on half of its perimeter was examined and the iterative approach was used to obtain the power-dependent transmission. For this nonlinear system, only the real part of graphene conductivity is power-dependent. The nonlinear CMT/full-wave FEM comparison revealed identical responses, as shown in Fig.~\ref{fig:CMTValidationFEM}(e). Only a single scan on the input power was necessary since no bistable behavior was expected. Two different SA models have been used for graphene and the agreement in both cases was excellent. Additionally, the simplified model of Eq.~\eqref{eq:DeltaOmegaGrapheneSASimplified} for SA was compared with the full one [Eq.~\eqref{eq:DeltaOmegaGrapheneSANormalized}] showing almost identical results due to the uniformity of the tangential electric field on graphene.\cite{Ataloglou2018}
    
    Finally, the case of DFWM has also been examined in a graphene standing-wave resonator that allows for strong light confinement due to the supported GSPs.\cite{Christopoulos2018} Since the cavity supports Fabry-P\'erot-like resonances which were almost equidistant (not exactly due graphene's dispersive linear conductivity), three consecutive modes were used to accommodate the idler, pump, and signal waves. For the nonlinear full-wave solution with FEM, first the electric field distributions of the pump and signal waves were retrieved at $\omega_1$ and $\omega_2$, respectively,  using two independent linear simulations. Subsequently, these distributions were used to specify the induced nonlinear current oscillating at $\omega_3$ [Eq.~\eqref{eq:GrapheneNonlinearSurfaceCurrentDFWM1}]. Specifically, this nonlinear current acts as the source in a third linear problem that is solved at $\omega_3 = 2\omega_1 - \omega_2$; the power of the produced idler wave at $\omega_3$ is measured to retrieve the CE shown in Fig.~\ref{fig:CMTValidationFEM}(f), where the full-wave simulation results are compared with those retrieved by a modified CMT where it has been momentarily assumed that $\tilde\beta_1 = \tilde\beta_2 = 0$ to emulate the same conditions as with the full-wave solution. This corresponds to the \emph{undepleted pump} approach which is accurate only for relatively small CEs (bellow $-20$~dB); when the produced idler at $\omega_3$ becomes stronger, back-conversion effects are significant and lead to a suppression of the initial DFWM process [see dashed curve in Fig.~\ref{fig:CMTValidationFEM}(f), retrieved with CMT using the complete \emph{depleted pump} model, Eqs.~\eqref{eq:CMTDFWMNormalized}]. 

\section{Conclusion}\label{sec:Concl}

    In this tutorial, we have attempted to thoroughly cover the main elements of temporal coupled-mode theory and showcase its massive capabilities for the analysis of contemporary linear and nonlinear resonant photonic systems. We have comprehensively presented how CMT should be applied to study a broad range of linear and nonlinear phenomena, taking care to adopt physically-accurate models in each case. Although the treatment cannot be exhaustive, we have striven to lucidly outline each step as well as  the overall procedure, so that the  reader is equipped with all the necessary skills to introduce any linear/nonlinear effect in the CMT framework and efficiently solve the resulting system of equations. 
    We have also discussed the validity range of CMT and provided confirmation of its accuracy by comparing with full-wave simulations.
    Our work highlights CMT as a powerful and efficient tool for studying the temporal response of contemporary photonic systems.

\begin{acknowledgments}
    This research work was supported by the Hellenic Foundation for Research and Innovation (H.F.R.I.) under the ``First Call for H.F.R.I. Research Projects to support Faculty members and Researchers and the procurement of high-cost research equipment grant.'' (Project Number: HFRI-FM17-2086, GRAINS) and under the ``2nd Call for H.F.R.I. Research Projects to support Post-doctoral Researchers'' (Project Number: 916, PHOTOSURF).

    The Authors would like to warmly thank
    Philippe Lalanne, 
    Alejandro W. Rodriguez, 
    Bjorn Maes, 
    Yannis Kominis, and 
    Lei Zhou, 
    for their valuable comments during the revision of the original manuscript.
\end{acknowledgments}

\section*{Data Availability Statement}
    
    The data that support the findings of this study are available within the article.
    
\appendix
\renewcommand\thefigure{\thesection\arabic{figure}}\setcounter{figure}{0}  

\section{First-Order Perturbation Theory} \label{App:PertTheory}

    Perturbation theory can be used to find approximate solutions to a range of problems that arise from small modification to a related, simpler problem for which the solution is already known or can be easily computed.\cite{Pozar,Joannopoulos} Throughout this tutorial we use the lowest-order variant, which is termed  \emph{first-order} perturbation theory and omits higher-order contributions as negligible. In what follows, we will present the application of perturbation theory on material perturbations (Sec.~\ref{AppSub:MaterialPertTheory}) and the coupling of cavities (Sec.~\ref{AppSub:CouplingPertTheory}).

    \subsection{Material perturbations} \label{AppSub:MaterialPertTheory}

        The broad term \emph{material perturbations} describes slight modifications in the electromagnetic properties of materials that comprise the resonant cavity. Typical examples of linear perturbations are the inclusion of ohmic loss in dielectrics or modifications to the dielectric properties of the material (e.g., strain-induced). Nonlinear effects which also weakly modify the macroscopic material properties can be described as well; notable examples are the nonlinear saturation of ohmic loss (saturable absorption) and the Kerr effect. Despite their diverse nature, all the aforementioned effects result in a perturbation of the polarization ($\mathbf{P}$ in C/m$^2$) and/or the induced current density ($\mathbf{J}$ in A/m$^2$) inside the cavity [Fig.~\ref{fig:PertTheoryAppendix}(a)]. Thus, they can be treated under a unified perturbation theory framework. 

        \renewcommand\thefigure{\arabic{figure}}
        \setcounter{figure}{14}
        \begin{figure}
            \centering
            \includegraphics{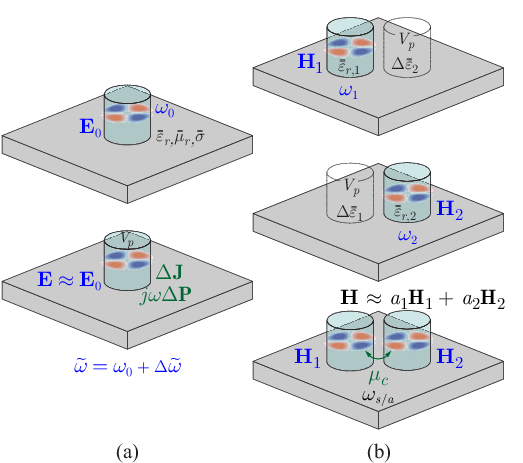}
            \caption{(a)~Material perturbation (induced by the generic quantities $j\omega\Delta\mathbf{P}$ and/or $\Delta\mathbf{J}$) and its impact on the resonance frequency of the cavity. (b)~Perturbation of a cavity due to coupling with an adjacent one.}
            \label{fig:PertTheoryAppendix}
        \end{figure}

        For the original, unperturbed cavity the Maxwell curl equations are
        \begin{subequations}
            \begin{align}
                \nabla\times\mathbf{E}_0 &= -j\omega_0\mu_0\dbar{\mu}_r(\omega_0)\mathbf{H}_0, \\
                \nabla\times\mathbf{H}_0 &= +j\omega_0\varepsilon_0\dbar{\varepsilon}_r(\omega_0)\mathbf{E}_0 + \dbar{\sigma}(\omega_0)\mathbf{E}_0.
            \end{align}
            \label{Eq:PertThUnperturbed}
        \end{subequations}
        In Eqs.~\eqref{Eq:PertThUnperturbed}, we used the subscript ``0'' to indicate unperturbed quantities; $\omega_0$ is the resonance frequency of the unperturbed cavity and $\mathbf{E}_0$, $\mathbf{H}_0$ are the respective electric and magnetic field distributions on resonance. We have suppressed the spatial dependence in material properties and field quantities for brevity and allowed for the case of diagonally anisotropic and dispersive materials. 

        Let us now assume that a small material perturbation is introduced in the system. For generality, this perturbation is introduced as a modification of the polarization and/or the induced current density, i.e., through terms of the form $j\omega\Delta\mathbf{P}$ and $\Delta\mathbf{J}$, respectively. Maxwell's equations then become
        \begin{subequations}
            \begin{align}
                \nabla\times\mathbf{E} &= -j\omega\mu_0\dbar{\mu}_r(\omega)\mathbf{H}, \\
                \nabla\times\mathbf{H} &= +j\omega\varepsilon_0\dbar{\varepsilon}_r(\omega)\mathbf{E} + \dbar{\sigma}(\omega)\mathbf{E} + j\omega\Delta\mathbf{P} + \Delta\mathbf{J},
            \end{align}
            \label{Eq:PertThPerturbed}
        \end{subequations}
        where the induced perturbations modify the resonance frequency of the cavity and the field distributions on resonance; this is indicated by removing the ``0'' subscript.

        To entangle the two versions of the problem (perturbed and unperturbed), the vector field $\mathbf{F}_c = \mathbf{E}_0^*\times\mathbf{H} + \mathbf{E}\times\mathbf{H}_0^*$ is defined. Although seemingly arbitrary, the definition of $\mathbf{F}_c$ is convenient since (i)~it has a form similar to the Poynting vector and (ii)~the evaluation of its divergence results in inner products between fields and their curls allowing to combine Eqs.~\eqref{Eq:PertThUnperturbed}~and~\eqref{Eq:PertThPerturbed}. 
        The divergence of $\mathbf{F}_c$ sometimes is referred to as an application of the conjugated Lorentz reciprocity theorem. This is quantitatively correct since the basic principles are the same but also somewhat inaccurate considering that in the textbook application of Lorentz reciprocity, the two problems share the same frequency. 
        Conjugated fields have been used in the definition of $\mathbf{F}_c$ because this allows to arrive at expressions with a clear physical interpretation. However, using the conjugated form of $\mathbf{F}_c$ means that the 
        the framework is strictly correct for Hermitian systems, i.e., for systems without ohmic loss or radiation.\cite{Joannopoulos,SnyderLove} Nevertheless, small levels of radiation and losses are tolerated, and thus quasi-Hermitian systems are equally well described.\cite{Bravo-abad2007} Thus, we can make the following replacements:  $\dbar{\varepsilon}_r = \RE{\dbar{\varepsilon}_r}$, $\dbar{\mu}_r = \RE{\dbar{\mu}_r}$, and $\dbar{\sigma} = \IM{\dbar{\sigma}}$. 
        Note that for non-Hermitian systems with high ohmic or radiation loss we need to use an unconjugated form of $\mathbf{F}_c$ and repeat a similar procedure with some extra complexity.\cite{Christopoulos2020OL} 

        The next step is to calculate the divergence of $\mathbf{F}_c$ which, via the vector identity $\nabla\cdot(\mathbf{A}\times\mathbf{B}) = (\nabla\times\mathbf{A})\cdot\mathbf{B} - \mathbf{A}\cdot(\nabla\times\mathbf{B})$, results in
        \begin{align}
            \nabla\cdot\mathbf{F}_c = &\nabla\cdot(\mathbf{E}_0^*\times\mathbf{H}) + \nabla\cdot(\mathbf{E}\times\mathbf{H}_0^*) \nonumber \\
            = &-j\mu_0[\omega\RE{\dbar{\mu}_r(\omega)}-\omega_0\RE{\dbar{\mu}_r(\omega_0)}]\mathbf{H}_0^*\cdot\mathbf{H} \nonumber \\
              &-j\varepsilon_0[\omega\RE{\dbar{\varepsilon}_r(\omega)}-\omega_0\RE{\dbar{\varepsilon}_r(\omega_0)}]\mathbf{E}_0^*\cdot\mathbf{E} \nonumber \\
              &-j[\IM{\dbar{\sigma}(\omega)}-\IM{\dbar{\sigma}(\omega_0)}]\mathbf{E}_0^*\cdot\mathbf{E} \nonumber \\
              &-j\omega\Delta\mathbf{P}\cdot\mathbf{E}_0^* - \Delta\mathbf{J}\cdot\mathbf{E}_0^*.
            \label{Eq:DivFcGeneral}
        \end{align}
        To simplify the terms inside the brackets, relative permittivity, permeability, and conductivity tensors at $\omega$ are expanded into Taylor series around $\omega_0$, i.e.,
        \begin{align}
            \omega\RE{\dbar{\varepsilon}_r(\omega)} = &\omega_0\RE{\dbar{\varepsilon}_r(\omega_0)} \nonumber \\
            &+ \left.(\omega-\omega_0)\partialFRAC{\{\omega\RE{\dbar{\varepsilon}_r(\omega)}\}}{\omega}\right|_{\omega=\omega_0} + \mathcal{O}^2,
            \label{Eq:TaylorPermittivity}
        \end{align}
        with similar expressions holding for $\omega\dbar{\mu}_r(\omega)$ and $\dbar{\sigma}(\omega)$. 
        Introducing Eq.~\eqref{Eq:TaylorPermittivity}~into~\eqref{Eq:DivFcGeneral} and ignoring second or higher order terms in the Taylor expansion we arrive at 
        \begin{align}
            \nabla\cdot\mathbf{F}_c = &-j\Delta\omega\mu_0\partialFRAC{\{\omega\RE{\dbar{\mu}_r(\omega)}\}}{\omega}\mathbf{H}_0^*\cdot\mathbf{H} \nonumber \\
              &-j\Delta\omega\varepsilon_0\partialFRAC{\{\omega\RE{\dbar{\varepsilon}_r(\omega)}\}}{\omega}\mathbf{E}_0^*\cdot\mathbf{E} \nonumber \\
              &- j\Delta\omega\partialFRAC{\{\IM{\dbar{\sigma}(\omega)}\}}{\omega}\mathbf{E}_0^*\cdot\mathbf{E} \nonumber \\
              &-j\omega\Delta\mathbf{P}\cdot\mathbf{E}_0^* - \Delta\mathbf{J}\cdot\mathbf{E}_0^*,
            \label{Eq:DivFcBeforePerturbation}
        \end{align}
        where $\Delta\omega=\omega-\omega_0$ and all the derivatives are calculated at $\omega = \omega_0$, with the notation dropped for brevity.

        The final step is to integrate Eq.~\eqref{Eq:DivFcBeforePerturbation} in an arbitrary volume $V$ that includes the cavity. We use the Gauss divergence theorem and exploit the zeroing of fields on the boundary $\partial V$ (zero or small radiation for the considered Hermitian or quasi-Hermitian system), so that
        \begin{equation}
            \iiint_V \nabla\cdot\mathbf{F}_c \mathrm{d}V = \oiint_{\partial V} \mathbf{F}_c\cdot \hat{\mathbf{n}} \mathrm{d}S = 0.
            \label{eq:divFcGauss}
        \end{equation}
        Under this condition, Eq.~\eqref{Eq:DivFcBeforePerturbation} takes the form
        \begin{align}
            \Delta\omega&\left(\iiint_V\mu_0\partialFRAC{\{\omega\RE{\dbar{\mu}_r(\omega)}\}}{\omega}\mathbf{H}_0^*\cdot\mathbf{H} \mathrm{d}V \right. \nonumber \\
            &+\iiint_V\varepsilon_0\partialFRAC{\left\{\omega\RE{\dbar{\varepsilon}_r(\omega)}\right\}}{\omega}\mathbf{E}_0^*\cdot\mathbf{E} \mathrm{d}V \nonumber \\
            &+\left.\iiint_V\partialFRAC{\{\IM{\dbar{\sigma}(\omega)}\}}{\omega}\mathbf{E}_0^*\cdot\mathbf{E} \mathrm{d}V \right) = \nonumber \\
            -\omega&\iiint_{V_p}\Delta\mathbf{P}\cdot\mathbf{E}_0^* \mathrm{d}V + j\iiint_{V_p}\Delta\mathbf{J}\cdot\mathbf{E}_0^* \mathrm{d}V.
            \label{Eq:PertTheoryWithEH}
        \end{align}
         Equation~\eqref{Eq:PertTheoryWithEH} is of limited practical use since it involves the perturbed quantities (fields and resonance frequency) which are unknown. To proceed, the fields are also expanded in Taylor series, i.e., $\mathbf{E} = \mathbf{E}_0 + \mathcal{O}^1$ and $\mathbf{H} = \mathbf{H}_0 + \mathcal{O}^1$, keeping only the zeroth order term; in physical terms, this means that the field distributions on resonance are left practically unchanged by the perturbation. We also use $\omega\approx \omega_0$ in the right-hand side of Eq.~\eqref{Eq:PertTheoryWithEH} and finally arrive at
        \begin{widetext}
            \begin{equation}
                \frac{\Delta\tilde\omega}{\omega_0} = -\frac{\displaystyle\iiint_{V_p} \Delta\mathbf{P}\cdot\mathbf{E}_0^* \mathrm{d}V - j\displaystyle\frac{1}{\omega_0}\displaystyle\iiint_{V_p} \Delta\mathbf{J}\cdot\mathbf{E}_0^* \mathrm{d}V}{\displaystyle\iiint_V \varepsilon_0\partialFRAC{\left\{\omega\RE{\dbar{\varepsilon}_r}\right\}}{\omega}\mathbf{E}_0^*\cdot\mathbf{E}_0 \mathrm{d}V + \displaystyle\iiint_V \mu_0\partialFRAC{\{\omega\RE{\dbar{\mu}_r}\}}{\omega}\mathbf{H}_0^*\cdot\mathbf{H}_0 \mathrm{d}V + \displaystyle\iiint_V \partialFRAC{\IM{\dbar{\sigma}}}{\omega}\mathbf{E}_0^*\cdot\mathbf{E}_0 \mathrm{d}V }. 
                \label{eq:GeneralPertTheoryAppendix}
            \end{equation}
        \end{widetext}
        which is the final result of the first-order perturbation theory, identical to Eq.~\eqref{eq:GeneralPertTheory} presented in Sec.~\ref{sec:CMTParamsCalculations}. To simplify the notation, one observes that the denominator of Eq.~\eqref{eq:GeneralPertTheoryAppendix} is proportional to the stored energy in the resonator
        \begin{align}
            W\rsub{res} = &\frac{1}{4}\iiint_V \varepsilon_0\partialFRAC{\left\{\omega\RE{\dbar{\varepsilon}_r}\right\}}{\omega}\mathbf{E}_0^*\cdot\mathbf{E}_0 \mathrm{d}V \nonumber \\
            &+ \frac{1}{4}\iiint_V \mu_0\partialFRAC{\{\omega\RE{\dbar{\mu}_r}\}}{\omega}\mathbf{H}_0^*\cdot\mathbf{H}_0 \mathrm{d}V \nonumber \\
            &+ \frac{1}{4}\iiint_V \partialFRAC{\IM{\dbar{\sigma}}}{\omega}\mathbf{E}_0^*\cdot\mathbf{E}_0 \mathrm{d}V,
            \label{eq:EnergyGeneral}
        \end{align}
        and, thus, Eq.~\eqref{eq:GeneralPertTheoryAppendix} can be written more compactly as
        \begin{equation}
            \frac{\Delta\tilde\omega}{\omega_0} = -\frac{\displaystyle\iiint_{V_p} \Delta\mathbf{P}\cdot\mathbf{E}_0^* \mathrm{d}V - j\displaystyle\frac{1}{\omega_0}\displaystyle\iiint_{V_p} \Delta\mathbf{J}\cdot\mathbf{E}_0^* \mathrm{d}V}{4W\rsub{res}}. 
            \label{eq:GeneralPertTheoryShortAppendix}
        \end{equation}

        In Eqs.~\eqref{eq:GeneralPertTheoryAppendix},~\eqref{eq:GeneralPertTheoryShortAppendix}, the integrals in the numerator are restricted to the perturbation volume $V_p$ where $\Delta\mathbf{P},\,\Delta\mathbf{J}\neq \mathbf{0}$.
        Furthermore, the resonance frequency shift induced by the perturbation, $\Delta\tilde{\omega}$, is assumed generally complex (notice the use of the tilde) to allow for the incorporation of any type of loss in $\Delta\mathbf{P}$ or $\Delta\mathbf{J}$. 
        We stress that the use of the conjugated $\mathbf{F}_c$ mentioned earlier places a requirement of Hermiticity on the \emph{unperturbed} problem.
        For instance, the addition of linear loss can be represented by setting $\Delta\mathbf{P} = j\IM{\Delta\dbar{\epsilon}}\mathbf{E}_0$ or $\Delta\mathbf{J} = \RE{\Delta\dbar{\sigma}}\mathbf{E}_0$ in Eq.~\eqref{eq:GeneralPertTheoryAppendix}.  
        Naturally,  $\Delta\mathbf{P}$ and/or $\Delta\mathbf{J}$ are not restricted to simple linear relations and can be used to incorporate nonlinear effects as well, as described in Secs.~\ref{sec:InstantNonLinearCMT}-\ref{sec:Gain}.

    \subsection{Coupling of cavities} \label{AppSub:CouplingPertTheory}

        Perturbation theory can be used to calculate the effects of coupling between cavities as well. We assume $N$ coupled cavities [see Fig.~\ref{fig:PertTheoryAppendix}(b) for a simplified system with $N = 2$]. The modes of the $N$-cavity structure can be specified by finding the extrema of the functional
        \begin{equation}
            \omega^2 = \frac{\displaystyle\frac{1}{\varepsilon_0}\displaystyle\iiint_V (\nabla\times\mathbf{H}^*)\cdot(\dbar{\varepsilon}_r^{-1}\nabla\times\mathbf{H})\mathrm{d}V}{\mu_0\displaystyle\iiint_V\mathbf{H}^*\cdot\dbar{\mu}_r\mathbf{H}\mathrm{d}V}.
            \label{eq:OmegaFunctional}
        \end{equation}
        This functional is found by dot multiplying the $H$-field vector-wave equation with $\mathbf{H}^*$ and integrating in an arbitrary enclosing volume $V$. The form of Eq.~\eqref{eq:OmegaFunctional} is reached only when the boundary term $\oiint_{\partial V}\mathbf{H}^*\cdot(\hat{\mathbf{n}}\times\mathbf{E})\mathrm{d}S = 0$ vanishes, a requirement which clearly holds in systems with negligible radiation loss.
        Strong radiation can be also accommodated by terminating the computational space with PMLs so that fields are still zeroed out at the outer boundary and re-deriving the functional by dot multiplying with the unconjugated field $\mathbf{H}$.\cite{Popovic2006,Tasolamprou2016} 
        
        Next, we suppose that the supermodes supported by the aggregate structure can be expressed as a linear combination of the $N$ isolated-cavity modes, i.e., $\mathbf{H}=\sum_m a_m\mathbf{H}_m$.\cite{Haus1991} In other words, we assume that coupling does not significantly perturb the individual modes. 
        We can now see that using the magnetic field as the working variable in the formulation is crucial in order for the supermode trial function to satisfy the divergence condition $\nabla\cdot\mathbf{D}=0$.\cite{Johnson2002} If the electric field is instead used then $\nabla\cdot\dbar{\varepsilon}\mathbf{E}=\sum_m a_m\nabla\cdot\Delta\dbar{\varepsilon}_m\mathbf{E}_m\neq0$, where $\dbar{\varepsilon}\equiv\dbar{\varepsilon}_m+\Delta\dbar{\varepsilon}_m$. 
        

        Obviously, each mode $\mathbf{H}_n$ satisfies a vector wave equation of its own. Taking the inner product with $\mathbf{H}_m^*$ and omitting the boundary term we can write
        \begin{equation}
            \frac{1}{\varepsilon_0}\iiint_V (\nabla\times\mathbf{H}_m^*)\cdot(\dbar{\varepsilon}_{r,n}^{-1}\nabla\times\mathbf{H}_n)\mathrm{d}V = \omega_n^2\mu_0\iiint_V\mathbf{H}_m^*\cdot\dbar{\mu}_{r,n}\mathbf{H}_n\mathrm{d}V.
            \label{eq:OmegaPerMode}
        \end{equation}
        Introducing the mode expansion in Eq.~\eqref{eq:OmegaFunctional} and using Eq.~\eqref{eq:OmegaPerMode} one can write Eq.~\eqref{eq:OmegaFunctional} in the matrix from
        \begin{equation}
            \omega^2 = \frac{\mathbf{a}^\dagger\mathbf{K}\mathbf{a}}{\mathbf{a}^\dagger\mathbf{W}\mathbf{a}},
            \label{eq:OmegaFunctionalCompact}
        \end{equation}
        where $\mathbf{a}$ is a column vector containing the expansion amplitudes and $\mathbf{K}$, $\mathbf{W}$ are $N\times N$ matrices defined through
        \begin{subequations}
            \begin{align}
                K_{mn} &= \frac{1}{\varepsilon_0}\iiint_V (\nabla\times\mathbf{H}_m^*)\cdot(\dbar{\varepsilon}_r^{-1}\nabla\times\mathbf{H}_n)\mathrm{d}V, \\
                W_{mn} &= \mu_0\iiint_V\mathbf{H}_m^*\cdot\dbar{\mu}_r\mathbf{H}_n\mathrm{d}V. \label{eq:Wdefinitions}
            \end{align}
            \label{eq:KWdefinitions}
        \end{subequations}
        Note that the integrals in Eqs.~\eqref{eq:KWdefinitions} should be performed in the entire computational space $V$ and involve the total material distributions $\dbar{\varepsilon}_r$ and $\dbar{\mu}_r$ (i.e., all cavities), but only two mode profiles. 
        To simplify matters we express the inverse relative permittivity (relative impermeability) as $\dbar{\varepsilon}_r^{-1} = \dbar{\varepsilon}_{r,n}^{-1} + \Delta\dbar{\varepsilon}_n^{-1}$ and restrict the presentation to cavities without magnetic properties. Then, the elements of the $\mathbf{K}$ matrix can be written as 
        \begin{align}
            K_{mn} = &\frac{1}{\varepsilon_0}\iiint_V (\nabla\times\mathbf{H}_m^*)\cdot(\dbar{\varepsilon}_{r,n}^{-1}\nabla\times\mathbf{H}_n)\mathrm{d}V \nonumber \\
            &+\frac{1}{\varepsilon_0}\iiint_{V_p} (\nabla\times\mathbf{H}_m^*)\cdot(\Delta\dbar{\varepsilon}_n^{-1}\nabla\times\mathbf{H}_n)\mathrm{d}V \nonumber \\
            = &\omega_n^2\mu_0\iiint_V\mathbf{H}_m^*\cdot\mathbf{H}_n\mathrm{d}V + \omega_m\omega_n\frac{1}{\varepsilon_0}\iiint_{V_p} \mathbf{D}_m^*\cdot\Delta\dbar{\varepsilon}_n^{-1}\mathbf{D}_n\mathrm{d}V \nonumber \\
            = &W_{mn}\omega_n^2 + \omega_m M_{mn} \omega_n.
            \label{eq:KdefinitionSpecial}
        \end{align}
        It is clear that the newly introduced matrix $\mathbf{M}$ describes coupling. However, the elements of $\mathbf{M}$ are not the coupling coefficients $\mu_{c,mn}$ of CMT since Eq.~\eqref{eq:OmegaFunctionalCompact} does not correspond to the standard form of a first-order CMT equation. 

        To proceed, Eq.~\eqref{eq:OmegaFunctionalCompact} is differentiated with respect to the complex $a_m^*$ (for details see Ref.~\onlinecite{Haus1991}) and becomes
        \begin{equation}
            \mathbf{W}^{-1}\mathbf{K}\mathbf{a} = \omega^2\mathbf{a}.
            \label{eq:OmegaEigenvalue}
        \end{equation}
        Equation~\eqref{eq:OmegaEigenvalue} describes an eigenvalue problem; the eigenvalues of the $\mathbf{W}^{-1}\mathbf{K}$ matrix are the (squared) resonance frequencies of the supported supermodes. Thus, by knowing the mode profiles and resonance frequencies of each isolated cavity one can calculate the matrices $\mathbf{K}$ and $\mathbf{W}$ through Eqs.~\eqref{eq:KdefinitionSpecial}~and~\eqref{eq:Wdefinitions}, respectively, and then  return to Eq.~\eqref{eq:OmegaEigenvalue} and compute the eigenvalues.
        As discussed in Sec.~\ref{subsec:LinerCMTCouplingParams}, it is then possible to use Eq.~\eqref{eq:CMTmutualCouplingFrequencies} to specify $\mu_c$. 
        This process can be performed for each pair of cavities in the complete system, assuming that coupling between two cavities is independent of the presence of the other $N-2$ remaining resonators.

\section{Handling of Perturbative Nonlinearities in Coupled-Mode Theory}\label{App:DeltaOmegaVSa}

    In this Appendix, we discuss how perturbation theory can be utilized to introduce some notable nonlinear effects in the CMT framework. This process actually consists of transforming the physical law that governs a nonlinearity into a term appropriate for introduction in the CMT framework, i.e., typically expressing the physical law as a functions of the cavity amplitude. The cornerstone of this process is the application of Eq.~\eqref{eq:GeneralPertTheoryShortAppendix} for a specific nonlinearity, i.e., for a specific $\Delta\mathbf{P}$ and/or $\Delta\mathbf{J}$ term. We will mainly discuss nonlinearities in 2D materials, i.e., we will present the framework in terms of the nonlinear induced \emph{surface} current, since this represents the most recent development. However, we will also provide with expressions for bulk nonlinear materials.

    \subsection{Single-channel instantaneous third-order nonlinearities: Kerr effect and two-photon absorption}\label{AppSub:DeltaOmegaVSaKerrTPA}

        Two-dimensional materials are typically treated as infinitesimally thin layers, effectively modeled by a surface conductivity and a consequent induced surface  current. For large intensities, nonlinearities can appear and
        \begin{equation}
            \mathbf{J}_s = \mathbf{J}_s^{(1)} + \mathbf{J}_s^{(3)} = \dbar{\sigma}^{(1)}\mathbf{E} + \mathbf{J}_s^{(3)},
        \end{equation}
        assuming a centrosymmetric material without second-order nonlinearities.\cite{Chatzidimitriou2015} Note here that the surface current, measured in $\mathrm{A/m}$, is connected with the current density of Maxwell's equation, measured in $\mathrm{A/m^2}$, through $\mathbf{J} = \mathbf{J}_s \delta_s(\mathbf{r})$, where $\delta_s(\mathbf{r})$ is a surface Dirac function. Obviously, when applying Eq.~\eqref{eq:GeneralPertTheoryShortAppendix}, $\Delta\mathbf{J} = \mathbf{J}_s^{(3)}\delta_s(\mathbf{r})$ and, similarly, if a bulk nonlinear material is involved, $\Delta\mathbf{P} = \mathbf{P}^{(3)}$.\cite{Christopoulos2017} The exact expression of $\mathbf{J}_s^{(3)}$ depends on the material and the possible nonlinear anisotropic properties that it possesses, expressed through the fourth order tensor $\dbar{\sigma}^{(3)}$ [and $\dbar{\chi}^{(3)}$ for polarization nonlinearities]. In the most general case where all 81 elements are nonzero, the $\mu\rsup{th}$ component of the induced nonlinear current is written as\cite{ButcherCotter}
        \begin{equation}
            J_{s,\mu}^{(3)}= \frac{3}{4}\sum_{\alpha\beta\gamma}\sigma^{(3)}_{\mu\alpha\beta\gamma}E_\alpha E_\beta^* E_\gamma, \label{eq:GeneralNonlinearSurfaceCurrent}
        \end{equation}
        where $\mu,\,\alpha,\,\beta,\,\gamma$ run through the components of the coordinate system used (e.g., $\{x,y,z\}$ for the Cartesian system).
    
        The cumbersome form of Eq.~\eqref{eq:GeneralNonlinearSurfaceCurrent} can be simplified in most materials. For example, in a graphene sheet placed normal to one Cartesian axis, all 81 elements of $\dbar{\sigma}^{(3)}$ either depend on a single parameter $\sigma_3$ or zero out,\cite{Chatzidimitriou2015} leading to the compact expression
        \begin{equation}
            \mathbf{J}_{s,\mathrm{gr}}^{(3)} = \frac{\sigma_3}{4}[2(\mathbf{E}_\parallel\cdot\mathbf{E}_\parallel^*)\mathbf{E}_\parallel + (\mathbf{E}_\parallel\cdot\mathbf{E}_\parallel)\mathbf{E}_\parallel^*], 
            \label{eq:GrapheneNonlinearSurfaceCurrent}
        \end{equation}
        with $\mathbf{E}_\parallel = \mathbf{\hat n}\times\mathbf{E}\times\mathbf{\hat n}$ representing the in-plane electric field components, parallel to the sheet, 
        Similar simplifications can be exercised for bulk materials described through a nonlinear polarization term. Silicon, for instance, possess an anisotropic nonlinear response and its third order nonlinear polarization is typically written as
        \begin{equation}
            \mathbf{P}\rsub{Si}^{(3)} = \varepsilon_0\frac{\chi_{xxxx}^{(3)}}{4}[(3-\eta)(\mathbf{E}\cdot\mathbf{E}^*)\mathbf{E} + \eta(\mathbf{E}\cdot\mathbf{E})\mathbf{E}^*], 
            \label{eq:SiliconNonlinearPolarization}
        \end{equation}
        where $\eta = 1.27$ and now the full $\mathbf{E}$-field is used.\cite{ButcherCotter} However, quite sometimes it is considered that $\eta=1$ which simplifies the analysis without introducing significant errors.\cite{Wang2013a}
    
        When the expression of the nonlinearity is specified, one can apply Eq.~\eqref{eq:GeneralPertTheoryShortAppendix} to calculate the induced resonance frequency shift by further assuming the approximate equality of the perturbed and unperturbed fields, i.e., $\mathbf{E}\approx\mathbf{E}_0$. In case of graphene, for example, by substituting Eq.~\eqref{eq:GrapheneNonlinearSurfaceCurrent} into \eqref{eq:GeneralPertTheoryShortAppendix}, we reach\cite{Christopoulos2016a,Christopoulos2017}
        \begin{equation}
            \Delta\tilde\omega = j\frac{\displaystyle\iint_{S_p} \frac{\sigma_3}{4}[2|\mathbf{E}_{0,\parallel}|^4 + |\mathbf{E}_{0,\parallel}\cdot\mathbf{E}_{0,\parallel}|^2] \mathrm{d}S}{4W\rsub{res}}. 
            \label{eq:DeltaOmegaGrapheneShortIntegrals}
        \end{equation}
        Note that in the numerator of Eq.~\eqref{eq:DeltaOmegaGrapheneShortIntegrals}, the volume integral is reduced to a surface integral on graphene
        due to the involvement of the $\delta_s(\mathbf{r})$ Dirac function. 
        Thus far we have not made any assumption regarding the nature of $\sigma_3$. Kerr-like nonlinearities in 2D materials induce self-phase modulation and are naturally described by a purely imaginary $\sigma_3$ which will result in a purely real $\Delta\omega$.\cite{Christopoulos2016a} However, by allowing for a complex $\sigma_3$ one can also consider third-order nonlinear losses, i.e., two-photon absorption.\cite{ButcherCotter,Cheng2014} This approach is allowed by the framework and simply leads to a complex $\Delta\tilde\omega$, with its (energy-dependent) imaginary part, which stems from $\RE{\sigma_3} \neq 0$, describing the nonlinear contribution of TPA.
    
        The fact that Eq.~\eqref{eq:DeltaOmegaGrapheneShortIntegrals} involves the stored energy highlights the path to express the nonlinear frequency shift $\Delta\tilde\omega$ in terms compatible with CMT. More specifically, using the fact that $W\rsub{res} \equiv |a|^2$, one can multiply the numerator and denominator of Eq.~\eqref{eq:DeltaOmegaGrapheneShortIntegrals} with $W\rsub{res}$, and reach
        \begin{align}
            \Delta\tilde\omega(a) &= \tilde\gamma_s|a(t)|^2 = (-\gamma_{s,\mathrm{SPM}}+j\gamma_{s,\mathrm{TPA}})|a(t)|^2 =\left(\frac{\omega_0}{c_0}\right)^4 \times \nonumber \\
            & \left(-\kappa_{s,\mathrm{SPM}}\frac{\IM{\sigma_3\rsup{max}}}{\varepsilon_0^2}+j\kappa_{s,\mathrm{TPA}}\frac{\RE{\sigma_3\rsup{max}}}{\varepsilon_0^2}\right) |a(t)|^2.
            \label{eq:DeltaOmegaGrapheneFull}
        \end{align}
        Although this manipulation seems arbitrary, it is necessary so that the newly introduced \emph{surface nonlinear parameter} $\tilde\gamma_s$ (measured in $\mathrm{1/Js}$) is independent of the stored energy in the cavity. An alternative, more strict, and general path to reach Eq.~\eqref{eq:DeltaOmegaGrapheneFull} will be presented in Appendix~\ref{AppSub:DeltaOmegaVSaXPMTHGFWM}. The (normalized) spatial overlap between the electric field and the nonlinear material is encapsulated by the dimensionless \emph{nonlinear feedback parameters} $\kappa$ which take the form\cite{Christopoulos2016a,Christopoulos2017}
        \begin{subequations}
            \begin{align}
                \kappa_{s,\mathrm{SPM}} &= \left(\frac{c_0}{\omega_0}\right)^4\frac{\displaystyle\iint_{S_p} \IM{\sigma_3}[2|\mathbf{E}_{0,\parallel}|^4 + |\mathbf{E}_{0,\parallel}\cdot\mathbf{E}_{0,\parallel}|^2] \mathrm{d}S}{\displaystyle\frac{16}{\varepsilon_0^2}W\rsub{res}^2\IM{\sigma_3\rsup{max}}}, \\
                \kappa_{s,\mathrm{TPA}} &= \left(\frac{c_0}{\omega_0}\right)^4\frac{\displaystyle\iint_{S_p} \RE{\sigma_3}[2|\mathbf{E}_{0,\parallel}|^4 + |\mathbf{E}_{0,\parallel}\cdot\mathbf{E}_{0,\parallel}|^2] \mathrm{d}S}{\displaystyle\frac{16}{\varepsilon_0^2}W\rsub{res}^2\RE{\sigma_3\rsup{max}}},
            \end{align}
            \label{eq:kappaGrapheneSPMTPA}
        \end{subequations}
        for SPM and TPA, respectively.\cite{Christopoulos2017} The definition of $\kappa$ is performed in such a way so that it is independent of\cite{Soljacic2002} (i)~the $\mathbf{E}$-field intensity used in the calculations, achieved via the normalization with $W\rsub{res}^2$, (ii)~the dimensionality of the system via the introduction of the $(\omega_0/c_0)^4$ term (see, e.g., the treatment of $\kappa$ in Ref.~\onlinecite{Christopoulos2016a}), and (iii)~the involved nonlinear material(s) and the strength of their nonlinearity, achieved via the normalization with $\sigma_3\rsup{max}$ as the maximum value of $\sigma_3$ in the considered system.
    
        A similar path can be followed with bulk polarization nonlinearities.\cite{Bravo-abad2007,Tsilipakos2014,Tsilipakos2016} We only provide here with the respective final expressions, introduced in such a way so that the respective nonlinear parameters $\tilde\gamma_{b/s}$  (both measured in $\mathrm{1/Js}$) can be algebraically added together to evaluate the total nonlinear frequency shift
        \begin{align}
            \Delta&\tilde\omega(a) = \tilde\gamma_b|a(t)|^2 = (-\gamma_{b,\mathrm{SPM}}+j\gamma_{b,\mathrm{TPA}})|a(t)|^2 \nonumber \\
            &=\left(\frac{\omega_0}{c_0}\right)^3 \left(-4\omega_0 c_0 \kappa_{b,\mathrm{SPM}}n_2\rsup{max} + j2c_0^2\kappa_{b,\mathrm{TPA}}\beta\rsub{TPA}\rsup{max}\right) |a(t)|^2,
            \label{eq:DeltaOmegaBulkFull}
        \end{align}
        
        \begin{subequations}
            \begin{align}
                \kappa_{b,\mathrm{SPM}} &= \left(\frac{c_0}{\omega_0}\right)^3\frac{\displaystyle\frac{1}{3}\iiint_{V_p} n_2n_0^2[2|\mathbf{E}_0|^4 + |\mathbf{E}_0\cdot\mathbf{E}_0|^2] \mathrm{d}V}{\displaystyle\frac{16}{\varepsilon_0^2}W\rsub{res}^2n_2\rsup{max}}, \\
                \kappa_{b,\mathrm{TPA}} &= \left(\frac{c_0}{\omega_0}\right)^3\frac{\displaystyle\frac{1}{3}\iiint_{V_p} \beta\rsub{TPA}n_0^2[2|\mathbf{E}_0|^4 + |\mathbf{E}_0\cdot\mathbf{E}_0|^2] \mathrm{d}V}{\displaystyle\frac{16}{\varepsilon_0^2}W\rsub{res}^2\beta\rsub{TPA}\rsup{max}}.
            \end{align}
            \label{eq:kappaBulk}
        \end{subequations}
        To reach Eqs.~\eqref{eq:kappaBulk}, we have expressed the (complex) nonlinear susceptibility $\chi_{xxxx}^{(3)}$ through a common transformation\cite{ButcherCotter} that relates it with the nonlinear index $n_2$ and the respective TPA parameter $\beta\rsub{TPA}$ of the nonlinear material, i.e.,
        \begin{equation}
            \chi_{xxxx}^{(3)} = \frac{4}{3}n_0^2\varepsilon_0c_0n_2 - j\frac{2}{3}\frac{\varepsilon_0c_0^2n_0^2}{\omega}\beta\rsub{TPA}.
            \label{eq:chi3ton2betaTPA}
        \end{equation}

        Equations~\eqref{eq:DeltaOmegaGrapheneFull}~and~\eqref{eq:DeltaOmegaBulkFull} are the end result of the described process and coincide with the respective terms that were heuristically introduced in CMT in Sec.~\ref{subsec:SPM} to describe SPA and TPA. The nonlinear feedback parameters are typically calculated using the on-resonance cavity mode, retrieved either by modal techniques (most commonly)\cite{Tsilipakos2014} or by solving a weakly coupled, on-resonance harmonic propagation problem. Finally, we should note that $\gamma\rsub{TPA}$ is  always positive to introduces loss in the framework ($\RE{\sigma_3},~\beta\rsub{TPA}>0$). On the contrary, $\gamma\rsub{SPM}$ can be either positive or negative, leading to a red- or blue-shift, respectively, of the resonance frequency. For instance, in silicon $n_2>0$ (self-focusing), red-shifting the ``cold'' resonance frequency, while in graphene the sign of $\IM{\sigma_3}$ depends on the frequency and on its chemical doping.

    \subsection{Multi-channel instantaneous nonlinearities: Cross-phase modulation, high-harmonic generation, and frequency mixing}\label{AppSub:DeltaOmegaVSaXPMTHGFWM}

        The form of Eq.~\eqref{eq:GeneralNonlinearSurfaceCurrent} refers to the nonlinear surface current at frequency $\omega$, induced by a single field at the same frequency. In the most general case, however, three fields at different frequencies can interact and induce a nonlinear surface current at a new, fourth frequency, which is, for instance, the algebraic sum of the previous three, i.e., $\omega_n = \omega_k+\omega_\ell+\omega_m$. In this sense, a more general version of Eq.~\eqref{eq:GeneralNonlinearSurfaceCurrent} can be written as\cite{ButcherCotter}
        \begin{equation}
            J_{s,n,\mu}^{(3)} = \frac{1}{4}\sum_{\alpha\beta\gamma}\sigma^{(3)}_{\mu\alpha\beta\gamma}E_{k,\alpha}E_{\ell,\beta} E_{m,\gamma}. \label{eq:GeneralNonlinearSurfaceCurrentArbOmega}
        \end{equation}
        Evidently, Eq.~\eqref{eq:GeneralNonlinearSurfaceCurrent} emerges from Eq.~\eqref{eq:GeneralNonlinearSurfaceCurrentArbOmega} for $\omega_k=\omega$, $\omega_\ell=-\omega$, and $\omega_m=\omega$ so that $\omega_n = \omega-\omega+\omega \equiv \omega$. Note that when a negative frequency is involved, the respective electric field component should be replaced by its complex conjugate, i.e., $\mathbf{E}_{-k}=\mathbf{E}^*_{k}$ and the notation $\mathbf{E}_{\pm k}$ is reserved for the field in the respective frequency $\pm\omega_k$.\cite{ButcherCotter,BoydBook}

        Generalizing the single-channel derivation of the previous section and allowing for a second, different frequency, more interesting nonlinear phenomena emerge. The two most notable are (i)~cross-phase modulation which appears for $\omega_k=\omega_1$, $\omega_\ell=-\omega_2$, $\omega_m=\omega_2$ so that $\omega_n = \omega_1-\omega_2+\omega_2 \equiv \omega_1$ and describes how a wave at $\omega_2$ nonlinearly affects the response of a wave at $\omega_1$ and (ii)~third-harmonic generation which appears for $\omega_k=\omega_\ell=\omega_m=\omega$ so that $\omega_n = \omega+\omega+\omega \equiv 3\omega$ and describes the generation of a new wave at the third-harmonic of the fundamental frequency $\omega$. When more frequencies are allowed, even more complex interactions emerge but only a handful of them are useful, describing interactions of waves in the same frequency band. All those interactions are collectively termed four-wave mixing processes, appearing for $\omega_k=\pm\omega_1$, $\omega_\ell=\pm\omega_2$, $\omega_m=\pm\omega_3$ so that $\omega_n = \pm\omega_1\pm\omega_2\pm\omega_3=\omega_4$. The most common and useful version of FWM involves two incident fields and produces a third one, it is termed degenerate four-wave mixing, and appears for $\omega_k=\omega_\ell=\omega_1$, $\omega_m=-\omega_2$ so that $\omega_n = \omega_1+\omega_1-\omega_2 \equiv 2\omega_1-\omega_2=\omega_3$.

        Equation~\eqref{eq:GeneralNonlinearSurfaceCurrentArbOmega}, similar to Eq.~\eqref{eq:GeneralNonlinearSurfaceCurrent}, can be simplified when a specific material with given symmetries and a specific nonlinear effect is considered. We will only give the simplified version of the induced nonlinear currents on a graphene sheet at the various involved frequencies but similarly expressions for $\mathbf{J}^{(3)}$ and $\mathbf{P}^{(3)}$ can be extracted for any other sheet or bulk nonlinear materials, respectively.\cite{Yanik2003a,Yanik2004,Rodriguez2007} Beginning from XPM, the nonlinear current at $\omega_k$, induced by the interaction with $\omega_\ell$ is given by
        \begin{align}
            &\mathbf{J}_{s,k,\mathrm{gr}}^{(3)} = \frac{\sigma_3}{4} \times \nonumber \\
            &~~~[2(\mathbf{E}_{\ell,\parallel}\cdot\mathbf{E}_{\ell,\parallel}^*)\mathbf{E}_{k,\parallel} +2(\mathbf{E}_{k,\parallel}\cdot\mathbf{E}_{\ell,\parallel})\mathbf{E}_{\ell,\parallel}^* +2(\mathbf{E}_{k,\parallel}\cdot\mathbf{E}_{\ell,\parallel}^*)\mathbf{E}_{\ell,\parallel}].
            \label{eq:GrapheneNonlinearSurfaceCurrentXPM}
        \end{align}
        To extract Eq.~\eqref{eq:GrapheneNonlinearSurfaceCurrentXPM}, the true two-dimensional nature of $\dbar{\sigma}^{(3)}$ is considered, as well as all the available field permutations in Eq.~\eqref{eq:GeneralNonlinearSurfaceCurrentArbOmega}.\cite{ButcherCotter} The influence of XPM on the resonance frequency of a cavity fed with two waves at $\omega_k$ and $\omega_\ell$ ($k\neq\ell$) is retrieved by substituting Eq.~\eqref{eq:GrapheneNonlinearSurfaceCurrentXPM} into \eqref{eq:GeneralPertTheoryShortAppendix} to obtain 
        \begin{align}
            \Delta&\omega_{\mathrm{XPM},k}(a_\ell) = -2\gamma_{s,\mathrm{XPM},k\ell}|a_\ell(t)|^2 \nonumber \\
            &= -2\left(\frac{\omega_k}{c_0}\right)^4 \kappa_{s,\mathrm{XPM},k\ell}\frac{\IM{\sigma_3\rsup{max}}}{\varepsilon_0^2} |a_\ell(t)|^2.
            \label{eq:DeltaOmegaGrapheneXPM}
        \end{align}
        The respective XPM nonlinear feedback parameter is given by
        \begin{align}
            &\kappa_{s,\mathrm{XPM},k\ell} = \left(\frac{c_0}{\omega_k}\right)^4 \times \nonumber \\ 
            &~~~\frac{\displaystyle\iint_{S_p} \IM{\sigma_3}[|\mathbf{E}_{k,\parallel}|^2|\mathbf{E}_{\ell,\parallel}|^2 + |\mathbf{E}_{k,\parallel}\cdot\mathbf{E}_{\ell,\parallel}|^2 + |\mathbf{E}_{k,\parallel}\cdot\mathbf{E}_{\ell,\parallel}^*|^2] \mathrm{d}S}{\displaystyle\frac{16}{\varepsilon_0^2}W_{\mathrm{res},k}W_{\mathrm{res},\ell}\IM{\sigma_3\rsup{max}}},
            \label{eq:kappaGrapheneXPM}
        \end{align}
        measuring, in a normalized manner, the interaction between the resonance modes at the two frequencies $\omega_k$ and $\omega_\ell$. $\mathbf{E}_{k,\parallel}$ denotes the unperturbed quantity, i.e., we have dropped the use of ``0'' for compactness.
        Finally, note that 
        there exists another resonance frequency shift for the other mode $\Delta\omega_{\mathrm{XPM},\ell}$ with the expressions readily derived  by mutually exchanging $k\leftrightarrow\ell$ in Eqs.~\eqref{eq:DeltaOmegaGrapheneXPM}~and~\eqref{eq:kappaGrapheneXPM}.

        THG is approached similarly, with the nonlinear current at $\omega_3 = 3\omega_1$ given by\cite{Theodosi2022}
        \begin{equation}
            \mathbf{J}_{s,3,\mathrm{gr}}^{(3)} = \frac{\sigma_3}{4}(\mathbf{E}_{1,\parallel}\cdot\mathbf{E}_{1,\parallel})\mathbf{E}_{1,\parallel}.
            \label{eq:GrapheneNonlinearSurfaceCurrentTHG3w}
        \end{equation}
        However, there is an additional nonlinear current at $\omega_1$ induced by the interaction of the produced probe wave at $\omega_3$ and the pump wave at $\omega_1$ through the interaction $\omega_3-2\omega_1=\omega_1$ which is actually a wave-mixing effect.\cite{Rodriguez2007,Hashemi2009} This nonlinear current is given by
        \begin{equation}
            \mathbf{J}_{s,1,\mathrm{gr}}^{(3)} = \frac{\sigma_3}{4}[2(\mathbf{E}_{1,\parallel}^*\cdot\mathbf{E}_{3,\parallel})\mathbf{E}_{1,\parallel}^*+(\mathbf{E}_{1,\parallel}^*\cdot\mathbf{E}_{1,\parallel}^*)\mathbf{E}_{3,\parallel}].
            \label{eq:GrapheneNonlinearSurfaceCurrentTHGw}
        \end{equation}
        Consequently, the pump wave at $\omega_1$ produces a probe wave at $3\omega_1$ but when the latter becomes strong enough, a back conversion process appears, as well.\cite{Christopoulos2018} 
        Substituting Eqs.~\eqref{eq:GrapheneNonlinearSurfaceCurrentTHG3w}~and~\eqref{eq:GrapheneNonlinearSurfaceCurrentTHGw} into \eqref{eq:GeneralPertTheoryShortAppendix}, we get the respective nonlinear frequency shifts
        \begin{subequations}
            \begin{align}
                \Delta&\tilde\omega_{\mathrm{THG},1}(a_1,a_3) = -3\tilde\beta_{s,\mathrm{THG},1}\frac{[a_1^*(t)]^2 a_3(t)}{a_1(t)} \nonumber \\
                &= -3\left(\frac{\omega_1}{c_0}\right)^4 \tilde\kappa_{s,\mathrm{THG},1}\frac{\IM{\sigma_3\rsup{max}}}{\varepsilon_0^2}\frac{[a_1^*(t)]^2 a_3(t)}{a_1(t)}, \\
                \Delta&\tilde\omega_{\mathrm{THG},3}(a_1,a_3) = -\tilde\beta_{s,\mathrm{THG},3}\frac{a_1^3(t)}{a_3(t)} \nonumber \\
                &= -\left(\frac{\omega_3}{c_0}\right)^4 \tilde\kappa_{s,\mathrm{THG},3}\frac{\IM{\sigma_3\rsup{max}}}{\varepsilon_0^2}\frac{a_1^3(t)}{a_3(t)}.
            \end{align}
            \label{eq:DeltaOmegaGrapheneTHG}
        \end{subequations}
        THG nonlinear feedback parameters are  given by
        \begin{subequations}
            \begin{align}
                \tilde\kappa_{s,\mathrm{THG},1} &= \left(\frac{c_0}{\omega_1}\right)^4 \frac{\displaystyle\iint_{S_p} \IM{\sigma_3}(\mathbf{E}_{1,\parallel}^*\cdot\mathbf{E}_{1,\parallel}^*) (\mathbf{E}_{1,\parallel}^*\cdot\mathbf{E}_{3,\parallel}) \mathrm{d}S}{\displaystyle\frac{16}{\varepsilon_0^2}W_{\mathrm{res},1}^{3/2}W_{\mathrm{res},3}^{1/2}\IM{\sigma_3\rsup{max}}}, \\
                \tilde\kappa_{s,\mathrm{THG},3} &= \left(\frac{c_0}{\omega_3}\right)^4 \frac{\displaystyle\iint_{S_p} \IM{\sigma_3}(\mathbf{E}_{1,\parallel}\cdot\mathbf{E}_{1,\parallel}) (\mathbf{E}_{1,\parallel}\cdot\mathbf{E}_{3,\parallel}^*) \mathrm{d}S}{\displaystyle\frac{16}{\varepsilon_0^2}W_{\mathrm{res},1}^{3/2}W_{\mathrm{res},3}^{1/2}\IM{\sigma_3\rsup{max}}},
            \end{align}
            \label{eq:kappaGrapheneTHG}
        \end{subequations}
        implying that $\tilde\beta_{s,\mathrm{THG},1} = \tilde\beta_{s,\mathrm{THG},3}^*$. This reveals that the forward and backward conversion processes are physically related and their relative strength depends only on the intensity of the two waves [$(a_1^*)^2 a_3$ versus $a_1^3$]. For the case of bulk materials, the respective equality is $\tilde\beta_{b,\mathrm{THG},1}/\omega_1 = \tilde\beta_{b,\mathrm{THG},3}^*/\omega_3$.\cite{Rodriguez2007,Hashemi2009} Finally, the notation $\tilde\beta$ with the use of tilde is deliberate since the parameters are complex. Naturally, an imaginary $\tilde\beta$ implies a loss/gain mechanism, physically modeling the process of, e.g., pump depletion (loss) to produce the probe wave (gain) or the backward process of frequency mixing between pump and probe waves. The actual effect of THG as a loss/gain mechanism is further determined by the exact amplitudes of the respective modes, see Eqs.~\eqref{eq:DeltaOmegaGrapheneTHG} and how $a_1$, $a_3$ are entangled to ultimately determine the respective $\Delta\tilde\omega\rsub{THG}$.
        
        We shall stop here and comment on a quite elegant handling that is required to reach Eqs.~\eqref{eq:DeltaOmegaGrapheneTHG}. As the reader may have realized, the resonance frequency shift of Eqs.~\eqref{eq:DeltaOmegaGrapheneTHG} does not depend on the stored energy in the cavity but rather on an interplay of the modes amplitudes. This implies that the approach of multiplying numerator and denominator with the respective stored energy cannot be followed here. However, there is an inherently different handling, which is actually more rigorous than the previously used one. Specifically, two reference fields $\mathbf{E}\rsub{ref}(\mathbf{r})$ and $\mathbf{H}\rsub{ref}(\mathbf{r})$ are defined, normalized so that their on-resonance stored energy equal unity, i.e., $W\rsub{ref} \equiv |a\rsub{ref}|^2 = 1$, as calculated through Eq.~\eqref{eq:EnergyGeneral}. This means that any other field can be expressed as $\mathbf{E}(\mathbf{r}) = a\mathbf{E}\rsub{ref}(\mathbf{r}) = \mathbf{E}\rsub{ref}(\mathbf{r})\sqrt{W\rsub{res}}$.
        Note that this concept can also be used in a time-evolving field quantity in order to disentangle its spatial and temporal dependencies on-resonance,  i.e., $\mathbf{E}\OF{r}{t} = a(t)\mathbf{E}\rsub{ref}(\mathbf{r})$. 
        Through this idea,  Eqs.~\eqref{eq:DeltaOmegaGrapheneTHG} [and~also~Eq.~\eqref{eq:kappaGrapheneXPM}] can be retrieved, albeit in a slightly different form involving the reference field $\mathbf{E}\rsub{ref}(\mathbf{r})$. For uniformity and to align with what is typically reported in the literature,\cite{Christopoulos2018,Christopoulos2020JOSAB,Rodriguez2007,Hashemi2009,Ramirez2011,Zeng2014,Lin2014} we opt to de-normalize the obtained expression once again [via $\mathbf{E}\rsub{ref}(\mathbf{r}) = \mathbf{E}(\mathbf{r})/\sqrt{W\rsub{res}}$] and then Eqs.~\eqref{eq:DeltaOmegaGrapheneTHG} are indeed retrieved.

        Finally, we present the respective nonlinear currents and resonance frequency shifts for the DFWM process. Assuming a strong pump wave at $\omega_1$ and a weak signal wave at $\omega_2$, the produced idler wave emerges at $\omega_3 = 2\omega_1-\omega_2$. The induced nonlinear currents at $\omega_3$, $\omega_2 = 2\omega_1-\omega_3$, and $\omega_1 = -\omega_1+\omega_2+\omega_3$ are
        \begin{subequations}
            \begin{align}
                \mathbf{J}_{s,3,\mathrm{gr}}^{(3)} = \frac{\sigma_3}{4}[&2(\mathbf{E}_{1,\parallel}\cdot\mathbf{E}_{2,\parallel}^*)\mathbf{E}_{1,\parallel}+(\mathbf{E}_{1,\parallel}\cdot\mathbf{E}_{1,\parallel})\mathbf{E}_{2,\parallel}^*], \label{eq:GrapheneNonlinearSurfaceCurrentDFWM1} \\
                \mathbf{J}_{s,2,\mathrm{gr}}^{(3)} = \frac{\sigma_3}{4}[&2(\mathbf{E}_{1,\parallel}\cdot\mathbf{E}_{3,\parallel}^*)\mathbf{E}_{1,\parallel}+(\mathbf{E}_{1,\parallel}\cdot\mathbf{E}_{1,\parallel})\mathbf{E}_{3,\parallel}^*], \\
                \mathbf{J}_{s,1,\mathrm{gr}}^{(3)} = \frac{\sigma_3}{4}[&2(\mathbf{E}_{2,\parallel}\cdot\mathbf{E}_{3,\parallel})\mathbf{E}_{1,\parallel}^*+2(\mathbf{E}_{1,\parallel}^*\cdot\mathbf{E}_{3,\parallel})\mathbf{E}_{2,\parallel} \nonumber \\
                +&2(\mathbf{E}_{1,\parallel}^*\cdot\mathbf{E}_{2,\parallel})\mathbf{E}_{3,\parallel}],
            \end{align}
            \label{eq:GrapheneNonlinearSurfaceCurrentDFWM}
        \end{subequations}
        respectively. Any other possible combination of the three involved frequencies induces a nonlinear current in a completely different band and is, thus, ignored. The nonlinear currents of Eqs.~\eqref{eq:GrapheneNonlinearSurfaceCurrentDFWM} result in the following resonance frequency shifts
        \begin{subequations}
            \begin{align}
                \Delta&\tilde\omega_{\mathrm{DFWM},1}(a_1,a_2,a_3) = -2\tilde\beta_{s,\mathrm{DFWM},1}\frac{a_1^*(t)a_2(t)a_3(t)}{a_1(t)} \nonumber \\
                &= -2\left(\frac{\omega_1}{c_0}\right)^4 \tilde\kappa_{s,\mathrm{DFWM},1}\frac{\IM{\sigma_3\rsup{max}}}{\varepsilon_0^2}\frac{a_1^*(t)a_2(t)a_3(t)}{a_1(t)}, \\
                \Delta&\tilde\omega_{\mathrm{DFWM},2}(a_1,a_2,a_3) = -\tilde\beta_{s,\mathrm{DFWM},2}\frac{a_1^2(t)a_3^*(t)}{a_2(t)} \nonumber \\
                &= -\left(\frac{\omega_2}{c_0}\right)^4 \tilde\kappa_{s,\mathrm{DFWM},2}\frac{\IM{\sigma_3\rsup{max}}}{\varepsilon_0^2}\frac{a_1^2(t)a_3^*(t)}{a_2(t)}, \\
                \Delta&\tilde\omega_{\mathrm{DFWM},3}(a_1,a_2,a_3) = -\tilde\beta_{s,\mathrm{DFWM},3}\frac{a_1^2(t)a_2^*(t)}{a_3(t)} \nonumber \\
                &= -\left(\frac{\omega_3}{c_0}\right)^4 \tilde\kappa_{s,\mathrm{DFWM},3}\frac{\IM{\sigma_3\rsup{max}}}{\varepsilon_0^2}\frac{a_1^2(t)a_2^*(t)}{a_3(t)},
            \end{align}
            \label{eq:DeltaOmegaGrapheneDFWM}
        \end{subequations}
        respectively, with the DFWM nonlinear feedback parameters given by
        \begin{widetext}
            \begin{subequations}
            \begin{align}
                \tilde\kappa_{s,\mathrm{DFWM},1} &= \left(\frac{c_0}{\omega_1}\right)^4\frac{\displaystyle\iint_{S_p} \IM{\sigma_3}[2(\mathbf{E}_{1,\parallel}^*\cdot\mathbf{E}_{2,\parallel}) (\mathbf{E}_{1,\parallel}^*\cdot\mathbf{E}_{3,\parallel}) + (\mathbf{E}_{1,\parallel}^*\cdot\mathbf{E}_{1,\parallel}^*) (\mathbf{E}_{2,\parallel}\cdot\mathbf{E}_{3,\parallel})] \mathrm{d}S}{\displaystyle\frac{16}{\varepsilon_0^2}W_{\mathrm{res},1}W_{\mathrm{res},2}^{1/2}W_{\mathrm{res},3}^{1/2}\IM{\sigma_3\rsup{max}}}, \\
                \tilde\kappa_{s,\mathrm{DFWM},2} &= \left(\frac{c_0}{\omega_2}\right)^4\frac{\displaystyle\iint_{S_p} \IM{\sigma_3}[2(\mathbf{E}_{1,\parallel}\cdot\mathbf{E}_{2,\parallel}^*) (\mathbf{E}_{1,\parallel}\cdot\mathbf{E}_{3,\parallel}^*) + (\mathbf{E}_{1,\parallel}\cdot\mathbf{E}_{1,\parallel}) (\mathbf{E}_{2,\parallel}^*\cdot\mathbf{E}_{3,\parallel}^*)] \mathrm{d}S}{\displaystyle\frac{16}{\varepsilon_0^2}W_{\mathrm{res},1}W_{\mathrm{res},2}^{1/2}W_{\mathrm{res},3}^{1/2}\IM{\sigma_3\rsup{max}}}, \\
                \tilde\kappa_{s,\mathrm{DFWM},3} &= \left(\frac{c_0}{\omega_3}\right)^4\frac{\displaystyle\iint_{S_p} \IM{\sigma_3}[2(\mathbf{E}_{1,\parallel}\cdot\mathbf{E}_{2,\parallel}^*) (\mathbf{E}_{1,\parallel}\cdot\mathbf{E}_{3,\parallel}^*) + (\mathbf{E}_{1,\parallel}\cdot\mathbf{E}_{1,\parallel}) (\mathbf{E}_{2,\parallel}^*\cdot\mathbf{E}_{3,\parallel}^*)] \mathrm{d}S}{\displaystyle\frac{16}{\varepsilon_0^2}W_{\mathrm{res},1}W_{\mathrm{res},2}^{1/2}W_{\mathrm{res},3}^{1/2}\IM{\sigma_3\rsup{max}}},
            \end{align}
            \label{eq:kappaGrapheneDFWM}
        \end{subequations}
        \end{widetext}
        As with THG, the nonlinear parameters $\tilde\beta\rsub{DFWM}$ are also connected through $\tilde\beta_{s,\mathrm{DFWM},1}=\tilde\beta_{s,\mathrm{DFWM},2}^*=\tilde\beta_{s,\mathrm{DFWM},3}^*$ and $\tilde\beta_{b,\mathrm{DFWM},1}/\omega_1=\tilde\beta_{b,\mathrm{DFWM},2}^*/\omega_2=\tilde\beta_{b,\mathrm{DFWM},3}^*/\omega_3$ for their bulk counterparts.

        Following the approach presented in this section, the influence of any other nonlinear effect emerging from either bulk or sheet materials (e.g., second harmonic generation or sum/difference-frequency generation in $\chi^{(2)}$ materials) can be similarly described in terms of a complex $\Delta\tilde\omega$, which can be readily incorporated in the CMT framework presented in Sec.~\ref{sec:InstantNonLinearCMT}.\cite{Christopoulos2018,Christopoulos2020JOSAB,Rodriguez2007,Hashemi2009,Ramirez2011,Zeng2014,Lin2014}
        
    \subsection{Non-instantaneous nonlinearities: Saturable absorption}\label{AppSub:DeltaOmegaVSaSA}

        In a quite similar manner we can use perturbation theory to introduce the effect of saturable absorption, i.e., the nonlinear saturation of material losses when it is illuminated with high intensity. There is, however, a key difference between SA and third-order effects that were introduced previously.  
        SA is \emph{not} a weak effect; due to SA, Ohmic loss are significantly suppressed and, theoretically, can even vanish for very high light intensities. Nevertheless, Ohmic loses, be it saturated or unsaturated, \emph{are} typically a perturbative effect for the resonant system, as discussed in Sec.~\ref{sec:LinearCMT}. Hence, assuming that in the original system losses are small, one can introduce them (and their nonlinear dependence on the light intensity) in the formulation by treating them collectively as a perturbative effect.

        Again, we will focus our interest on 2D materials, but the analogies are obvious and a similar framework can emerge for bulk materials.\cite{Ataloglou2018,Ma2014} In the most simple scenario, saturable absorption is described through an instantaneous saturation term in the real part of the surface conductivity, acquiring the form\cite{Ataloglou2018}
        \begin{equation}
            \sigma_1(\mathbf{E}_\parallel) = \frac{\RE{\sigma_0}}{1+|\mathbf{E}_\parallel|^2/E\rsub{sat}^2} + j\IM{\sigma_0}.
            \label{eq:SurfaceSA}
        \end{equation}
        Several simplifying notations have been introduced in Eq.~\eqref{eq:SurfaceSA}. First, instead of the conductivity tensor $\dbar{\sigma}$, its sole nonzero term $\sigma_1$ is used, a valid assumption for isotropic 2D materials as is, for example, a single graphene sheet in the absence of  magnetic bias.\cite{Alexander2018} The $\sigma_1$ parameter corresponds to the in-plane conductivity, with the out-of-plane components of $\dbar{\sigma}$ being zero. Secondly, the complex surface conductivity $\sigma_0$ represents the surface conductivity of the material for low intensities, i.e., $\sigma_1(\mathbf{E}_\parallel\rightarrow 0) = \sigma_0$. Thirdly, the 2D nature of the material is additionally captured by the involvement of the $\mathbf{E}_\parallel$ term in the denominator, a fact that will become clearer toward the end of the section. Finally, $E\rsub{sat}$ is the field saturation strength, which is connected with the more commonly used saturation intensity via $E\rsub{sat}^2 = 2\eta_0I\rsub{sat}$\cite{Chatzidimitriou2020} and represents the light intensity that is required to saturate Ohmic loss to 50\%.

        Using Eq.~\eqref{eq:SurfaceSA}, one can return to \eqref{eq:GeneralPertTheoryShortAppendix} and replace $\Delta\mathbf{J}\rsub{NL}$ with $\sigma_1(\mathbf{E}_\parallel)\mathbf{E}_\parallel \approx \sigma_1(\mathbf{E}_{0,\parallel})\mathbf{E}_{0,\parallel}$ to reach
        \begin{align}
            \Delta\tilde\omega\rsub{SA} = &-\frac{\displaystyle\iint_{S_p} \IM{\sigma_0}|\mathbf{E}_{0,\parallel}|^2\mathrm{d}S}{4W\rsub{res}} \nonumber \\ 
            &+ j\frac{\displaystyle\iint_{S_p} \frac{\RE{\sigma_0}}{1+|\mathbf{E}_{0,\parallel}|^2/E\rsub{sat}^2}|\mathbf{E}_{0,\parallel}|^2\mathrm{d}S}{4W\rsub{res}}.
            \label{eq:DeltaOmegaGrapheneSA}
        \end{align}
        The first term on the right-hand side of Eq.~\eqref{eq:DeltaOmegaGrapheneSA} describes a power independent shift on the resonance frequency, induced by the dielectric properties of the 2D material. The second term, on the other hand, is power-dependent and it is the one that represents the SA process. For the rest of the analysis, we will focus our interest only on this term; in the context of CMT, this term is equivalent to an amplitude dependent decay rate $\gamma\rsub{SA}(a)$. Due to its more complex dependence on the electric field, we will use the reference field approach of Appendix~\ref{AppSub:DeltaOmegaVSaXPMTHGFWM} to transform Eq.~\eqref{eq:DeltaOmegaGrapheneSA} into a CMT-appropriate equation.
        Through this normalization, the imaginary part of Eq.~\eqref{eq:DeltaOmegaGrapheneSA} is simplified to
        \begin{equation}
            \gamma\rsub{SA}(a) = \frac{1}{4}\iint_{S_p} \frac{\RE{\sigma_0}}{1+|a(t)|^2|\mathbf{E}\rsub{ref,\parallel}|^2/E\rsub{sat}^2}|\mathbf{E}\rsub{ref,\parallel}|^2\mathrm{d}S,
            \label{eq:DeltaOmegaGrapheneSANormalized}
        \end{equation}
        and it is measured in $\mathrm{1/s}$, as expected.

        An obvious complication of Eq.~\eqref{eq:DeltaOmegaGrapheneSANormalized} compared to e.g. $\gamma\rsub{SPM}$ of the Kerr effect is that the nonlinear parameter is power dependent. However, this is not a problem, due to the introduced normalization which renders $\mathbf{E}\rsub{ref,\parallel}(\mathbf{r})$ independent of $a(t)$. Any numerical integration scheme can efficiently handle the integral in Eq.~\eqref{eq:DeltaOmegaGrapheneSANormalized} and the resulting CMT ODE [see, e.g., Eqs.~\eqref{eq:GammaSA}~and~\eqref{eq:CMTSAfull}] can be solved with standard numerical techniques (e.g. an appropriate Runge-Kutta method).

        In the simple case that $\mathbf{E}\rsub{ref,\parallel}(\mathbf{r})$ is spatially uniform, Eq.~\eqref{eq:DeltaOmegaGrapheneSANormalized} can acquire the more intuitive form\cite{Ataloglou2018}
        \begin{equation}
            \gamma\rsub{SA}(a) = \frac{\gamma\rsub{SA,0}}{1+|a(t)|^2/W\rsub{sat}},
            \label{eq:DeltaOmegaGrapheneSASimplified}
        \end{equation}
        where $\gamma\rsub{SA,0}$ is the decay rate induced by the 2D material for low intensities (i.e., in the absences of SA) and $W\rsub{sat} = E\rsub{sat}^2/|\mathbf{E}\rsub{ref,\parallel}|^2$ (measured in $\mathrm{J}$) is the  equivalent saturation energy (the value of the total stored energy for which the SA decay rate is halved). 
        We should note here that the saturation model of Eq.~\eqref{eq:SurfaceSA} is not unique. For example, a more complex model for graphene SA has been proposed in the literature,\cite{Sahoo2021} and has been introduced in the CMT formalism following similar steps.\cite{Christopoulos2020JAP}

        We now proceed to the study of a non-instantaneous SA mechanism. In physical terms, saturable absorption is a consequence of the Pauli blocking principle. When the conduction band of a 2-level medium is relatively full with electrons (excited, e.g., by linear absorption), new electrons are less likely to be transferred from the valence to 
        the conduction band.
        Macroscopically, this implies a carrier-dependent conductivity of the form\cite{Chatzidimitriou2020}
        \begin{equation}
            \sigma_1(N_c) = \RE{\sigma_0}\left(1-\frac{N_c}{2N\rsub{sat}}\right) + j\IM{\sigma_0},
            \label{eq:SurfaceSAwithCarriers}
        \end{equation}
        where $N_c = N_c\OF{r}{t}$ is the surface carrier density, measured in $\mathrm{m}^{-2}$, and $N\rsub{sat}$ is the carrier saturation density, which is related with the saturation intensity $I\rsub{sat}$.\cite{Chatzidimitriou2020} 
        In the considered 2-level medium, the surface carrier density over time is governed by a PDE of the form\cite{Chatzidimitriou2020,Nousios2022}
        \begin{equation}
            \partialFRAC{N_c\OF{r}{t}}{t} = -\frac{N_c\OF{r}{t}}{\tau_c} + R_s\OF{r}{t} + D\nabla^2N_c\OF{r}{t}.
            \label{eq:CarrierRateEquationSA}
        \end{equation}
        The first term on the right-hand side of Eq.~\eqref{eq:CarrierRateEquationSA} describes the carrier recombination process with a carrier relaxation lifetime $\tau_c$. The second term is related with the carrier excitation process and is written in a general form using a source term $R_s$ measured in $\mathrm{1/(m^2s)}$. In the case of linear absorption, $R_s\OF{r}{t} = (|\mathrm{d}P\rsub{lin}/\mathrm{d}S|)/(\hbar\omega) = (1/2)\RE{\sigma_1(N_c)}|\mathbf{E}\rsub{ref,\parallel}(\mathbf{r})|^2|a(t)|^2/(\hbar\omega)$ measuring the absorbed power per surface unit, divided by the energy of a single photon, $\hbar\omega$. Finally, the third term represents spatial diffusion of carriers ($D$ is the diffusion coefficient, measured in $\mathrm{m^2/s}$); although its contribution is typically ignored or captured by an effective relaxation lifetime,\cite{Moille2016} it can be incorporated in the CMT framework and has been shown to significantly affect the response in compact nanophotonic structures.\cite{Nousios2022} Equation~\eqref{eq:CarrierRateEquationSA} captures rigorously the spatial dependence of the carrier distribution and, given the knowledge of $\mathbf{E}\rsub{ref,\parallel}(\mathbf{r})$, it can be solved concurrently with the CMT amplitude equation(s). The final system of ODEs are connected through Eq.~\eqref{eq:SurfaceSAwithCarriers} which is transformed into CMT terms using perturbation theory arguments, i.e., via the nonlinear parameter
        \begin{equation}
            \gamma\rsub{SA}(N_c) = \frac{1}{4}\iint_{S_p} \RE{\sigma_0}\left[1-\frac{N_c}{2N\rsub{sat}}\right]|\mathbf{E}\rsub{ref,\parallel}|^2\mathrm{d}S.
            \label{eq:DeltaOmegaGrapheneSANormalizedCarriers}
        \end{equation}
        The final system of ODEs consists a coupled system when the involved time scales (i.e., $\tau_\ell$ and $\tau_c$) are comparable. In the absence of diffusion, Eqs.~\eqref{eq:SurfaceSA}~and~\eqref{eq:DeltaOmegaGrapheneSANormalized} can be retrieved by ignoring the time derivative in Eq.~\eqref{eq:CarrierRateEquationSA} (instantaneous response, i.e. $\tau_c \ll \tau_\ell$) and additionally using the equality $I\rsub{sat} = 2N\rsub{sat}(\hbar\omega)/(\eta_0\RE{\sigma_0}\tau_c)$, which connects the carrier saturation density $N\rsub{sat}$ with the saturation intensity $I\rsub{sat}$.
        
        As a final remark, we must note that loss saturation can occur in multi-channel schemes, as well. For instance, a strong pump wave can saturate the resistive loss that a weak probe (at a different wavelength) experiences, allowing for all-optical control. In perturbation theory terms this cross-saturable absorption effect can be incorporated by assuming that only the pump wave (being much stronger than the probe) contributes to the carrier excitation, i.e., $R_s$ in Eq.~\eqref{eq:CarrierRateEquationSA} solely depends on the pump wave distribution. On the contrary, suppressed losses are perceived by both pump (self-SA) and probe waves (cross-SA), through two distinct $\gamma\rsub{SA}$ parameters, each corresponding to the (carrier-dependent) decay rate for each wave.\cite{Christopoulos2020JAP,Nousios2022} 

    \subsection{Non-instantaneous nonlinearities: Free-carrier effects and thermo-optic effect}\label{AppSub:DeltaOmegaVSaFCEsTOE}

        The presence of \emph{free carriers} in the conduction band of a material might impact the response of a cavity by inducing additional effects different from SA. For example, in silicon, an indirect bandgap semiconductor, the presence of free carriers induces absorption and also affects its refractive index. These phenomena are termed \emph{free-carrier effects}, and more specifically \emph{free-carrier absorption} and \emph{free-carrier dispersion}, respectively. In this section, we will focus on FCEs and other related phenomena in Si. Initially, it was assumed that the free carrier density was linearly affecting the properties of Si (refractive index and additional losses), i.e., $\Delta n\rsub{FCD} = -\sigma_n N_c$ and $\Delta\alpha\rsub{FCA} = \sigma_a N_c$, with $\sigma_n$ and $\sigma_a$ being characteristic constants of the material. Although measurements by Soref and Bennett have showed that neither of these two relations is truly linear,\cite{Soref1987,Nedeljkovic2011} here for simplicity we will use the linear version.  The  model with the actual dependence can also be incorporated into the CMT formulation using perturbation theory.\cite{Tsilipakos2016,Chen2012,Sethi2014} FCEs can be described by a modification in the linear susceptibility  
        \begin{equation}
            \Delta\chi\rsub{FCE}^{(1)} = 2n_0\Delta n\rsub{FCD} - j\frac{c_0 n_0}{\omega} \Delta\alpha\rsub{FCA},
            \label{eq:DeltaChiFCEs}
        \end{equation}
        and can be readily incorporated in the CMT framework by a perturbative polarization of the form $\Delta\mathbf{P} = \varepsilon_0\Delta\chi\rsub{FCE}^{(1)}\mathbf{E}$.

        Before proceeding with the introduction of $\Delta\mathbf{P}$ in Eq.~\eqref{eq:GeneralPertTheoryShortAppendix} 
        we must determine the equation that governs the free carrier density in Si. The general form of Eq.~\eqref{eq:CarrierRateEquationSA} is valid for FCEs as well, with the difference that the carrier density now is measured in $\mathrm{m}^{-3}$ since it corresponds to a bulk material. As with graphene, the spatial diffusion of carriers can be rigorously included,\cite{Moille2016} but is typically captured by an effective recombination lifetime, which significantly simplifies the framework. For this tutorial, we will assume that the equation governing carrier evolution in Si reads
        \begin{equation}
            \partialFRAC{N_c\OF{r}{t}}{t} = -\frac{N_c\OF{r}{t}}{\tau_c} + R_s\OF{r}{t}.
            \label{eq:CarrierRateEquationSi}
        \end{equation}

        Since silicon is transparent in the communication wavelength of 1550~nm, carriers are only excited in the conduction band due to TPA (nonlinear absorption). Thus, the source term in this case is $R_s = (|\mathrm{d}P\rsub{TPA}/\mathrm{d}V|)/(2\hbar\omega)$. Alternatively, carriers may be injected through the external biasing of a $pn$-junction,\cite{Sinatkas2021} but we will not consider such a possibility here. Returning to Eqs.~\eqref{eq:SiliconNonlinearPolarization}~and~\eqref{eq:chi3ton2betaTPA}, we arrive at
        \begin{align}
            R_s = &\left| \fullFRAC{P\rsub{TPA}}{V} \right| \frac{1}{2\hbar\omega} = \left|\frac{1}{2} \RE{\mathbf{E}^*\cdot j\omega \mathbf{P}\rsub{Si}^{(3)}}\right|\frac{1}{2\hbar\omega} \nonumber \\
            = &\frac{1}{12}\frac{\varepsilon_0^2c_0^2n_0^2\beta\rsub{TPA}}{2\hbar\omega}\left(2|\mathbf{E}|^4 + |\mathbf{E}\cdot\mathbf{E}|^2\right),
            \label{eq:TPAlosses}
        \end{align}
       for the case of FCEs in Si.

        The next step is to efficiently handle the spatial dependence of free carriers and this is typically done by introducing a spatially-averaged carrier density $\bar N_c$ through\cite{Tsilipakos2016,Daniel2010}
        \begin{equation}
            \bar N_c (t) = \frac{\displaystyle\iiint_{V_p}N_c\OF{r}{t}|\mathbf{E}\rsub{ref}(\mathbf{r})|^2\mathrm{d}V}{\displaystyle\displaystyle\iiint_{V_p}|\mathbf{E}\rsub{ref}(\mathbf{r})|^2\mathrm{d}V}.
            \label{eq:NcAverage}
        \end{equation}
        The normalized field $\mathbf{E}\rsub{ref}$ is used for the averaging although using $\mathbf{E}_0 \approx \mathbf{E}$ would have resulted in the same quantity due to the integral in the denominator. Furthermore, note that the ``perturbation'' volume $V_p$ is used for the integrations, restricting the averaging only in silicon. The introduction of $\bar N_c$ modifies Eq.~\eqref{eq:CarrierRateEquationSi}, which now reads
        \begin{equation}
            \fullFRAC{\bar N_c(t)}{t} = -\frac{\bar N_c(t)}{\tau_c} + \bar R_s(t),
            \label{eq:CarrierRateEquationAveragedSi}
        \end{equation}
        with the spatially-averaged source term being
        \begin{align}
            \bar R_s(t) &= \frac{1}{12}\frac{\varepsilon_0^2c_0^2}{2\hbar\omega} \frac{\displaystyle\iiint_{V_p} n_0^2\beta\rsub{TPA}\left(2|\mathbf{E}|^4 + |\mathbf{E}\cdot\mathbf{E}|^2\right)|\mathbf{E}\rsub{ref}|^2 \mathrm{d}V}{\displaystyle\displaystyle\iiint_{V_p}|\mathbf{E}\rsub{ref}|^2\mathrm{d}V} \nonumber \\ 
            &= \frac{1}{48}\frac{\varepsilon_0^2c_0^2}{2\xi\hbar\omega} \displaystyle\iiint_{V_p} n_0^2\beta\rsub{TPA}\left(2|\mathbf{E}|^4 + |\mathbf{E}\cdot\mathbf{E}|^2\right)|\mathbf{E}\rsub{ref}|^2 \mathrm{d}V.
            \label{eq:RsFCEs}
        \end{align}
        In Eq.~\eqref{eq:RsFCEs}, we have introduced the confinement parameter 
        \begin{equation}
            \xi = \frac{1}{4} \iiint_{V_p} |\mathbf{E}\rsub{ref}|^2 \mathrm{d}V
            \label{eq:xiDefinition}
        \end{equation}
        for brevity and uniformity with results that will follow.  Equation~\eqref{eq:CarrierRateEquationAveragedSi} comprises both normalized and full fields. However, this can be easily handled by applying first-order perturbation theory and introducing the transformation $\mathbf{E}\OF{r}{t} \approx \mathbf{E}_0\OF{r}{t}=a(t)\mathbf{E}\rsub{ref}(\mathbf{r})$. One can then reach the more useful form
        \begin{equation}
            \fullFRAC{\bar N_c(t)}{t} = -\frac{\bar N_c(t)}{\tau_c} + \gamma_N|a(t)|^4,
            \label{eq:CarrierRateEquationAveragedWithAlphaSi}
        \end{equation}
        where the nonlinear parameter $\gamma_N$ (measured in $\mathrm{1/J^2sm^3}$), independent of the stored energy of the cavity, is defined as
        \begin{equation}
            \gamma_N = 2\left(\frac{\omega_0}{c_0}\right)^6 \frac{c_0^2}{\hbar \omega_0} \kappa_N \beta\rsub{TPA}\rsup{max}.
            \label{eq:GammaNFCEs}
        \end{equation}
        An appropriate nonlinear feedback parameter is also defined here, expressed in terms of the reference field $\mathbf{E}\rsub{ref}(\mathbf{r})$ instead of the full field $\mathbf{E}_0(\mathbf{r})$ that was used in the Sec.~\ref{AppSub:DeltaOmegaVSaKerrTPA}. The two representations are equivalent.
        \begin{align}
            \kappa_N = &\left(\frac{c_0}{\omega_0}\right)^6\times \nonumber \\
            &\frac{\displaystyle\frac{1}{3}\iiint_{V_p} n_0^2\beta\rsub{TPA}(2|\mathbf{E}\rsub{ref}|^4 + |\mathbf{E}\rsub{ref}\cdot\mathbf{E}\rsub{ref}|^2)|\mathbf{E}\rsub{ref}|^2 \mathrm{d}V}{\displaystyle\frac{64}{\varepsilon_0^2}\xi\beta\rsub{TPA}\rsup{max}}.
            \label{eq:kappaN}
        \end{align}
        Note that the source term in Eq.~\eqref{eq:CarrierRateEquationAveragedWithAlphaSi} is proportional to the square of the stored energy in the cavity. This is a direct consequence of the physical mechanism that excites the carriers (TPA process), which leads to power absorption $\propto|\mathbf{E}|^4$.

        In terms of the CMT ODE, the influence of free carries should be incorporated via a nonlinear resonance frequency shift $\Delta\tilde\omega\rsub{FCE}$. Since free carriers are the cause of FCEs, $\Delta\tilde\omega\rsub{FCE}$ should be a function of $\bar N_c$, i.e., $\Delta\tilde\omega\rsub{FCE} = \Delta\tilde\omega\rsub{FCE}(\bar N_c)$. We introduce Eq.~\eqref{eq:DeltaChiFCEs}  into Eq.~\eqref{eq:GeneralPertTheoryShortAppendix} and reach
        \begin{align}
            \Delta\tilde\omega\rsub{FCE}(\bar N_c) &= \tilde\gamma_{N,\mathrm{FCE}} \bar N_c(t) = (\gamma_{N,\mathrm{FCD}}+j\gamma_{N,\mathrm{FCA}})\bar N_c(t) \nonumber \\ 
            &= \left( 2\omega_0\varepsilon_0n_0\sigma_n\xi + j\varepsilon_0c_0 n_0\sigma_a\xi \right) \bar N_c(t),
            \label{eq:DeltaOmegaDynamicFCEs}
        \end{align}
        with $\tilde\gamma_{N,\mathrm{FCE}}$ here measured in $\mathrm{m^3/s}$.
        Equation~\eqref{eq:DeltaOmegaDynamicFCEs} together with Eq.~\eqref{eq:CarrierRateEquationAveragedWithAlphaSi} allows to introduce the dynamic response of FCEs in the CMT framework.
        
        The so far developed framework is accurate regardless of the actual carrier recombination lifetime $\tau_c$ and introduces an additional ODE that describes free carriers evolution over time. However, when $\tau_c \ll \tau_\ell$ one can simplify the framework by eliminating the carrier rate equation. Effectively, this means that FCEs act quite faster than the timescate at which the cavity responds. Thus, one can set $\mathrm{d}\bar N_c/\mathrm{d}t \rightarrow 0$, meaning that $\bar N_c$ instantaneously follows the evolution of the cavity energy, i.e., $\bar N_c = \tau_c\gamma_N |a|^4$. Then, Eq.~\eqref{eq:DeltaOmegaDynamicFCEs} becomes
        \begin{widetext}
            \begin{equation}
                \Delta\tilde\omega\rsub{FCE}(a) = \tilde\gamma\rsub{FCE} |a(t)|^4 = (\gamma\rsub{FCD}+j\gamma\rsub{FCA})|a(t)|^4 = \left(\frac{\omega_0}{c_0}\right)^6 \left( 4\kappa\rsub{FCE}\frac{\omega_0\sigma_nc_0^2\tau_c}{\hbar\omega_0}\beta\rsub{TPA}\rsup{max} + j2\kappa\rsub{FCE}\frac{\sigma_ac_0^2\tau_c}{\hbar\omega_0}\beta\rsub{TPA}\rsup{max} \right) |a(t)|^4,
                \label{eq:DeltaOmegaCWFCEs}
            \end{equation}
        \end{widetext}
        where $\kappa\rsub{FCE}$ is defined through\cite{Tsilipakos2016}
        \begin{align}
            \kappa\rsub{FCE} = &\left(\frac{c_0}{\omega_0}\right)^6\times \nonumber \\ 
            &\frac{\displaystyle\frac{1}{3}\iiint_{V_p} n_0^3\beta\rsub{TPA}(2|\mathbf{E}\rsub{ref}|^4 + |\mathbf{E}\rsub{ref}\cdot\mathbf{E}\rsub{ref}|^2)|\mathbf{E}\rsub{ref}|^2 \mathrm{d}V}{\displaystyle\frac{64}{\varepsilon_0^3}\beta\rsub{TPA}\rsup{max}},
            \label{eq:kappaFCE}
        \end{align}
        and $\tilde\gamma_{\mathrm{FCE}}$ measured in $\mathrm{1/J^2s}$.
        We have opted here to express $\kappa\rsub{FCE}$ in terms of $\mathbf{E}\rsub{ref}(\mathbf{r})$, albeit in the literature it is typically given in terms of $\mathbf{E}_0(\mathbf{r})$.\cite{Tsilipakos2016,Zhang2013a,Sethi2014} The two representations are ultimately equivalent but we use the former for uniformity with the rest of the tutorial. Finally, it should be noted that the shift introduced by FCD in the resonance frequency is of different sign with that of the Kerr effect [cf. the real parts of respective $\tilde\gamma$'s].\cite{Tsilipakos2016} This is a well known behavior in Si where the Kerr effect red-shifts the resonance frequency (self-focusing material with $n_2>0$) while FCD induces a blue-shift. FCD typically dominates unless appropriate engineering approaches are followed.\cite{Turner-Foster2010,Forst2007}

        In the same spirit, other linear or nonlinear phenomena that are governed by additional rate equations can be introduced in the CMT framework. We will briefly discuss here the case of the \emph{thermo-optic effect} (TOE), i.e., how the refractive index of a material changes with temperature, and in the next section we will discuss gain. 
        For most materials and for reasonable temperature increases up to a few hundred degrees Kelvin, the TOE-induced change in the refractive index is linear. Furthermore, it only affects the real part of the refractive index, i.e., it does not induce additional losses.\cite{Chen2012,Zhang2013a,Sethi2014} In mathematical terms
        \begin{equation}
            \Delta\chi\rsub{TOE}^{(1)} = 2n_0\Delta n\rsub{TOE} = 2n_0\kappa_\theta\Delta T,
            \label{eq:DeltaChiTOE}
        \end{equation}
        where $\Delta T$ is the temperature increase from room temperature, 
        and $\kappa_\theta = \mathrm{d}n/\mathrm{d}T$ is the \emph{thermo-optic} coefficient of the material, measured in $1/\mathrm{K}$; obviously, Eq.~\eqref{eq:DeltaChiTOE} stands as $\Delta n\rsub{TOE} \ll n_0$.

        The second ingredient that is necessary for describing TOE is an equation that governs the temporal and spatial evolution of temperature; this is the standard heat diffusion equation 
        \begin{equation}
            C_p\rho\partialFRAC{T\OF{r}{t}}{t} + \nabla\cdot[-k\nabla T\OF{r}{t}] =  Q\OF{r}{t},
            \label{eq:HeatTransferGeneral}
        \end{equation}
        where $C_p$ [in $\mathrm{J/(kgK)}$] is the thermal capacity, $\rho$ (in $\mathrm{kg/m^3}$) the mass density, and $k$ [in $\mathrm{W/(mK)}$] the thermal conductivity of the material. Finally, $Q$ (in $\mathrm{W/m^3}$) is the heat source and can describe any heating mechanism. 
         Equation~\eqref{eq:HeatTransferGeneral} is typically simplified before introduced in the CMT framework, by representing heat diffusion through an effective thermal lifetime $\tau_\theta$. Also assuming a constant $T_0$ as the initial condition for the system, one can write
        \begin{equation}
            \partialFRAC{\Delta T\OF{r}{t}}{t} = - \frac{\Delta T\OF{r}{t}}{\tau_\theta} + \frac{1}{C_p\rho}Q\OF{r}{t}.
            \label{eq:HeatTransferSimple}
        \end{equation}
        Now, Eq.~\eqref{eq:HeatTransferSimple} has the exact same form as Eq.~\eqref{eq:CarrierRateEquationSi} [with $R_p = (1/C_p\rho)Q$] and, thus, can be handled similarly. Although Eq.~\eqref{eq:HeatTransferSimple} has been proven sufficient for most practical applications, more complex alternatives have been proposed in the literature.\cite{Borghi2021}
        
        In the interest of space, we will not present the lengthy equations that define the respective nonlinear parameters. 
        We will discuss, however, the source term $Q$. The cause of heating can be internal or external, with the former being the more interesting one since it is typically a consequence of losses. Thus far, we have presented three distinct loss mechanisms, namely Ohmic loss (\emph{Joule heating} induced by linear absorption), two-photon absorption, and free-carrier absorption. Each one acts as an independent heat source in Eq.~\eqref{eq:HeatTransferSimple} and thus one can write $Q = Q\rsub{J} + Q\rsub{TPA} + Q\rsub{FCA}$. Since $Q$ represents the power that is absorbed by the material and transformed into heat, it is straightforward to write
        \begin{subequations}
            \begin{align}
                Q\rsub{J} &= \left|\fullFRAC{P\rsub{J}}{V}\right| = \frac{1}{2}\RE{\mathbf{E}^*\cdot \sigma\mathbf{E}}, \\
                Q\rsub{TPA} &= \left|\fullFRAC{P\rsub{TPA}}{V}\right| = \frac{1}{2}\RE{\mathbf{E}^*\cdot j\omega\mathbf{P}^{(3)}}, \\
                Q\rsub{FCA} &= \left|\fullFRAC{P\rsub{FCA}}{V}\right| = \frac{1}{2}\RE{\mathbf{E}^*\cdot j\omega\mathbf{P}^{(1)}\rsub{FCA}}.
            \end{align}
            \label{eq:HeatQdefinitions}
        \end{subequations}
        The polarizations in Eq.~\eqref{eq:HeatQdefinitions} express the physical effect that induces heating, i.e., $\mathbf{P}^{(3)}$ is directly taken from Eq.~\eqref{eq:SiliconNonlinearPolarization} and $\mathbf{P}^{(1)}\rsub{FCA}$ equals $\varepsilon_0\Delta\chi\rsub{FCE}^{(1)}\mathbf{E}$. By introducing the polarization expression into Eqs.~\eqref{eq:HeatQdefinitions} and defining a spatially averaged temperature difference [through an expression equivalent to Eq.~\eqref{eq:NcAverage}], one can lift the spatial dependence of $\Delta T$ and reach
        \begin{align}
            \fullFRAC{\Delta\bar T(t)}{t} = -\frac{\Delta\bar T(t)}{\tau_\theta} &+ \gamma_{T,\mathrm{J}}|a(t)|^2 + \gamma_{T,\mathrm{TPA}}|a(t)|^4 \nonumber \\
            &+ \gamma_{T,\mathrm{FCA}}\bar N_c(t) |a(t)|^2,
            \label{eq:TemperatureRateEquationAveragedWithAlpha}
        \end{align}
        which is complemented by
        \begin{equation}
            \Delta\omega\rsub{TOE}(\Delta\bar T) = -\gamma_{T,\mathrm{TOE}} \Delta\bar T(t).
            \label{eq:DeltaOmegaDynamicTOE}
        \end{equation}
        describing  the TOE-induced resonance frequency shift and, thus, connecting the ODE that governs the temperature change with the CMT equation. 
        $\gamma_{T,\mathrm{TOE}}$ here is measured in $\mathrm{1/Ks}$ and $\gamma_{T,\mathrm{J/TPA/FCA}}$ in $\mathrm{K/Js}$, $\mathrm{K/J^2s}$, and $\mathrm{Km^3/Js}$, respectively.
        Note the minus sign in Eq.~\eqref{eq:DeltaOmegaDynamicTOE}, indicating a red-shift of the resonance frequency when $\gamma\rsub{TOE}>0$ or, equivalently, when $\kappa_\theta>0$.

        Finally, Eq.~\eqref{eq:TemperatureRateEquationAveragedWithAlpha} can be omitted if one assumes that the temperature is only slowly varying.
        Then, by setting $\mathrm{d}\Delta\bar T/\mathrm{d}t \rightarrow 0$, one gets $\Delta\bar T = \tau_\theta\gamma_{T,\mathrm{J}}|a|^2 + \tau_\theta\gamma_{T,\mathrm{TPA}}|a|^4 + \tau_\theta\gamma_{T,\mathrm{FCA}}\bar N_c |a|^2$, and we ultimately reach
        \begin{align}
            \Delta\omega\rsub{TOE}(a) = &-\gamma\rsub{TOE,J} |a(t)|^2 - \gamma\rsub{TOE,TPA} |a(t)|^4 \nonumber \\ 
            &- \gamma\rsub{TOE,FCA} \bar N_c(t) |a(t)|^2.
            \label{eq:DeltaOmegaCWTOE}
        \end{align}
        Note that each heating mechanism has a different contribution with respect to the stored energy in the cavity, i.e., FCA-induced TOE is expected to dominate for higher intensities while Ohmic loss-induced TOE is the main heating mechanism at low intensities.    

    \subsection{Non-instantaneous gain}\label{AppSub:DeltaOmegaVSaGain}

        We finally present the treatment of gain. To rigorously capture the physics of carrier interactions that induce gain, a slightly different approach on evaluating perturbation theory is in order.\cite{Chua2011,Nousios2023} This is because carriers do not affect the properties of the lasing medium in a straightforward way as was the case with FCEs or SA; rather, in the most general case they induce a time-dependent polarization field. For this section, we will not use Eq.~\eqref{eq:GeneralPertTheoryShortAppendix} but derive a similar one starting from the Maxwell's curl equations in the time domain
        \begin{subequations}
            \begin{align}
                \nabla\times\bmcal{E}\OF{r}{t} &= -\mu_0\dbar{\mu}_r\frac{\partial\bmcal{H}\OF{r}{t}}{\partial t}, \label{eq:MaxTimeE} \\
                \nabla\times\bmcal{H}\OF{r}{t} &= \varepsilon_0\dbar{\varepsilon}_r\frac{\partial\bmcal{E}\OF{r}{t}}{\partial t} + \frac{\partial\bmcal{P}\OF{r}{t}}{\partial t}, \label{eq:MaxTimeH}
            \end{align}
            \label{eq:MaxTime}
        \end{subequations}
        where $\bmcal{P}\OF{r}{t}$ is the polarization field induced by the emission process. The dielectric properties of the involved materials are represented by $\dbar{\varepsilon}_r$ and their dispersion is ignored for compactness of the formulation. Equations~\eqref{eq:MaxTime} describe the \emph{perturbed} version of the problem. To bring the equations in a more useful form, it is assumed that the field consists of a slowly varying envelope that oscillates at an arbitrary reference frequency $\omega\rsub{ref}$ in the optical spectrum meaning that, e.g., the electric field can be written as $\bmcal{E}\OF{r}{t} = \mathrm{Re}\{\mathbf{\breve E}\OF{r}{t}\exp\{j\omega\rsub{ref} t\}\}$ and the symbol $\mathbf{\breve E}\OF{r}{t}$ is used for this envelop quantity in the time domain. Equations~\eqref{eq:MaxTime} are then transformed into
        \begin{subequations}
            \begin{align}
                \nabla\times\mathbf{\breve E} = &-\mu_0\dbar{\mu}_r\left(\frac{\partial\mathbf{\breve H}}{\partial t}+j\omega\rsub{ref}\mathbf{\breve H}\right), \label{eq:MaxSVEAE} \\
                \nabla\times\mathbf{\breve H} = &\varepsilon_0\dbar{\varepsilon}_r\left(\frac{\partial\mathbf{\breve E}}{\partial t}+j\omega\rsub{ref}\mathbf{\breve E}\right) + \left(\frac{\partial\mathbf{\breve P}}{\partial t}+j\omega\rsub{ref}\mathbf{\breve P}\right), \label{eq:MaxSVEAH}
            \end{align}
            \label{eq:MaxSVEA}
        \end{subequations}
        where we have included only terms oscillating at the positive frequency $\omega\rsub{ref}$. We use the arbitrary $\omega\rsub{ref}$ instead of the ``hot'' cavity resonance frequency $\omega$, which coincides with the lasing frequency $\omega_L$ but is actually unknown and cannot emerge from linear full-wave simulations.\cite{Nousios2023} An expression to approximately calculate $\omega_L$ is given in Sec.~\ref{sec:Gain} [Eq.~\eqref{eq:LasingFrequencyGain}] and can be used as $\omega\rsub{ref}$ when evaluating the framework. 
        
        A Fourier transform around the baseband frequency $\zeta$ is then applied in Eqs.~\eqref{eq:MaxSVEA} so that, e.g., the electric field's envelope in the frequency domain is $\mathbf{E}\OF{r}{\zeta} = \int \mathbf{\breve E}\OF{r}{t} \exp\{-j\zeta t\} \mathrm{d}{t}$. Equations~\eqref{eq:MaxSVEA} then become
        \begin{subequations}
            \begin{align}
                \nabla\times\mathbf{E} &= -j\mu_0\dbar{\mu}_r(\omega\rsub{ref}+\zeta)\mathbf{H}, \label{eq:MaxFreqE} \\
                \nabla\times\mathbf{H} &= j\varepsilon_0\dbar{\varepsilon}_r(\omega\rsub{ref}+\zeta)\mathbf{E} + j(\omega\rsub{ref}+\zeta)\mathbf{P}. \label{eq:MaxFreqH}
            \end{align}
            \label{eq:MaxFreq}
        \end{subequations}
         Equations~\eqref{eq:MaxFreq} are complemented by the following set describing the \emph{unperturbed} system,
        \begin{subequations}
            \begin{align}
                \nabla\times\mathbf{E}_0 &= -j\mu_0\dbar{\mu}_r(\omega_0+\zeta)\mathbf{H}_0, \label{eq:MaxFreq0E} \\
                \nabla\times\mathbf{H}_0 &= j\varepsilon_0\dbar{\varepsilon}_r(\omega_0+\zeta)\mathbf{E}_0. \label{eq:MaxFreq0H}
            \end{align}
            \label{eq:MaxFreq0}
        \end{subequations}
        obtained by simply ignoring the gain-induced polarization. Note that for Eqs.~\eqref{eq:MaxFreq0}, we make use of the ``cold'' resonance frequency $\omega_0$, since the on-resonance field will oscillate at that frequency; $\omega_0$ is calculated through modal simulations or can emerge after appropriate measurements (see Sec.~\ref{sec:CMTParamsCalculations}).

        To proceed, it is necessary to disentangle the spectral and spatial dependencies of the involved fields using the concept of reference fields introduced in Appendix~\ref{AppSub:DeltaOmegaVSaXPMTHGFWM}, so that the polarization is written as $\mathbf{P}\OF{r}{\zeta} = P(\zeta)\mathbf{E}\rsub{ref}(\mathbf{r})$, the electric field as $\mathbf{E}\OF{r}{\zeta} = A(\zeta)\mathbf{E}\rsub{ref}(\mathbf{r})$, and the magnetic field as $\mathbf{H}\OF{r}{\zeta} = A(\zeta)\mathbf{H}\rsub{ref}(\mathbf{r})$. The only conjecture here is that polarization follows the spatial distribution of the electric field, which is a rather rational assumption as it will become evident below. It is then straightforward to calculate the divergence of the vector field $\mathbf{F}_c$ (defined in Appendix.~\ref{AppSub:MaterialPertTheory}) as
        \begin{align}
            \nabla\cdot\mathbf{F}_c = &-j(\omega\rsub{ref} - \omega_0)A A_0^* \left[\mu_0\dbar{\mu}_r\mathbf{H}\rsub{ref}\cdot\mathbf{H}\rsub{ref}^* + \varepsilon_0\dbar{\varepsilon}_r\mathbf{E}\rsub{ref}\cdot\mathbf{E}\rsub{ref}^*\right] \nonumber \\
            &-j(\omega\rsub{ref} + \zeta)PA_0^*|\mathbf{E}\rsub{ref}|^2.
            \label{eq:divFcGain}
        \end{align}
        Assuming low light leakage, one can integrate over a volume $V$ enclosing the laser cavity, use Eq.~\eqref{eq:divFcGauss}, the definition of $\xi$ through Eq.~\eqref{eq:xiDefinition}, and the fact that $\Delta\tilde\omega = \omega\rsub{ref} - \omega_0$ to reach
        \begin{equation}
            \Delta\tilde\omega A(\zeta)= -\xi(\omega\rsub{ref} + \zeta) P(\zeta).
            \label{eq:DeltaOmegaGainFreq}
        \end{equation}
        Equation~\eqref{eq:DeltaOmegaGainFreq} is the final product of the perturbation analysis but it is more useful to express it in the time domain so that it is compatible with the CMT ODE. Applying an inverse Fourier transform as $\tilde a(t) = (1/2\pi) \int A(\zeta) \exp\{j\zeta t\} \mathrm{d}\zeta$, we reach
        \begin{equation}
            \Delta\tilde\omega (\tilde p) = j\xi\frac{1}{\tilde a(t)}\left[ j\omega\rsub{ref} \tilde p(t) + \fullFRAC{\tilde p(t)}{t} \right].
            \label{eq:DeltaOmegaGainTime}
        \end{equation}
        Note  that $\Delta\tilde\omega$ is a function not only of the polarization amplitude $\tilde p(t)$ used to describe the gain process, but also of the corresponding time derivative. This is a distinct characteristic compared to the nonlinear phenomena studied previously, which necessitated a modified approach.
        Furthermore, $\Delta\tilde\omega$ is generally complex, with its imaginary part representing gain, i.e., $\IM{\Delta\tilde\omega}<0$ when light is emitted. Finally, it may possess a nonzero real part to accommodate a possible resonance frequency shift, moving the ``cold'' resonance frequency $\omega_0$ toward the peak emission frequency $\omega_m$, a characteristic quantity of the gain material.

        Having calculated $\Delta\tilde\omega$, we need another equation that describes the polarization field amplitude. We start from a homogeneously broadened Lorentzian oscillator equation that typically describes such a polarization field in gain media,\cite{Siegman}
        \begin{equation}
            \frac{\partial^2\bmcal{P}\OF{r}{t}}{\partial t^2} + \Gamma_m \frac{\partial\bmcal{P}\OF{r}{t}}{\partial t} + \omega_m^2\bmcal{P}\OF{r}{t} = -\sigma_m \Delta N\OF{r}{t} \bmcal{E}\OF{r}{t}.
            \label{eq:Pgeneral}
        \end{equation}
        Parameters $\omega_m$ and $\Gamma_m$ are the central frequency and the linewidth of the atomic transition of the gain medium, respectively, while $\Delta N\OF{r}{t}$ is the carrier population inversion, necessary for light emission. Finally, $\sigma_m$ is a coupling parameter that characterizes the gain medium.\cite{Siegman} Note that the driving term in Eq.~\eqref{eq:Pgeneral} is proportional to $\bmcal{E}\OF{r}{t}$; this justifies the hypothesis that the spatial distribution of the polarization follows that of the electric field. 
        Using the same approximations followed earlier with the Maxwell's equations, it is not difficult to reach\cite{Nousios2023}
        \begin{equation}
            \fullFRAC{\tilde p(t)}{t} + \frac{\omega_m^2-\omega\rsub{ref}^2+j\omega\rsub{ref}\Gamma_m}{\Gamma_m+j2\omega\rsub{ref}}\tilde p(t) = -\frac{\sigma_m}{\Gamma_m+j2\omega\rsub{ref}} \Delta \bar N(t) \tilde a(t),
            \label{eq:Pfinal}
        \end{equation}
        where the population inversion is spatially averaged using the approach of Eq.~\eqref{eq:NcAverage}.

        Equation~\eqref{eq:Pfinal} can describe any gain medium. The introduction of the carrier population inversion in the form of $\Delta\bar N$ was deliberate to allow for the description of any medium with an arbitrary number of energy levels and interactions.\cite{Chua2011,Chua2011a,Chua2014,Rasmussen2017,Hu2019,Nousios2023} For the simplest scenario of a system with two energy levels, an equation similar to \eqref{eq:CarrierRateEquationAveragedSi} can be utilized for the carrier evolution, albeit with an extra term to describe stimulated (coherent) light emission.\cite{Chua2011,Nousios2023} This term is given by\cite{Siegman} $(1/\hbar\omega) \bmcal{E}\OF{r}{t} \cdot [\partial\bmcal{P}\OF{r}{t}/\partial t]$, which comprises full-field quantities. Using the same arguments and additionally utilizing the rotating wave approximation to ignore terms oscillating at $\pm 2\omega\rsub{ref}$ emerging from the dot product,\cite{Chua2011} one can reach the final equation describing population inversion
        \begin{align}
            \fullFRAC{\Delta\bar N(t)}{t} = &\bar R_p(t) -\frac{\Delta\bar N(t)}{\tau_g} \nonumber \\
            &- \frac{\xi_N}{\hbar\omega_m}\frac{1}{2}\RE{\left[j\omega\rsub{ref}\tilde p(t)+\fullFRAC{\tilde p(t)}{t}\right]\tilde a^*(t)}.
            \label{eq:CarrierPopulationInversion}
        \end{align}
        The terms on the right-hand side of Eq.~\eqref{eq:CarrierPopulationInversion} describe (i)~the pumping rate, (ii)~the recombination of radiative carriers with a lifetime $\tau_g$, and (iii)~the stimulated emission process. A notable difference  is that the spatially averaged pumping rate $\bar R_p$ is not related with the mode amplitude $a(t)$ as was the case in FCEs or SA. On the contrary, it is left arbitrary here and can describe either optical\cite{Fang2010,Nousios2023} or electrical pumping.\cite{Romeira2018} Finally, the overlap parameter $\xi_N$ is defined through
        \begin{equation}
            \xi_N = \frac{\displaystyle\iiint_{V_p}|\mathbf{E}\rsub{ref}|^4\mathrm{d}V}{\displaystyle\iiint_{V_p}|\mathbf{E}\rsub{ref}|^2\mathrm{d}V}.
            \label{eq:xiNDefinition}
        \end{equation}

        Equations~\eqref{eq:DeltaOmegaGainTime},~\eqref{eq:Pfinal},~and~\eqref{eq:CarrierPopulationInversion}, together with the cavity amplitude ODE are the minimum number of equations required to describe gain in a resonant cavity assuming the general case of Class~C lasers.\cite{Nousios2023} Other effects, such as spontaneous emission or photobleaching can be incorporated as additional terms on the right-hand side of Eq.~\eqref{eq:CarrierPopulationInversion} and/or with more energy levels and the respective carrier densities in each level.\cite{Chua2011a,Chua2014} On the other hand, the number of required equations can be reduced in media where the polarization instantaneously follows the electric field [i.e., $\mathrm{d}\tilde p/\mathrm{d}t\rightarrow0$ in Class~B lasers] or when the carrier dynamics are fast [i.e., $\mathrm{d}\Delta\bar N/\mathrm{d}t\rightarrow0$ in Class~A lasers], leading to simpler expressions 
        (see for example the supplemental material of Ref.~\onlinecite{Nousios2023}). As a final remark, we stress that the framework can also describe contemporary 2D gain media (such as single layer TMDs or TMDs hetero-bilayers) with only minor modifications. Specifically, the units of polarization and carrier density are affected (per unit surface instead of per unit volume) and the spatial overlap parameters $\xi$ and $\xi_N$ should now be evaluated through the following surface integrals
        \begin{subequations}
            \begin{align}
                \xi_s &= \frac{1}{4}\iint_{S_p} |\mathbf{E}\rsub{ref,\parallel}|^2 \mathrm{d}S, \\
                \xi_{s,N} &= \frac{\displaystyle\iint_{S_p} |\mathbf{E}\rsub{ref,\parallel}|^4 \mathrm{d}S}{\displaystyle\iint_{S_p} |\mathbf{E}\rsub{ref,\parallel}|^2 \mathrm{d}S}.
            \end{align}
            \label{eq:xixiNSurface}
        \end{subequations}
        


%

\end{document}